\documentclass[useAMS,usenatbib]{mn2e}
\usepackage{graphicx}
\usepackage{amssymb, amsmath}
\usepackage{times}
\usepackage{color}
\usepackage{lscape}

\newcommand{\hMpc}{h^{-1}{\rm\;Mpc}}

\newcommand{\sperp}{\Sigma_{\perp}}
\newcommand{\spar}{\Sigma_{\parallel}}

\newcommand{\alphapar}{\alpha_{||}}
\newcommand{\alphaper}{\alpha_{\perp}}

\newcommand{\comment}[1]{\relax}		    

\begin{document}
\title[BAO Fitting in SDSS-III BOSS galaxies] {SDSS-III Baryon Oscillation Spectroscopic Survey: Analysis of Potential Systematics in Fitting of Baryon Acoustic Feature}


\author[Vargas-Magana et al.]{\parbox{\textwidth}{\Large
Mariana Vargas Maga\~na$^{1}$,
Shirley Ho$^{1}$,
Xiaoying Xu$^{1}$,
Ariel G. S\'anchez$^{2}$,
Ross O'Connell$^{1}$,
Daniel J. Eisenstein$^{3}$,
Antonio J. Cuesta$^{4}$,
Will J. Percival$^{5}$,
Ashley J. Ross$^{5}$,
Eric Aubourg$^{6}$,
St\'ephanie Escoffier$^{7}$ 
David Kirkby$^{8}$,
Marc Manera$^{5,9}$,
Donald P. Schneider$^{10,11}$,
Jeremy L. Tinker$^{12}$,
Benjamin A. Weaver$^{12}$,
 } \vspace*{4pt} \\ 
 \scriptsize $^{1}$ Department of Physics, Carnegie Mellon University, 5000 Forbes Avenue, Pittsburgh, PA 15213, USA\vspace*{-2pt} \\
\scriptsize $^{2}$ Max-Planck-Institut f\"ur extraterrestrische Physik, Postfach 1312, Giessenbachstr., 85748 Garching, Germany\vspace*{-2pt} \\ 
\scriptsize $^{3}$ Harvard-Smithsonian Center for Astrophysics, 60 Garden St., Cambridge, MA 02138, USA\vspace*{-2pt} \\ 
\scriptsize $^{4}$ Institut de Ci\`encies del Cosmos, Universitat de Barcelona, IEEC-UB, Mart\'i Franqu\`es 1, E08028 Barcelona, Spain\vspace*{-2pt} \\ 
 \scriptsize $^{5}$ Institute of Cosmology \& Gravitation, Dennis Sciama Building, University of Portsmouth, Portsmouth, PO1 3FX, UK\vspace*{-2pt} \\   
 \scriptsize $^{6}$ APC, Astroparticule et Cosmologie, Universit\'e Paris Diderot, CNRS/IN2P3, CEA/Irfu, Observatoire de Paris, Sorbonne Paris Cit\'e, 10, rue Alice Domon \& L\'eonie Duquet, 75205 Paris Cedex 13, France\vspace*{-2pt} \\  
 \scriptsize $^{7}$ CPPM, Aix-Marseille Universit\'e, CNRS/IN2P3, Marseille, France \vspace*{-2pt}\\
  \scriptsize $^{8}$ Department of Physics and Astronomy, UC Irvine, 4129 Frederick Reines Hall, Irvine, CA 92697, USA\vspace*{-2pt} \\  
 \scriptsize $^{9}$ Department of Physics \& Astronomy, University College London,Gower Street, London, WC1E 6BT,  UK\vspace*{-2pt} \\ 
  \scriptsize $^{10}$ Department of Astronomy and Astrophysics, The Pennsylvania State University, University Park, PA 16802, USA\vspace*{-2pt} \\ 
 \scriptsize $^{11}$ Institute for Gravitation and the Cosmos, The Pennsylvania State University, University Park, PA 16802, USA\vspace*{-2pt} \\
\scriptsize $^{12}$ Center for Cosmology and Particle Physics, New York University, New York, NY 10003, USA\vspace*{-2pt} \\ 
}

\date{\today} 
\pagerange{\pageref{firstpage}--\pageref{lastpage}} \pubyear{2013}
\maketitle
\label{firstpage}

\begin{abstract}

Extraction of the Baryon Acoustic Oscillations (BAO) to percent level accuracy is challenging and demands an understanding of many potential systematic to an accuracy well below 1 per cent, in order ensure that they do not combine significantly when compared to statistical error of the BAO measurement.
In this paper, we analyze the potential systematics in BAO fitting methodology. 
We use the data and mock galaxy catalogs from Sloan Digital Sky Survey (SDSS)-III Baryon Oscillation Spectroscopic Survey (BOSS) included in the SDSS Data Release Ten and Eleven (DR10 and DR11) to test the robustness of various fitting methodologies. 
In DR11 BOSS reaches a distance measurement with $\sim 1\%$ statistical error and this prompts an extensive search for all possible sub-percent level systematic errors which could be safely ignored previously. 
In particular, we concentrate on the fitting of anisotropic clustering, following multipole methodology from Xu et al. 2012 as the fiducial methodology. 
We demonstrate the robustness of the fiducial multipole fitting methodology to be at $0.1\%-0.2\%$ level with a wide range of 
tests using mock galaxy catalogs in both DR10 and DR11 pre and post-reconstruction.
We also find the DR10 and DR11 data from BOSS to be robust against changes in methodology at similar level. This systematic error budget is incorporated into the the error budget of Baryon Oscillation Spectroscopic Survey (BOSS) DR10 and DR11 BAO measurements.
Of the wide range of changes we have investigated, we find that when fitting pre-reconstructed data or mocks, the following changes have the largest effect on the best fit values of distance measurements both parallel and perpendicular to the line of sight: (a) Changes in  non-linear 
correlation function template; (b) Changes in fitting range of the correlation function; (c) Changes to the non-linear damping model parameters. 
The priors applied do not matter in the estimates of the fitted errors as long as we restrict ourselves to physically meaningful 
fitting regions.
Finally, we compare an alternative methodology denoted as Clustering Wedges 
with  Multipoles, and find that they are consistent with each other.

\end{abstract}


%
%

\maketitle

\section{Introduction}\label{sec:intro}

The baryon acoustic oscillations (BAO) method has been proven to be a powerful geometrical probe to study the expansion history of the universe. 
The BAO provide us a characteristic scale that can be used as an standard ruler for measuring its apparent size at certain redshift and comparing it with the physical size we know from first principles. The measurement of this standard ruler at different redshifts enables us to map the expansion history of the universe.

Furthermore,  measuring the BAO feature along the line of sight (LOS) and in the perpendicular directions constrains the Hubble parameter $H(z)$ and the angular diameter distance $D_A(z)$ at redshift $z$ separately. 
Because of the low signal to noise ratio of first large scale surveys, most of the previous studies have focused on the spherically averaged  analysis yielding measurement of the spherical average distance $D_{\rm V}(z)=((1+z)^2 D_{\rm A}(z)^{2/3}(cz/H(z))^{1/3} $, which have a strong degeneracy between $D_A(z)$ and $H(z)$. 
The two dimensional analysis enables us to break the degeneracy between $D_A(z)$ and $H(z)$ as it measures the clustering in different directions (i.e. along the line-of-sight [LOS] and perpendicular to the LOS) \citep{AP}. 

Early works on anisotropic clustering were performed with SDSS-II data. 
However, given the relative low redshift of this samples, the constraints were similar to those issues from isotropic analysis, as at redshift $z \rightarrow 0$ distances are degenerate. 
Different methodologies for fitting BAO have been explored over last few years. For example, \cite{Oku08}  proposed fitting the radial and transverse correlation functions,  and \cite{{PadWhi09}}
 proposed fitting the multipoles directly. More recently, \cite{Kaz12}
 suggest splitting the full correlation function based on the angle of the pair to the LOS, resulting in a correlation function in each of two angular wedges.

Based on the multipoles methodology, 
\cite{{Xeaip}} 
studied extensively the fitting procedure focusing on monopole and isotropic shifts.
In \cite{Xu12b},  this methodology was extended to anisotropic clustering including the effect of reconstruction on the anisotropic BAO signal as well as the application to SDSS-Data Release 7 for cosmological constraints. 

With subsequent surveys such as SDSS-III Baryon Oscillation Spectroscopic Survey (BOSS) Data Release 9 analysis \citep{And13}, 
two different fitting methods have been applied \cite {Xu12b}, \cite{Kaz12}, \cite{Kaz13}, producing consistent cosmological constraints on $D_A(z)$ and $H(z)$.
As the survey volume increases, and thus the precision of the BAO measurement, we extend previous studies by \cite {Xu12b} on multipoles fitting. In particular, BOSS has doubled its survey volume from Data Release 9 to Data Release 11, requiring much higher precision and understanding of the systematic error as the statistical error shrinks.  In particular, we analyze mock galaxy catalogs and data  from BOSS DR10 and DR11, and determine the effects of various choices in our fitting methodology on the final fitting result. We can then determine the systematic error budget from fitting methodologies that is included in the BOSS DR10 and DR11 BAO measurements in \citet{Aar13}. 

The layout of our paper is as follows. We introduce the anisotropic analysis techniques in Section~\ref{sec:anisosec}. In Section~\ref{sec:analysis}, we describe our methodology used in the correlation function analysis, the covariance matrices estimation, the simulations and the reconstruction procedure. In Section~\ref{sec:Fitting},  we present the fiducial fitting procedure and in Section~\ref{sec:sysfit},  we describe the systematics tests performed on the fitting of the BAO anisotropic clustering signal. Section~\ref{sec:resultsmocks} presents the results of our systematic tests using Sloan Digital Sky Survey III- Baryonic Oscillation Sky Survey Data Release 10 and Data Release 11, CMASS mocks galaxy catalogs. In Section~\ref{sec:other_method}, we compare the fitting results with results using other methodologies. In Section~\ref{sec:data},  we explore the consequences of the systematics in the fitting to the BOSS DR11 CMASS data before and after the application of reconstruction. Finally, we conclude and discuss our results in Section~\ref{sec:discussion}.

\section{Anisotropic Clustering Methods}\label{sec:anisosec}

\subsection{Parametrization}  \label{sec:parametrization} 

The angular-averaged clustering analysis assumes that the clustering is isotropic, and the BAO feature is shifted in an isotropic manner if we consider an incorrect cosmology. Any deviation from the true cosmology is parametrized by an isotropy shift $\alpha$: 
\begin{equation}
\alpha=\frac{(D_v/r_s)}{(D_v/r_s)_{fid}}
\end{equation}
where the distance is quoted relative to the sound horizon $r_s$ at the drag epoch. 
The angle average analysis, extensively used in galaxy-clustering analyses, has provided important constraints in the angle averaged distance.
However, as the clustering of galaxies is not isotropic, to optimize the extraction of information of BAO, we must perform an anisotropic analysis.  
There are two sources of anisotropies: the redshift distortions (RSD) and the anisotropies generated from assuming an incorrect cosmology.

The anisotropies arising from redshift distortions can be separated depending on the scale. At small scale, the peculiar velocities generate the Fingers Of God (FoG) effect at large scales,  the coherent flows towards over-dense regions generate Kaiser Effect \citep{Kai87}. Both cases uniquely affect the LOS separations generating a smooth change with scale. 
The second source of anisotropy arises from assuming an incorrect cosmology  via the Alcock-Paczynski test (AP). As $D_A(z)$ and $H(z)$ depend differently on cosmology, computing incorrectly the separations generates artificial anisotropies in the clustering along LOS and perpendicular directions \citep{Xu12b}. 

To distinguish anisotropy due to RSD from the Alcock-Paczynski effect due to wrongly assumed cosmological model, we consider simple RSD models.
We will present in the fitting model section the details of the model for RSD. Even if the simple models are not suffiently accurate to modulate redshift distortions, any residual from inadequate matching with the models and  broadband shape data could be compensated by additional marginalization terms.
 
For analyzing the anisotropic BAO signal, we need a model with a parametrization of the anisotropic signal.There are in the literature different ways of parametrizing the anisotropy in the BAO signal.
In Xu et al. (2013), and Anderson et al. (2013), the anisotropic signal is parametrized by 
$\alpha$  for isotropic dilation and $\epsilon$ for the anisotropic warping between true and fiducial cosmology. 
\begin{equation}
1+\epsilon=\left [\frac{H^{fid}(z) D_{A}^{fid}(z)}{H(z) D_A(z)}\right]^{1/3}
\end{equation}
Since we include in the model the anisotropy produced by RSD, $\epsilon$ parametrizes the amount of Alcock-Paczynski anisotropy. 

An alternative parametrization considers the shift parallel to the LOS ($\alpha_{||}$) and the shift  perpendicular to the LOS ($\alpha_{\perp}$). The constraints in $\alpha_{||}$ and $\alpha_\perp$, as well  as in  $ \alpha$ and $\epsilon$, translate to constraints in $D_A(z)$, and $H(z)$ separately.
\begin{equation}
\begin{array}{ll}
\alpha_{\perp}=&\frac{D_A(z)r_s^{fid}}{D_A^{fid}(z) r_s}\\
\alpha_{||}=&\frac{H(z)^{fid}r_s^{fid}}{H(z) r_s}\\
\end{array}
\end{equation}
The relations between both parametrizations are given by:
\begin{equation}\label{ae2parper}
\begin{array}{ll}
\alpha=\alpha_\perp^{2/3}\alpha_{||}^{1/3}\\
1+\epsilon=\left (\frac{\alpha_{||}}{\alpha_{\perp}}\right)^{1/3}
\end{array}
\end{equation}

\subsection{Clustering Estimators}
Measuring both $D_A$ and $H$ requires an estimator of the 2D correlation function $\xi(s,\mu)$, where $s$ is the separation between two galaxies of the angle between $s$ and the line of sight.  Working with the full 2D correlation function is not practical in the case of galaxy-clustering as  we estimate our covariance matrix directly from the sample covariance of mock catalogs. To calculate the covariance matrix for a full 2D correlation function it will requires a much larger number of mock catalogs. We therefore compress our 2D correlation function into a small number of angular moments and use these for our analysis. 

In particular, we will describe the following two clustering estimators: Multipoles \citep{Xu12b}  and Wedges \citep{Kaz13}. 
As \cite{Kaz13} has discussed extensively the systematics of fitting using Wedges, this paper will concentrate mostly on the Multipoles method \citep{Xu12b}, but will include comparisons with the Wedges method.  

\subsubsection{Multipoles} \label{sec:multipoles_description}

The formalism for the 2D-correlation function in terms of the multipole analysis is detailed in \cite{Xu12b} and  \cite{And13}. 
We will only briefly summarize the methodologies as a reminder to the readers. 

We start with the Legendre moments of the 2D correlation function: 
\begin{equation} 
\xi_l(r) = \frac{2l+1}{2} \int_{-1}^{+1} d\mu \xi(r,\mu) L_{\ell}(\mu) 
\end{equation} 
where $L_{\ell}(\mu)$ is the $\ell$th Legendre polynomial. 

For the multipole analysis, we focus only on the monopole and quadrupole. 
\footnote{The multipoles analysis focuses in the monopole and quadrupole; even if the higher order also provides information, their influence is quite negligible.}.
We refer the readers to \cite{And13} for more details.

\subsubsection{Clustering Wedges} \label{sec:wedges_description} 

We briefly review the alternate clustering estimator: clustering wedges, as we will later present comparisons between the Multipoles estimator and the clustering wedges \citep{Kaz12}: 
\begin{equation} 
\xi_{\Delta_\mu}(r) = \frac{1}{\Delta \mu} \int_{\mu_{min}}^{\mu_{min}+ \Delta \mu} d\mu \xi(r,\mu)
\end{equation} 
In our analysis of the comparison between the clustering wedges and the multipoles, we choose $\Delta \mu = 0.5$ such that we have a basis composed of a ``radial'' component $\xi_{||} (s) \equiv \xi(s, \mu \textgreater 0.5)$, and a ``transverse'' component $\xi_{\perp} (s) \equiv \xi(s, \mu \textless 0.5)$. A full description of the method and systematics tests can be found in \cite{Kaz13}.

\section{Analysis}\label{sec:analysis}

\subsection{Fiducial cosmology} \label{sec:fid_cosmo}

Throughout  we assume a fiducial $\Lambda$CDM cosmology with $\Omega_M=0.274$, $\Omega_b = 0.0457$, $h = 0.7$ and $n_s = 0.95$. The angular diameter distance to $z = 0.57$ for our fiducial cosmology is $D_A(0.57) = 1359.72$ Mpc, while the Hubble parameter is $H(0.57) = 93.56$ kms$^{-1}$Mpc$^{-1}$. The sound horizon for this cosmology is $r_s =153.19$ Mpc, where we adopt the conventions in \cite{Eis98}. 

\subsection{Measuring Correlation Function }\label{sec:corr_func} 

In anisotropic clustering analysis, we compute the 2-D correlation function 
decomposing the separation $r$ between two galaxies into the parallel $r$ and the perpendicular $r$ direction to the LOS.

\begin{equation}\label{eq:r}
r^2=r^2_{||}+r^2_{\perp}
\end{equation}
where $\theta$ is the angle  between the galaxy pair separation and the LOS direction; we define: 
\begin{equation}\label{eq:mu}
\mu=cos \theta = \frac{r_{||}}{r}
\end{equation}
The 2D-correlation function $\xi(r,\mu)$ (for the pre-reconstructed case) is computed using Landy-Szalay \citep{LanSza93} estimator:
\begin{equation}
\xi(r, \mu)=\frac{DD(r, \mu)-2*RD(r, \mu) +RR(r, \mu)}{RR(r, \mu)}
\end{equation}
where $DD(r,\mu), RR(r,\mu)$, and $RD(r, \mu)$ are the number of pairs of galaxies which are separated by a radial separation $r$ and angular separation $\mu$ from data-data samples, random-random and data-random samples, respectively.  The correlation function is computed in bins of $\Delta r = 4,8h^{-1}$Mpc and $\Delta \mu$= 0.01.

\subsection{Data} \label{sec:cmass} 

\begin{figure}
  \centering
\includegraphics[width=84mm]{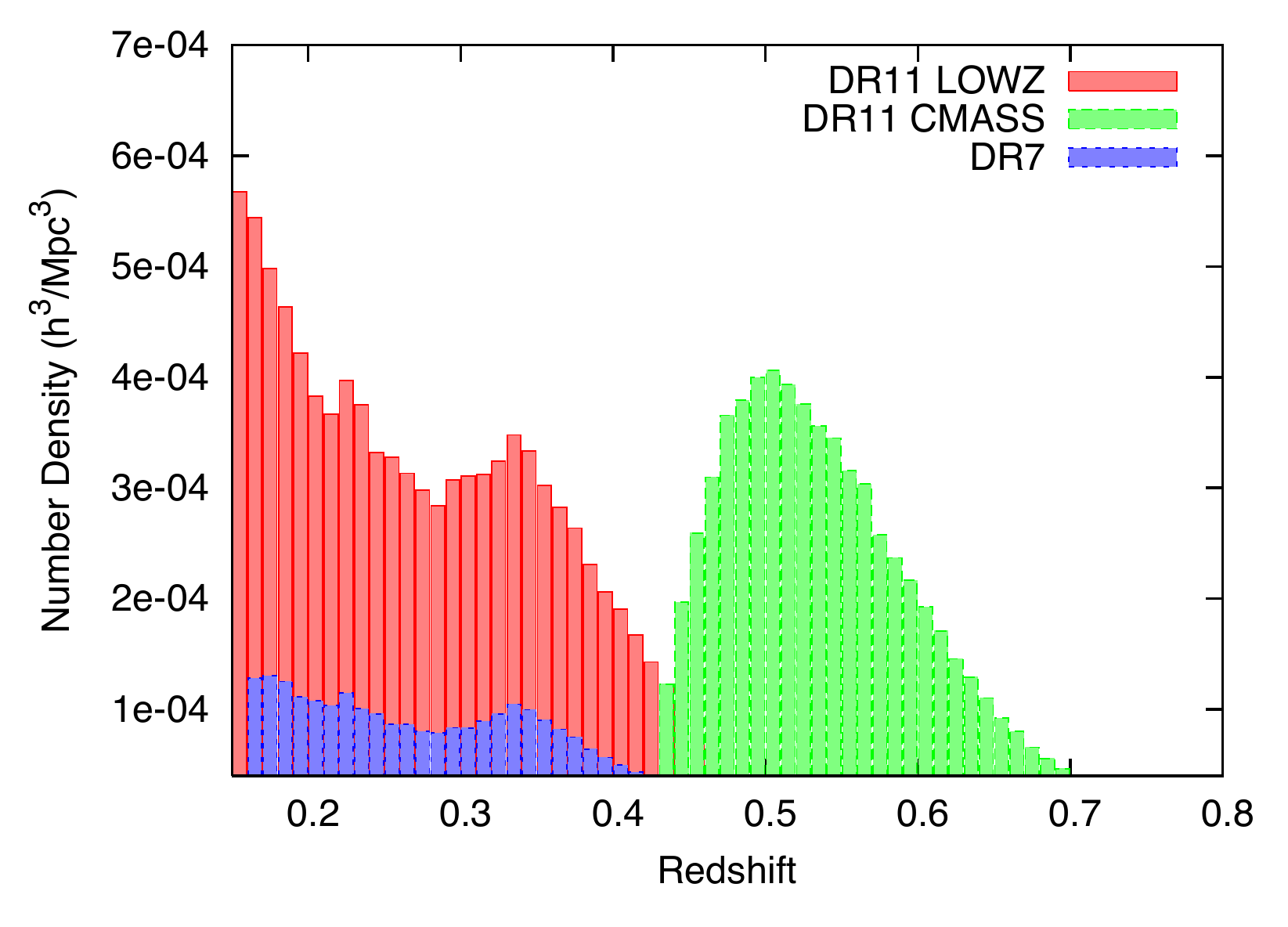}
\caption{Histograms of the galaxy number density as a function of
  redshift for LOWZ (red) and CMASS (green) samples. We analyze the CMASS sample in this paper.  
  We also display the number density of the SDSS-II DR7 LRG sample in
  order to illustrate the increase in sample size for BOSS at $z <
  0.43$.  }
\label{fig:redshift}
\end{figure}

We use data included in Data Releases 10 (DR10) and 11 (DR11) of the
Sloan Digital Sky Survey \citep[SDSS;][]{Yor00}. Together, SDSS I-II
\citep{Abazajian09}, and III \citep{Eis11} used a drift-scanning
mosaic CCD camera \citep{Gun98} to image over one third of the sky
($14\,055$ square degrees) in five photometric bandpasses
\citep{Fuk96,Smi02,Doi10} to a limiting magnitude of $r\simeq 22.5$
using the dedicated 2.5-m Sloan Telescope \citep{Gun06} located at
Apache Point Observatory in New Mexico. The imaging data were
processed through a series of pipelines that perform astrometric
calibration \citep{Pie03}, photometric reduction \citep{Photo}, and
photometric calibration. All of the imaging data were re-processed as part
of SDSS Data Release 8 \citep{DR8}.

BOSS is designed to obtain spectra and redshifts for 1.35 million
galaxies over an extragalactic footprint covering $10\,000$ square
degrees. These galaxies are selected from the SDSS DR8 imaging and are
being observed together with $200\,000$ quasars and approximately
$100\,000$ ancillary targets.  The targets are assigned to tiles of
diameter $3^\circ$ using a tiling algorithm that is adaptive to the
density of targets on the sky \citep{Tiling}. Spectra are obtained
using the double-armed BOSS spectrographs \citep{Smee13}. Each
observation is performed in a series of $900$-second exposures,
integrating until a minimum signal-to-noise ratio is achieved for the
faint galaxy targets. This requirement ensures a homogeneous data set with a 
redshift completeness of more than $97$ per cent over the full survey
footprint. Redshifts are extracted from the spectra using the methods
described in \cite{Bolton12}. A summary of the survey design appears
in \citet{Eis11}, a full description is provided in \citet{Daw12}.

\subsection{Simulations} \label{sec:sims}

In this paper, we use SDSS III-BOSS PTHalos mock galaxy catalogs
\citep{Man12, Man13}
exclusively to test both the systematics
of BAO fitting, and to generate the sample covariance matrices.
Inspired by the methodology in \citep{ScoShe02}, the mocks
are based on Second order Lagrangian Perturbation Theory (2LPT) for
the density fields. The PTHalos mocks galaxy   catalogues were
generated at z = 0.55, in boxes of size L = 2400 $h^{-1}$Mpc using $1280^3$
dark matter particles. The halos were found using a friends-of-friends
algorithm with an appropriate linking length and masses calibrated
with N-body simulations. For populating the halos with galaxies, the
Halo Occupation Distribution prescription was used, previously
calibrated to match the observed clustering at small scales [30,80]
$h^{-1}$ Mpc following  \cite{Whi11}. The angular and radial masks
from DR10/DR11 were applied to subsample the galaxies from their
original boxes. The mocks include redshift space distortions, but do
not include other systematic corrections such as stellar correlation,
or evolution with redshift. A full description of the PTHalos galaxy
mocks could be found in \citep{Man12, Man13}.

\subsection{Covariance} \label{sec:covariance}

The sample covariance is computed with the 600 mocks, as follows:
\begin{equation}
C_{i,j}=\frac{1}{n_{s}-1}\Sigma_{m=1}^{n_s} \left (\bar{\xi}_{{[2s]}_i}-\xi_{{[2s]}_i}^m \right )\left (\bar{\xi}_{{[2s]}_j}-\xi_{{[2s]}_j}^m \right )
\end{equation}
where, $\xi_{2s}=[\xi_0, \xi_2]$.\\
We rescale the inverse covariance matrix, unless otherwise noted, following \citep{Hartlap2007}

\begin{equation}
C^{-1}=C_s^{-1}\frac{n_{s}-n_b-2}{n_{s}-1}
\end{equation}
where $n_{s}$ is the number of simulations, and $n_{b}$ is the number of bins of parameters we are estimating\footnote{For example, in power-spectrum, this quantity would be the number of the binned power-spectra.}. This correction arises from the fact that the inverse of the sample covariance matrix computed from a finite number of mocks is a biased estimator of the inverse covariance matrix. A recent analysis of the covariance corrections 
extends this discussion and provides a prescription to propagate the correction to the inferred parameters \citep{Per13}. A complete section in our paper (Section~\ref{sec:cov_corr}) is devoted to summarize the corrections 
applied to the covariance and the sigmas inferred following \cite{Per13}, and their consequences on the final anisotropic fits.

\subsection{Reconstruction}\label{sec:reconstruction}

We include in this analysis the effect of reconstruction of the density field in the correlation function fitting procedure.
The reconstruction algorithm has proved to be effective in  correcting the effects of nonlinear evolution, thus increasing the statistical sensitivity of measurements \citep{Noh09, Pad12} 

The main idea of reconstruction is to use information encoded in the density field to estimate the displacement field and use this displacement field to reverse the peculiar motion of the particles and partially remove the effect of the nonlinear growth of structure. 
This is possible because  the nonlinear evolution of the density field is dominated by the infall velocities, and these bulk flows are approach by the same structures observed in the density field. The algorithm used here is described in \cite{Pad12}; the particular implementation is the same as in \cite{Pad12, And12, Aar13} and where detailed descriptions could be found. 
To summarize, the effects of  reconstruction on the correlation functions are the sharpening of the peak in the monopole and a decrease in amplitude of the quadrupole. At large scales, the quadrupole approaches zero. 
If reconstruction were perfect, the quadrupole will go to zero and the isotropy in the  2-point correlation will be restored for a correct cosmology.

\section{Fitting Methodology}\label{sec:Fitting}
In this section, we will describe the various components of our anisotropic clustering fitting methodology in detail. 
This subject has been discussed extensively in \cite{Xu12b}, we still detail the methodology here as we will test each component and assumption using BOSS mocks for the CMASS galaxy catalog. We will also test the effect of each of these components on the fit to the BOSS CMASS data. 
In particular, we will detail the model of the correlation function, the fitting methodology comparing the model and the data, various model parameter selection and priors included in the fitting procedure in this section. 

\subsection{ Model for the Correlation Function}\label{sec:NLM}

In order to extract the cosmological information from the BOSS data, a careful modeling of the correlation function is required.
We start with a nonlinear power-spectrum template in 2D:  
\begin{equation}
P(k, \mu) =(1+\beta \mu^2 )^2 F(k, \mu, \Sigma_s)P_{NL} (k, \mu)
\end{equation}
where the term $(1+\beta \mu^2)$  corresponds to the Kaiser model \citep{Kai87}  for large scale redshift space distortions, which produces anisotropy in an otherwise isotropic 2D correlation function.\\
$F(k, \mu, \Sigma_s)$ is the streaming model for the Finger-of-God \citep{PeacockDodds1994}   given by: 

\begin{equation}
\label{eq:streaming}
F(k, \mu, \Sigma_s)=\frac{1}{(1+k^2\mu^2 \Sigma_s^2)}
\end{equation}
where $\Sigma_s$ is the streaming scale, which we set to $1.4 h^{-1}$Mpc, and the $P_{NL}(k)$ is the non linear power spectrum.

In this work, we consider two templates for the non-linear power-spectrum $P_{NL}(k)$: the ``De-Wiggled" power spectrum $P_{dw}$  \citep{Xeaip, Xu12b, And12, And13}
and a template {\it inspired} by Renormalized Perturbation Theory (RPT), $P_{pt}$, also used in several galaxy clustering analyses \citep{Kaz10, Kaz12, And13}.  We will describe these two templates in more detail in Section~\ref{sec:dewiggled} and Section~\ref{sec:rpt}. 

We decompose the full 2D power-spectrum into its Legendre moments:
\begin{equation}
P_{\ell,t}(k)=\frac{2\ell+1}{2}\int_{-1}^{1}P_t(k, \mu)L_l(\mu)d\mu
\end{equation}
which can then be transformed to configuration space using

\begin{equation}
\label{eq:finalcorr} 
\xi_{\ell,t}(r) = i^\ell \int \frac{k^3 d log(k)}{2 \pi^2}P_{l,t} j_\ell(kr)
\end{equation}
where, $j_\ell(kr)$ is the $\ell$-th spherical Bessel function and $L_\ell(\mu)$ is the $\ell$-th Legendre polynomial.

\subsubsection{De-Wiggled Template} \label{sec:dewiggled}

The De-Wiggled template is a non-linear power spectrum prescription widely used in clustering analysis \citep{Xeaip,Xu12b,And12, And13, BCBL}.
This phenomenological prescription takes a linear power spectrum template to which we add the nonlinear growth of structure.
The De-Wiggled power spectrum is defined as:
\begin{equation}
\begin{array}{ll}
P_{\rm{dw}}(k, \mu)=&[P_{\rm{lin}}(k) -P_{\rm{nw}}(k)] \\
&\times \rm{\exp}\left[ -(k^2 \mu^2 \Sigma_{||}^2+k^2(1-\mu^2)\Sigma_{\perp}^2)/2 \right ] +P_{nw} 
\end{array}
\end{equation}
where $P_{lin}(k)$ is the linear theory power spectrum and $P_{nw}(k)$ is a power spectrum without the acoustic oscillations (\cite{Eis98}). $\Sigma_{||}$ and $\Sigma_\perp$ are the radial and transverse components of the standard Gaussian damping of BAO $\Sigma_{NL}$, 
\begin{equation}
\label{eqn:SigmaNL}
\Sigma_{NL}^2=(\Sigma_{||}^2+\Sigma_\perp^2)/2
\end{equation}
 $\Sigma_{NL}$ models the degradation of signal due nonlinear structure growth (\cite{EisSeoWhi07}).

We can consider the different parts of the model in monopole and quadrupole: 
\begin{itemize}
\item linear model, 
\item linear model+ Kaiser effect, 
\item linear model + nonlinear growth of structure and Kaiser effect, 
\item full nonlinear model including the Kaiser effect and FoG effect + Nonlinear growth of structure. 
\end{itemize}

The Kaiser effect produces a bump at $\approx 110 h^{-1}$Mpc in i.e. quadrupole. The quadrupole is much weaker in the full non-linear model when we include the Kaiser effect, the FoG effect and the nonlinear growth of structure when compared to  "Linear model + Kaiser effect". 
$\Sigma_{NL}$ and Finger-of-God effect introduce extra structures in the quadrupole.  A dip appears at the center and thus creating a crest-trough-crest structure.

\subsubsection{RPT Inspired Template}  \label{sec:rpt}

Several BAO galaxy clustering analyses  \citep{Beutler11, Kaz12, Sanchez13}  considered, instead of the De-Wiggled template, a template inspired by Renormalized Perturbation Theory. 
The main argument for this choice is that a RPT template provides an unbiased measurement of the dark energy equation of state \citep{CroSco08, Sanchez08}. 

However, the template described here is only ``inspired by RPT'', but its form is not the functional form one would obtain from RPT. 
The RPT-inspired template is given by \citet{Kaz13}:
\begin{equation}
\label{eqn:RPT}
 P_{\rm pt}(k)=P_{\rm lin}(k) e^{-k^2\sigma_v^2} + A_{\rm MC} P_{\rm MC} (k) 
\end{equation}
where $P_{\rm MC}$ is given by:
\begin{equation}
P_{\rm MC}=\frac{1}{4 \pi ^3} \int dq |F(k-q,q)|^2 P_{lin}(|k-q|)P_{lin}(q)
\end{equation}
The template of equation~\ref{eqn:RPT} is different from the one used in \citet{And13}. In particular, \citet{And13} has used the De-Wiggled Template, while we use $P_{pt}(k)$ in \citet{Aar13}. 
It was previously used by \citet{Kaz13}  in the analysis of the CMASS DR9 multipoles and
clustering wedges and is described in detail in \citet{Sanchez13}. 
The parameter $\sigma_v$ accounts for the damping of
the baryonic acoustic feature by non-linear evolution and $A_{MC}$ for the
induced coupling between Fourier modes.
We fit to the mocks with these parameters free and use the mean value of
the best-fits pre-reconstruction and post-reconstruction. 
In particular, $\sigma_v$ is  fixed to 4.85 (1.9) $h^{-1}$Mpc and $A_{\rm MC}$ is  fixed to $1.7$ ($0.05$) reconstruction.

For the post-reconstruction template we expect a significantly sharpening peak thus the value of $\sigma_{v}$ is set to $1.9 h^{-1}$Mpc that corresponds to the linear power spectrum. 
The term $A_{MC}$ retains the same value as in the pre-reconstruction case.  
The ideal scenario suggests that 
 post-reconstruction $A_{MC}=0$ but the test performed in \cite{Kaz13} indicates that this choice induces a bias of 0.7-1\% in $czH^{-1}r_s^{-1}$ (compare to 0.5\% when using the same value for $A_{MC}$ pre-reconstruction). Details about this template can be found in \cite{CroSco08, Sanchez08, Kaz13}
Finally, $\Sigma_s$ is set to  $3.64168h^{-1}$Mpc and $\beta$ is set to 0.39728. 
In the post-reconstruction case, $P(k,\mu)$ is assumed to be  isotropic, i.e. $P(k,\mu) = P_{NL}(k)$ 
(which corresponds to setting $\Sigma_s = \beta = 0$). In other words, the quadrupole is set to 0 ($\xi_2(s) = 0$) in the post-reconstruction template. 

\subsection{Overall Fitting Methodology} \label{sec:fitting_methodology_details}

We now take the moments of model correlation function as described in equation~\ref{eq:finalcorr} and synthesize the 2D correlation function by: 
\begin{equation}
\xi(r,\mu) = \sum_{\ell=0}^{\ell_{\rm max}} \xi_{\ell}(r) L_{\ell}(\mu) 
\end{equation} 
In this work, we actually truncate this sum at $\ell_{\rm max} = 4$. 

We now need to map the 2D correlation function to the data, i.e., mapping the observed $\xi(r,\mu)$ for each $(r,\mu)_{\rm fid} $ pair to their true $\xi(r,\mu)_{\rm true}$. 
This transformation can be compactly written by working in the transverse ($r_{\perp}$) and radial ($r_{\parallel}$) separations defined by equations~\ref{eq:r} and ~\ref{eq:mu}. 
We now have $r_{\perp} = \alpha_{\perp} r_{\perp,\rm obs}$ and $r_{||} = \alpha_{||} r_{||,\rm obs}$. 
Further details can be found in \cite{Xu12b}. 
We will now be able to compute  $\xi_{\rm obs}(r,\mu)$ from the data and project them into multipole basis for comparison to the 
model $\xi_{0,2}^{\rm obs} (r)$.

Finally, we include nuisance parameters to absorb the imperfect modeling of our broad-band model due to both mismatches in cosmology, uncertain theoretical modeling or potential smooth systematic effects. 
In particular, the multipole fits are performed simultaneously over the monopole and quadrupole for 10 parameters with four nonlinear parameters $log(B_0^2)$, $\beta$, $\alpha$, $\epsilon$ and  six linear nuisance parameters $A_l(r)$: 

\begin{equation}
\label{eqn:nuisance}
\begin{array}{ll}
\xi_{0}(r)=&B_0^2 \xi_0^t(r, \alpha, \epsilon)+A_0(r)\\
\xi_2(r)=&\xi_2^t(r, \alpha, \epsilon)+A_2(r)
\end{array}
\end{equation}

The monopole $\xi_{0}^t(r)$ and quadrupole $\xi_{2}^t(r)$ templates are estimated in a fiducial cosmology following the model described in Section~\ref{sec:NLM}. The $\alpha$ parameter measures the position of the peak of the data relative to the model, and  $\epsilon$ measures the degree to which it is anisotropic.

The nuisance terms $A_l$ are included for marginalizing the broad band effects : 
\begin{equation}
A_l(r)=\frac{a_{l,1}}{r^2}+\frac{a_{l,2}}{r}+a_{l,3}; l=0,2
\end{equation}

The quantity $B_0^2 $ is a bias-like term that  adjusts amplitude of monopole template $\xi_0^t$. 
Before fitting, $B_0$ is inferred from the multiplicative offset $B_0^2$ between the model and the measured correlation function at $r=50 h^{-1}Mpc$. This offset is then used to normalize the monopole and quadrupole. 

This procedure ensures $B_0\sim 1$ and $\beta$ is allowed to vary as it effectively allows the amplitude of the quadruple to change. The fits are performed over $\log(B_0^2)$ using a Gaussian prior with standard deviation of $0.4$ to prevent unphysical negative values. 

The best-fit values of $\alpha$ and $\epsilon$ are obtained from minimizing the $\chi^2$ given by:
\begin{equation}
\chi^2(p)=(m(p)-d)^TC^{-1}(m(p)-d)
\end{equation}
where $m(p)$ corresponds to the model vector for the $\xi_0$ and $\xi_2$ given the parameters $p$, $d$ is the data respective vector and $C$ is the covariance described in the Section~\ref{sec:covariance}. For a detailed description, see \cite{Xeaip, Xu12b}. 

\subsection{Fiducial Fitting Methodology Model Parameters}\label{sec:fiducial}

Since we will be investigating the effects of changing each of the model parameters, assumptions and prior choices in our fitting methodology, we  list the model parameters choices, assumptions and prior choices explicitly for the convenience of the reader.

In our fiducial fitting methodology we use the following model:
\begin{enumerate}
\item De-Wiggled template
\item $\alpha$-$\epsilon$ parametrization
\item 8 $h^{-1}$ Mpc binning. The binning scheme of 8 $h^{-1}$ Mpc binning differs from 4 $h^{-1}$ Mpc binning of previous analyses \citep{Xeaip, Xu12b}. This choice is based on the recent work on covariance matrix systematics performed by \cite{Per13}.
\item Fitting range, x=[46, 200]  $h^{-1}$Mpc, corresponding to 20 bins for each multipole
\item Nuisance terms: 3-term $A_l(r)$
\end{enumerate}

We apply the following priors on various model parameters and further discuss the motivation for each prior in Section~\ref{sec:priors}
\begin{enumerate}
\item Prior on $\log(B_0^2)$ centered on 1, with standard deviation of 0.4 to prevent $B_0$ wandering too far from 1.
\item Prior on $\beta$ centered at $f/b\sim\Omega_m ^{0.55}/b=0.4$, with a standard deviation of 0.2. This prior serves to limit the model from selection ing unphysical values of $\beta$. The central value is set to zero after reconstruction as we expect reconstruction to remove large scale redshift distortions. The central value of  $\beta$  is chosen to be $0.4$ in this analysis. This value differs from fiducial case of \cite{And13} which adopted a value of $\beta_c=0.6$. 
\item Prior on $1+\epsilon$ centered on zero, 15\% top hat prior (10\% Gaussian prior), limiting noise  from dragging $\epsilon$ to unrealistic values. The cosmological implications of this prior were also tested in \cite{Xu12b}, they estimate that the epsilon distribution is nearly Gaussian with standard deviation 0.026.  
\end{enumerate}

We also fix the following parameters values:
\begin{enumerate}
\item Streaming scale from equation~\ref{eq:streaming}:  $\Sigma_s=1.4 h^{-1} $Mpc.
\item Non linear damping before reconstruction: $\Sigma_\perp = 6h^{-1}$Mpc and $\Sigma_{||} = 11h^{-1}$Mpc . 
\item Non linear damping after reconstruction.: $\Sigma_{\perp} =\Sigma_{||} =3h^{-1}$Mpc. These $\Sigma_{\perp, ||}$ values used in  pre and post reconstruction were all fit from the average of the mocks in DR9 and we do not expect them to change drastically for DR10. 
\end{enumerate}

\subsection{Sigma Estimation}\label{sec:sigma_estimation} 

To estimate the errors, we calculate the probability distribution $p(\alpha, \epsilon)$ in a grid ($\alpha, \epsilon$). For each grid point ($\alpha,\epsilon$), we fit the remaining parameters using the best fit from $\chi^2$. Assuming the likelihood is Gaussian $p(\alpha, \epsilon) \propto \exp (-\chi(p)^2/2)$ and using the corresponding normalization:
\begin{equation}
\begin{array}{c}
p(\alpha)=\int p(\alpha, \epsilon) d\epsilon\\
p(\epsilon)=\int p(\alpha, \epsilon) d\alpha
\end{array}
\end{equation}
Under the hypothesis of Gaussian posteriors we can  take the widths of the distributions $\sigma_\alpha$ and $\sigma_\epsilon$ as the measurements of the errors, given by the following expressions:
\begin{equation}
\begin{array}{c}
\sigma_\alpha=\int p_\alpha (\alpha-\langle\alpha\rangle)^2 d\alpha\\
\sigma_\epsilon=\int p_\epsilon (\alpha-\langle\alpha\rangle)^2 d\epsilon
\end{array}
\end{equation}
where $<x>$ is the mean of the distribution p(x):
\begin{equation}
\langle x \rangle=\int p(x) \mathrm{x dx}
\end{equation}
We calculate the covariance between $\alpha$ and $\epsilon$ 
\begin{equation}
C_{\alpha, \epsilon}=\int \int p(\alpha, \epsilon) (\alpha-\langle\alpha \rangle)(\epsilon-\langle\epsilon \rangle)d\alpha, d\epsilon\\
\end{equation}
and the correlation coefficient $\rho_{\alpha, \epsilon}$:
\begin{equation}
\rho_{\alpha, \epsilon}=\frac{C_{\alpha, \epsilon}}{\sigma_\alpha \sigma_\epsilon}
\end{equation}
The fiducial parameters for the error ($\sigma$) estimation are:
\begin{itemize}
\item  The ranges for the error estimation are: $\alpha=[0.7, 1.3]$ and $\epsilon=[-0.3, 0.3]$
\item The spacing in grids are: $\Delta_\alpha=$ 0.6/121=0.005 and $\Delta_\epsilon=$0.6/61=0.01
\item A Gaussian prior on $\log(\alpha)$ with a width 0.15 is applied in the likelihood surface to suppress unphysical downturns in the $\chi^2$ distribution at small $\alpha$.
\comment{We probably need to mention the changes of integration interval that is used in Aardwolf here. Something like the following} 
We have adopted a slightly different methodology in the calculation of the best fit parameter uncertainties than \cite{Aar13}; our approach is detailed in Section~\ref{sec:resultsmocks}.
\end{itemize}

\section{$\xi(\lowercase{r})$ Systematics on Multipoles Fitting}\label{sec:sysfit}

Given our fiducial case we explore the robustness of the fitting method to different choices in the methodology. The choices explored are in similar order as described in Section~\ref{sec:fiducial}, except the additional two items at the end of the list: 

\begin{enumerate}
\item Model Templates and Parametrization of Anisotropic Clustering in (Section~\ref{sec:model_temp})
\item Fitting range and bin sizes in (Section~\ref{sec:frange_bsize})
\item Nuisance Terms Model  in Section~\ref{sec:nuisance}  
\item Priors on various parameters: $\log(B_0^2)$,$\beta$,$\epsilon$,$\alpha$ in (Section~\ref{sec:priors})
\item Streaming models in (Section~\ref{sec:streaming_sys})
\item Nonlinear damping model parameters in (Section~\ref{sec:damping})
\item Covariance matrix corrections in (Section~\ref{sec:cov_corr})
\item Grid sizes in likelihood surfaces in (Section~\ref{sec:grids}) 
\end{enumerate}

In this section, we describe the tests performed, as well as the predicted behavior in terms of variations of $\alpha$, $\epsilon$ and their respective errors. 

\subsection{Model Templates and Parametrization of Anisotropic Clustering} \label{sec:model_temp} 

There are multiple ways to define a theoretical correlation function, especially considering non-linear correlation function in redshift space for galaxies. 
In this paper, we consider two templates:  the de-wiggled template defined in Section~\ref{sec:dewiggled} and the RPT-inspired template defined in Section~\ref{sec:rpt}. 

There are also multiple ways to parameterize anisotropic clustering. In this paper, we concentrate on the multipoles only, and 
even within the multipoles, we have two parameterizations of the anisotropic clustering, one with $\alpha-\epsilon$, one with $\alpha_{||}-\alpha_{\perp}$, both described in Section~\ref{sec:parametrization}.

\subsection{Fitting Range and Bin Size} \label{sec:frange_bsize}

The choice of fitting ranges can be influenced by two factors, one based on our confident in our theoretical templates, and whether the broad-band polynomial terms can remove all effects of our uncertainty in our theoretical templates. In \cite{And13}, an optimal range for fitting was been found to be $[50,200]$ $h^{-1}$Mpc when we fit for anisotropic clustering signals, while when fitting for isotropic clustering signals, the optimal range was [30, 200] $h^{-1}$Mpc.  We test these two different 
scenarios in our analysis. 

The choice of bin sizes also must to be tested given that we look for a signal that has a width of $\approx 10h^{-1}$Mpc. A too large a bin size can lead to miss the peak entirely, while our signal is diluted when we select too small a bin size.
\cite{Xeaip} tested the effect for various bin sizes on the isotropic clustering and no significant differences were found when they fit for either  $4h^{-1}$Mpc or  $7h^{-1}$Mpc bin sizes. 
\cite{Per13} also examined at bin size choices in isotropic correlation function and power-spectrum. 
\cite{Per13} tested different bin sizes and found that the optimal choice was achieved with an 8 $h^{-1}$Mpc bin. He also had shown a $\lesssim 0.5\% $ difference in errors bars of inferred $\alpha_{\perp}$, $\alpha_{||}$ when we include the covariance corrections they proposed. 
The fiducial binning chosen for our study was $8 h^{-1}$Mpc as in \cite{Aar13}. However, a valid issue to be explored is as the BAO is only about $10h^{-1}$Mpc wide, and the bins are only slightly smaller than the width of the BAO, then perhaps a no-BAO model will be able to fit the data nearly as well as than a BAO model. 
Thus the main purpose of this test will be to verify that with a wider bin size, we continue to resolve the BAO feature without losing information. We compare the $8h ^{-1}Mpc$ fiducial results with a  4$h^{-1}Mpc$ bin size used in \cite{Xeaip, Xu12b}

\subsection{Nuisance Terms Model} \label{sec:nuisance} 

In order to remove broad band effects that are deficient to model,  the fiducial methodology adds  third-order polynomials (denoted 
nuisance terms $A_{\ell}(r)$)  to the theoretical monopole and quadrupole as described in equation~\ref{eqn:nuisance}. 
We test  variations from this choice by varying the order of the polynomials used \citep{Xeaip,Xu12b}. 
In principle, given the same type of broad-band features, either from mis-modeling of non-linearities, or observational systematics,  the 
polynomial order should not affect the fitting results. 
However, if we expect different types of observational systematics (such is the case in Lyman-$\alpha$ forest for example), the 
order of polynomials may need to change.

\subsection{Priors} \label{sec:priors}

\cite{Xu12b} have shown that variations  in parameters $B_0$ and $\beta$ affect mostly the shape of quadrupole, and do not influence the BAO position. The only parameters which can shift the BAO position are $\alpha$ and $\epsilon$.
The different behavior could be explained in terms of the derivatives of the parameters near the BAO scale. 
The $\beta $ and $B_0$ derivatives are symmetric compared with derivatives of $\alpha$ and $\epsilon$, which are antisymmetric. 
This different behavior, isotropy versus anisotropy, reflects ability to shift  (or not) BAO position \citep{Xu12b}. 
However, the structures in the derivatives of $\alpha$ and $\epsilon$ are partially degenerated with other parameters  ($B_0$, $\beta$), thus the roles of these priors are to limit the models from exploring these degeneracies.

We test the effect of each prior alone in the best  fit values of $\alpha$ and $\epsilon$ and in their corresponding errors. We compare the results against the extreme cases when we place no priors and when all priors applied.  
In the following subsections, we comment on what we expect when each priors are used individually  and the related degeneracies.

\subsubsection{Prior on $B_0$}  \label{sec:prior_B0}

$B_0$ is a bias-like term that adjusts the amplitude of the model to fit the data.
The prior on $\log(B_0^2)$ should not significantly affect $\alpha$; it should, however, have some affect on $\epsilon$ , $\sigma_\alpha$ and
$\sigma_\epsilon$ because $\epsilon$ and $B_0$ are slightly
degenerate.
\cite{Xu12b} has extensive discussions on the degeneracies between these parameters, which we summarize here. The  $\epsilon$ dependence arises in three places : the derivative of the monopole, the quadrupole and its derivative.
The quadrupole does not have a strong BAO feature, thus the dominant information when we marginalize the shape is produced by from the derivative of the monopole. Figure 3 of \cite{Xu12b}, demonstrates that there is a clear degeneracy between $\epsilon$ and $B_0$ for this term. 

In the case of $\sigma_\alpha$, we expect that this prior to provide tighten constraits. 
This is particularly valid for extreme $\alpha$ values. 
Without the prior the fitter is allowed to set the normalization $B_0^2$ of the monopole to any value that produces
the smallest $\chi^2$ including completely unphysical values or
even negative ones. However, if we have the prior, then the normalization
of the model will be limited to being close to the central value of
the prior (1.0 in this case), which is a reasonable assumption (i.e., we
are assuming that the model should resemble the data). In
this case, however, the prior penalizes values of $B_0^2$ that are substantially
different from 1.0, so the minimum $\chi^2$ occur closer to
$B_0^2=1$. This new minimum $\chi^2$ will, by definition, be larger than
the global minimum without the prior, so the $\chi^2$ versus $\alpha$
curve will be deeper and hence $\sigma_\alpha$ should become smaller.

\subsubsection{Prior on $\beta$} \label{sec:prior_beta}

The $\beta$ parameter modulates the amplitude of quadrupole; this parameter is degenerate with  $\epsilon$  since large values of $\epsilon$ can also modulate the amplitude of the quadrupole just as $\beta$. Because of this degeneracy the prior on $\beta$ should change the value of $\epsilon$, especially if $\epsilon$ is large. 
The $\beta$ prior and the $\log(B^0)$ prior have similar effects. Basically one suppresses extreme tails in $\alpha$ and the other suppresses tails in $\epsilon$. Without any prior on $\beta$, the fitter could push $\epsilon$ to an extreme  value to  lower $\chi^2$ in some cases. Thus, imposing a prior you effectively force $\chi^2$ to be larger at the tails, thus producing a smaller $p$ which translates to narrower likelihood surface i.e., smaller $\sigma_\epsilon$.

\subsubsection{Prior on $\epsilon$} \label{sec:prior_epsilon}

The prior on $\epsilon$ is basically a top-hat prior, so it will limit
all values of $\epsilon$ to be between $-0.15$ to 0.15. The prior is not exactly a top-hat; the edges are tapened with a 
Gaussian to make the likelihood surfaces more smooth. 
If $\epsilon$ is beyond this range before the prior is applied, then after the prior is applied, it will
equal to either $-0.15$ or 0.15.  This restriction also  decreases
$\sigma_\epsilon$ in the cases where $\epsilon$ is poorly measured. 
Outside the tophat, $\chi^2$ quickly approaches infinity.

\subsubsection{Prior on $\alpha$} \label{sec:prior_alpha} 

The prior on $\alpha$ is \emph{different} from that on the others parameters. This prior is
\emph{only} applied to the likelihood surface, so it does not actually
affect the best-fit values of $\alpha$ or $\epsilon$. If the likelihood
surface is highly irregular (non-Gaussian), it will tighten the constraints on the error
on $\alpha$ (see Figure 6 of \cite{Xu12b}).

\subsection{Streaming Models} \label{sec:streaming_sys} 

We explored the effect of changing the streaming model by testing
three streaming models 
\begin{itemize}   
\item 
$\frac{1}{(1+k^2\mu^2 \Sigma_s^2)^2}$ (fiducial)

\item 
$exp [\frac{1}{1+(k\mu\Sigma_s)^2} ] $ (exp)

\item  
$exp(-(k\mu\Sigma_s)^2/2)$ (Gaussian)
\end{itemize}

We choose the first model as our fiducial one and we also investigate the effects of  changing  the value of $\Sigma_s$ value. 
Variations on $\Sigma_s$ would damp the BAO in the monopole, as a large value broaden the BAO peak in monopole. 
In the quadrupole the effects of $\Sigma_{s}$ are partially degenerate with $\Sigma_{NL}$. 
The changes in this parameter affects the crest-trough contrast and can even eliminate the trough when $\Sigma_s=0$. The effects are stronger in small scales, since the FoG is much stronger at small scales. 
We tested the effect of using a larger  $\Sigma_{s}$ from the fiducial case, $\Sigma_{s}=1.4\rightarrow 3.0$ in pre- and post reconstruction cases.

\subsection{Nonlinear Damping Model Parameters}\label{sec:damping} 

The $\Sigma_{NL}$ parameter models the smearing of the BAO due to the non linear structure growth as defined in equation~\ref{eqn:SigmaNL}. 
It is designed to damp the BAO feature because of non linear evolution.
Varying the parameters $\Sigma_{||, \perp}$ changes the structure of the peaks and troughs and reduces the structure crest-trough contrast. Using isotropic  values for $\Sigma_{||, \perp}$ eliminates through-feature at the BAO scale in the quadrupole. Thus, with large values of $\Sigma_{||, \perp}$ we can significantly change the fitting values and also adjusting with isotropic/anisotropic values.
Therefore, in this paper, we test the effects of changing $\Sigma_{||}$ and $\Sigma_{\perp}$. We tested two cases: pre-reconstruction, we change anisotropic values, $\Sigma_{||}=11$Mpc$h
^{-1}$ and $\Sigma_{\perp}=6$Mpc$h
^{-1}$ to isotropic values, $\Sigma_{||}=\Sigma_{\perp}=8$Mpc$h^{-1}$, and post-reconstruction, we changed the isotropic fiducial values, $\Sigma_{||}=\Sigma_\perp=3$Mpc$h
^{-1}$ to anisotropic values $\Sigma_{||}=4$Mpc$h
^{-1}$ and $\Sigma_{\perp}=2$Mpc$h^{-1}$.

\subsection{Covariance Matrix Corrections} \label{sec:cov_corr}

 In this work we adopt the covariance matrix correction as suggested in \cite{Per13}  as an additional systematic in the fitting procedure. 
This approach includes the corrections introduced due to  specificity of  BOSS mocks. 
We begin by describing the two different kinds of corrections applied.

\subsubsection{Covariance Corrections and their propagation}
\cite{Per13} extended previous work (Taylor et al. 2012; Dodelson \& Schneider 2013) on the contribution of covariance matrix errors to the parameter errors.  \cite{Per13} suggested the following corrections in particular. 

To correct the bias caused by the limited number of mocks, a correction factor must be applied to the inverse covariance matrix:
\begin{equation}
\Psi^t=(1-D) C^{-1}
\end{equation}
where
\begin{equation}
D=\frac{n_b +1}{n_s-1}
\end{equation}
this factor accounts for the skewness of the inverse Wishart distribution that describes $L(\Psi|\Psi_t) $. 

\cite{Per13} also provided the correction needed to propagate the errors in covariance matrix through to parameter errors.
Given a measurement of the sample variance (from mocks),  we need to multiply the sample variance by  $m_1$ given by 
\begin{equation}
\label{eqn:m1}
m_1=\frac{1+B(n_b-n_p)}{1+A+B(n_p+1)}
\end{equation}
where
\begin{equation}
\begin{array}{ll}
A=&\frac{2}{(n_s-n_b-1)(n_s-n_b-4)}\\
B=&\frac{(n_s-n_b-2)}{(n_s-n_b-1)(n_s-n_b-4)}
\end{array}
\end{equation}
where $n_b$ is the number of data measurements such as band powers in $P(k)$, and $n_s$ is the number of simulations used to calculate the sample variance. 
This correction produces an unbiased estimate of the full variance on parameter $p$.

Since we construct the expected error by analyzing the distribution of best-fit values recovered from the same mock dataset used to estimate the covariance matrix, the distribution of best-fit parameter values recovered from data that was also used to estimate the covariance matrix is biased in a different way to that of an independent set of data, and from the covariance estimate made from the measured likelihood. 
 Therefore, to  recover the distribution we must in this case to apply a corrective factor $m_2$ to the expected errors by:
 \begin{equation}
 m_2=(1-D)/m_1
 \end{equation}

 \subsubsection{Corrections of covariance in SDSS-3-BOSS mocks} 
 
The mock galaxy catalogs used in \cite{And13} were generated by sampling from the same density field, althought we separate them into  the Northern Galactic Cap (NGC) and Southern Galactic Cap (SGC) to match the BOSS observations.  There is therefore an overlap between the mocks in North and South as they are drawn for the same box. This overlapped region was  relatively small for DR9 , but for DR10 and DR11, there is 75 (100) per cent of overlap in the area covered by SGC. 
 
To construct two sets of independent mocks,  \cite{And13} use a set of NGC mocks different from the SGC mocks. However,  there is still a correlation between them. The sample covariance matrix is thus defined as follows:
\begin{equation}
\begin{array}{ll}
C_{i,j}=\frac{1}{2}\left [ \right.&\frac{1}{299} \sum_{m< 300} (\xi_i-\bar{\xi})(\xi_j-\bar{\xi}) \\
&\left ..+\frac{1}{299}\sum_{m> 300} (\xi_i-\bar{\xi})(\xi_j-\bar{\xi})\right ]
\end{array}
\end{equation}
A correlation coefficient is defined as inverse of the effective volume:
 \begin{equation}
 r=\frac{2V_{overlap}}{V_{NGC}+V_{SGC}}
 \end{equation}
 producing a value of $r=0.33$ for DR10 and $r=0.49$ for DR11.
We propagate this correlation coefficient to covariance errors by rescaling the terms $A,B$ and $D$, by a factor $(1+r^2)/2$.
 
For clarity in our analysis, we single out the covariance corrections, as it only changes the covariance matrix we applied in the fitting. 
Therefore, we perform the fitting robustness tests without considering any corrections of the covariance matrix for the overlapping regions. Then we apply this correction to some cases and we measured the effect but without propagate the error to the inferred parameters. Finally we estimate the errors with the full corrections.

In the fiducial methodology (which  closely follows \citep{Xu12b}),  a correction has been applied to achieve unbiased estimates of the covariance matrix. We retain this correction so that the methodology is closest to \cite{Xu12b}, unless otherwise specified. 

\subsection{Grid Sizes in Likelihood Surfaces} \label{sec:grids}

To estimate  errors,  we calculate likelihood surfaces on a grid  (Section~\ref{sec:Fitting}). Exploring the grids is time-consuming as the investigations we performed require fitting a large number of mock galaxy correction functions, so that there is a trade off between width of the grids and the number of tests to be performed for this work. 
In the ideal case, as the error on $\alpha$ is expected to be $\sim 1$\% in current analysis, the optimal width for $\alpha$ grid would be $\Delta_\alpha=$0.001, producing 10 grid points sampled
within $1\%$. However, this binning requires a huge amount of time; using this grid will restrict the number of test performed. Thus, we tested the effect of using smaller grid widths such as: 0.0025 and 0.005. 
Smaller grids should work if the likelihood surface is smooth. The wider grid 
0.005 may be too coarse 
for $\alpha$, but for $\epsilon$,
where the error is maybe closer to  0.005 
might be sufficient. 
We study the effect of grid size of calculating $\sigma_\alpha$, $\sigma_\epsilon$ using
various $\Delta\alpha$ and $\Delta\epsilon$ values (0.001,0.0025,0.005).
We also vary the range on $\alpha$ explored by examining at the following:   [0.7, 1.3] and [0.8, 1.2]  and in $\epsilon$, [-0.3, 0.3] and [-0.2, 0.2].

\section{Results from the Mocks}\label{sec:resultsmocks}
In this section,  we present  results of applying various robustness tests described in previous section (Section~\ref{sec:sysfit}) to full set of mock galaxy catalogs from BOSS as described in Section~\ref{sec:sims}. 
We analyze the choices with which we find a significant impact on the results. We apply our robustness tests on both DR10 and DR11 mock galaxy catalogs, but will focus on DR11 results in the paper.  We will also concentrate on one parameterization of$\alpha-\epsilon$ unless stated otherwise. This choice is due to the fact that the effects on $\alpha$ and $\epsilon$ naturally propagates to $\alphaper$ and $\alphapar$, and so there is a significant redundancy if we discuss both parameterizations. 

\subsection{Fiducial Results}

\begin{figure}
   \centering     
   \includegraphics[width=0.47\columnwidth]{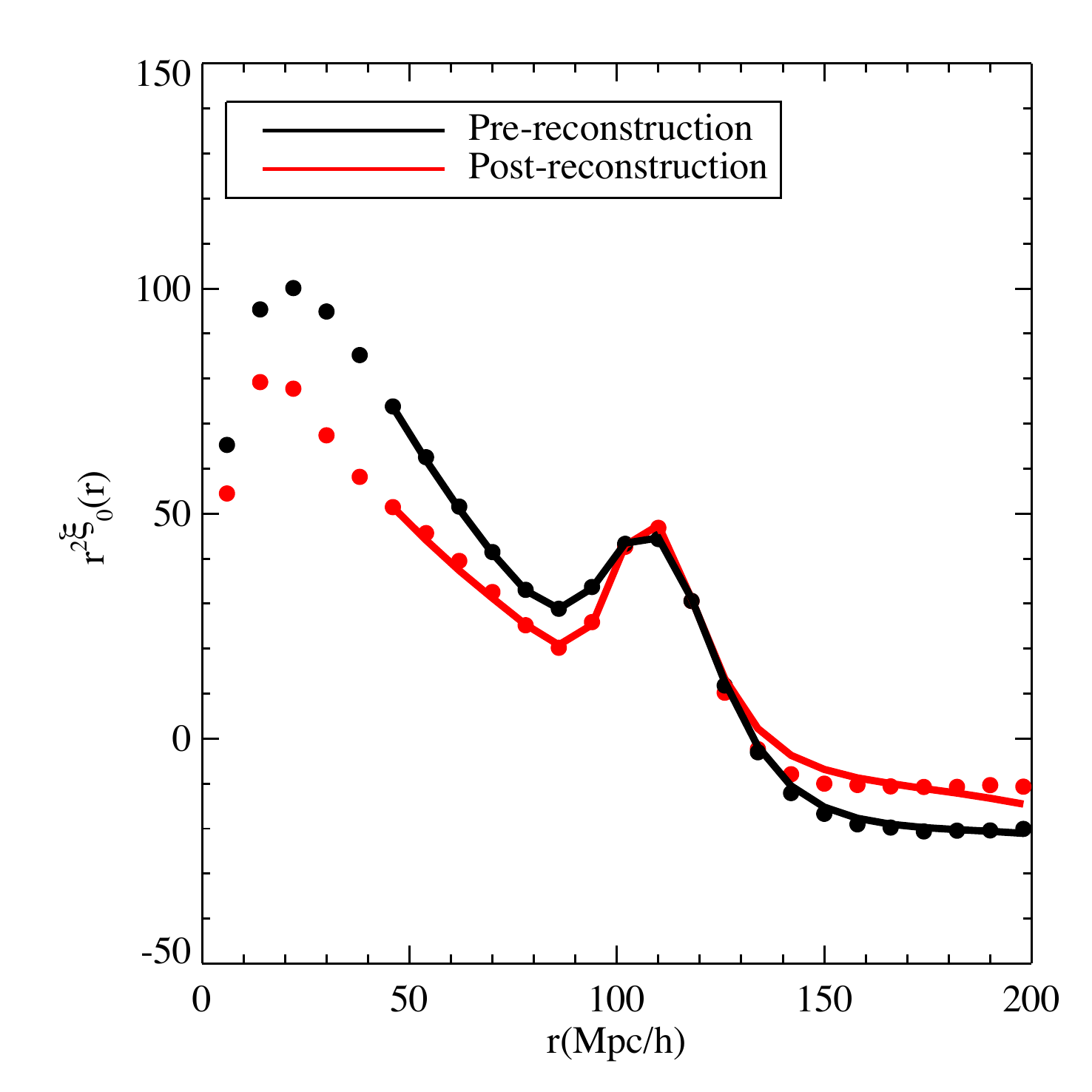}
   \includegraphics[width=0.47\columnwidth]{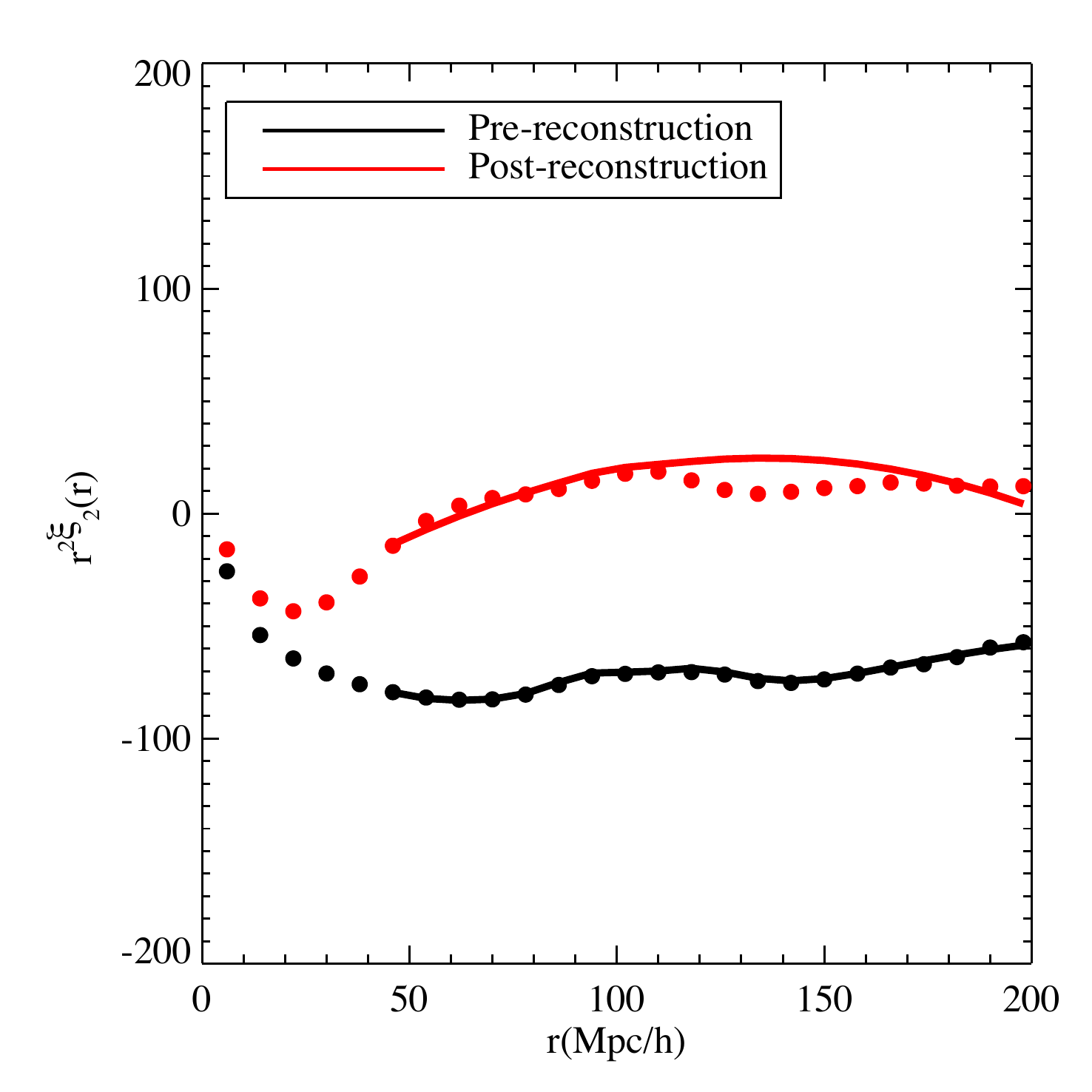}
   \caption{Mean monopole [left]  and quadrupole [right] of all mock galaxy catalogs of  DR11 pre-reconstruction [black] and post-reconstruction[red]. The fit shown in figure is found by applying the fiducial fitting methodology.}
  \label{fig:monoquadfid}
\end{figure}

\begin{figure}
   \centering     
   \includegraphics[width=1.0\columnwidth]{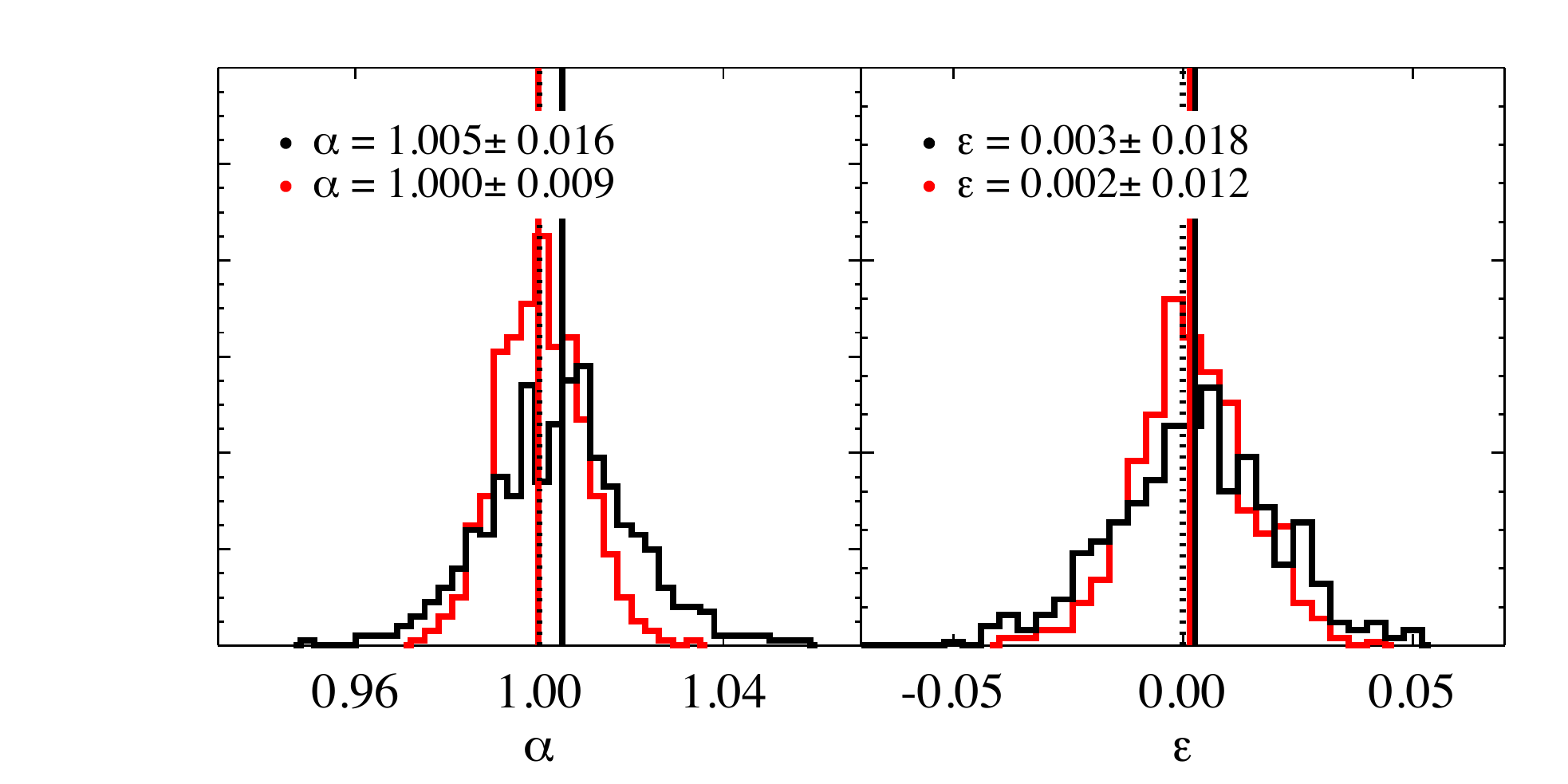}
   \caption{Histograms of $\alpha$ [left]  , $\epsilon$ [right] for fiducial case pre-reconstruction [black] and post-reconstruction [red]. The legend indicates the median and RMS of the distributions. We can see that reconstruction narrows the distribution of both $\alpha$ and $\epsilon$, which is to be expected.}
   \label{DistributionFiducialRec}
\end{figure}

Fig.~\ref{fig:monoquadfid}  shows the average monopole and quadrupole of the simulations and their corresponding fits using our fiducial methodology pre- and post 
reconstruction for DR11. 
There is a perfect match at large scale to the fitting template, especially in monopoles. 
The match between the observed multipoles and the fit improves after reconstruction.  
 We also observe the sharpening of the baryonic acoustic feature on the reconstructed monopole and a quadrupole consistent with zero at larges scales suggesting that reconstruction does indeed
undo the smearing of the peak generated by the non linear evolution and partially restoring the isotropy of the 2-point correlation function.
The distribution of the best fit values in the $\alpha-\epsilon$ parametrization are presented in Fig.~\ref{DistributionFiducialRec} pre-[black] and post- reconstruction[red]. The labels indicate the mean and standard error on the mean.

Since we assume the input cosmology of the simulations in our analysis we analyze them,  we expect to achieve a distribution of $\alpha$  and $\epsilon$ center around $1$  and $0$, respectively. 
For ease of discussion, we first  define the fitting-bias as :
\begin{equation}
\begin{array}{l}
b_{\alpha}=[\tilde{\alpha}-1]*100\\
b_{\epsilon}=[\tilde{\epsilon}-0]*100. 
\end{array}
\end{equation}
In pre-reconstruction, we expect an order of ~0.5\% shift from 1, since there is non-linear structure growth, even if we do not expect the 2LPT mocks tocapture all the non-linearities.  
When we apply the fiducial methodology,  we find  $b_\alpha=0.5$\% and $b_\epsilon=0.3\%$ pre-reconstruction. These values reduce to $b_\alpha<1\%$ and $b_\epsilon=0.2\%$ with reconstruction. 
As expected, reconstruction reduces the bias in both $\alpha$ and $\epsilon$ and also decreases the dispersion of the best fit values significantly. 
The bias on $\alpha$ and $\epsilon$, however, are both within the RMS dispersion of the mocks. 
This result suggests that  the fiducial fitting methodology does not produce a significantly biased result.

\subsection{Potential systematics on the fitting results}

\subsubsection{The effects of methodology change on best fit parameters} 

\begin{table*}
\caption{Fitting results with numerous variations of our fiducial fitting methodology for DR11 mock galaxy catalogs pre-reconstruction without covariance corrections corrections for the overlapping mock regions. Each methodology tests have been described earlier in Section~\ref{sec:sysfit};  we clarify some of the less obvious ones here: ``$P_{pt}(k)$ floating'' refers to using RPT-inspired $P_{\rm pt}(k)$  template with $\beta$ floating and ``$P_{pt}(k)$ fixed" with $\beta$ fixed to $0.4$ for pre-reconstructed and $\beta=0$ post-reconstruction, ``exp'' and ``Gaussian'' refers to various Finger-of-God models as described in Section~\ref{sec:streaming_sys}. The first line shows the median and 16th and 84th percentiles for the $\alpha, \epsilon, \alpha_{||, \perp}$. The rest of the lines lists the median bias and median  variations $\Delta v=v_{i}-v_{f}$ with their corresponding 16th and 84th percentiles. The median bias $\widetilde{b}$, median variations $\widetilde{\Delta v}$, and percentiles are multiplied by 100.}
\label{tab:dr11norec}
\begin{tabular}{@{}lrrrrrrrr}

\hline
Model&
$\widetilde{\alpha}$&
-&
$\widetilde{\epsilon}$&
-&
$\widetilde{\alpha_{\parallel}}$&
-&
$\widetilde{\alpha_{\perp}}$&
-\\
\hline
\multicolumn{9}{c}{DR11 Pre-Reconstruction}\\
\hline

Fiducial&
$1.0049^{+0.0142}_{-0.0156}$&-&
$0.0027^{+0.0166}_{-0.0181}$&-&
$1.0107^{+0.0404}_{-0.0413}$&-&
$1.0018^{+0.021}_{-0.020}$&-\\
\hline
Model&
$\widetilde{b_\alpha}$&
$\widetilde{\Delta\alpha}$&
$\widetilde{b_\epsilon}$&
$\widetilde{\Delta\epsilon}$&
$\widetilde{b_{\parallel}}$&
$\widetilde{\Delta\alpha_{\parallel}}$&
$\widetilde{b_{\perp}}$&
$\widetilde{\Delta\alpha_{\perp}}$\\
\hline
\\

$r_{bin}\rightarrow 4\mathrm{Mpc}/h$&
$0.46^{+1.24}_{-1.46}$&
$0.03^{+0.40}_{-0.48}$&
$-0.27^{+1.59}_{-1.70}$&
$0.02^{+0.47}_{-0.47}$&
$1.14^{+3.61}_{-4.24}$&
$0.09^{+0.89}_{-0.99}$&
$0.11^{+1.99}_{-1.73}$&
$-0.03^{+0.68}_{-0.66}$\\
\\[-1.5ex]

$30<r<200\mathrm{Mpc}/h$&
$0.17^{+1.42}_{-1.46}$&
$-0.26^{+0.24}_{-0.28}$&
$-0.09^{+1.71}_{-1.76}$&
$-0.15^{+0.15}_{-0.19}$&
$0.39^{+4.07}_{-3.90}$&
$-0.58^{+0.54}_{-0.65}$&
$0.03^{+2.19}_{-1.97}$&
$-0.08^{+0.11}_{-0.17}$\\
\\[-1.5ex]

$\mathrm{2-term} \; A_\ell(r)$&
$0.49^{+1.49}_{-1.48}$&
$0.05^{+0.12}_{-0.11}$&
$-0.31^{+1.66}_{-1.66}$&
$0.03^{+0.31}_{-0.23}$&
$1.15^{+4.11}_{-3.75}$&
$0.10^{+0.73}_{-0.51}$&
$0.17^{+2.11}_{-2.03}$&
$-0.02^{+0.21}_{-0.23}$\\
\\[-1.5ex]

$\mathrm{4-term} \; A_\ell(r)$&
$0.47^{+1.44}_{-1.57}$&
$-0.05^{+0.16}_{-0.13}$&
$-0.27^{+1.64}_{-1.89}$&
$-0.03^{+0.24}_{-0.23}$&
$0.96^{+4.02}_{-4.18}$&
$-0.11^{+0.58}_{-0.55}$&
$0.13^{+2.16}_{-2.04}$&
$-0.01^{+0.17}_{-0.17}$\\
\\[-1.5ex]

$\mathrm{Fixed \;} \beta=0.4$&
$0.48^{+1.42}_{-1.57}$&
$-0.00^{+0.01}_{-0.02}$&
$-0.26^{+1.65}_{-1.83}$&
$0.00^{+0.05}_{-0.07}$&
$1.01^{+4.08}_{-4.02}$&
$0.01^{+0.10}_{-0.15}$&
$0.16^{+2.17}_{-1.99}$&
$-0.00^{+0.07}_{-0.07}$\\
\\[-1.5ex]

$\spar=\sperp=8\mathrm{Mpc}/h$&
$0.49^{+1.51}_{-1.52}$&
$0.05^{+0.22}_{-0.21}$&
$0.05^{+1.75}_{-1.92}$&
$-0.33^{+0.28}_{-0.25}$&
$0.62^{+3.88}_{-4.46}$&
$-0.62^{+0.68}_{-0.64}$&
$0.54^{+2.12}_{-1.96}$&
$0.39^{+0.24}_{-0.30}$\\
\\[-1.5ex]

$\Sigma_s \rightarrow 3.0\mathrm{Mpc}/h$&
$0.53^{+1.40}_{-1.53}$&
$0.03^{+0.04}_{-0.04}$&
$-0.33^{+1.70}_{-1.83}$&
$0.07^{+0.06}_{-0.06}$&
$1.20^{+4.18}_{-4.16}$&
$0.18^{+0.16}_{-0.15}$&
$0.13^{+2.13}_{-2.03}$&
$-0.04^{+0.04}_{-0.03}$\\
\\[-1.5ex]

$\mathrm{No \; Priors}$&
$0.54^{+1.52}_{-1.55}$&
$0.02^{+0.13}_{-0.06}$&
$-0.27^{+1.73}_{-1.79}$&
$-0.01^{+0.21}_{-0.20}$&
$1.19^{+3.97}_{-4.23}$&
$-0.02^{+0.50}_{-0.37}$&
$0.21^{+2.24}_{-2.02}$&
$0.02^{+0.25}_{-0.17}$\\
\\[-1.5ex]

$\mathrm{Only} \; \log (B_0^2) \; \mathrm{prior}$&
$0.54^{+1.42}_{-1.54}$&
$0.01^{+0.06}_{-0.02}$&
$-0.25^{+1.70}_{-1.78}$&
$-0.01^{+0.18}_{-0.18}$&
$1.12^{+3.98}_{-4.23}$&
$-0.02^{+0.37}_{-0.31}$&
$0.17^{+2.21}_{-2.00}$&
$0.01^{+0.22}_{-0.18}$\\
\\[-1.5ex]

$\mathrm{Only} \; \beta \; \mathrm{prior}$&
$0.50^{+1.49}_{-1.56}$&
$0.00^{+0.10}_{-0.05}$&
$-0.28^{+1.68}_{-1.78}$&
$0.00^{+0.04}_{-0.04}$&
$1.12^{+4.03}_{-4.15}$&
$0.01^{+0.16}_{-0.11}$&
$0.19^{+2.20}_{-2.00}$&
$0.01^{+0.07}_{-0.03}$\\
\\[-1.5ex]

$\mathrm{P_{pt}(k) \;with \; floating \;} \beta$&
$-0.04^{+1.47}_{-1.52}$&
$-0.54^{+0.16}_{-0.15}$&
$-0.21^{+1.82}_{-1.89}$&
$-0.04^{+0.21}_{-0.18}$&
$0.57^{+4.12}_{-4.29}$&
$-0.64^{+0.47}_{-0.35}$&
$-0.35^{+2.19}_{-2.02}$&
$-0.51^{+0.24}_{-0.21}$\\
\\[-1.5ex]

$\mathrm{P_{pt}(k)\; with \;} \beta=0.0$&
$-0.07^{+1.48}_{-1.51}$&
$-0.55^{+0.16}_{-0.15}$&
$-0.20^{+1.80}_{-1.85}$&
$-0.06^{+0.20}_{-0.20}$&
$0.53^{+4.08}_{-4.42}$&
$-0.67^{+0.46}_{-0.44}$&
$-0.33^{+2.16}_{-2.04}$&
$-0.49^{+0.25}_{-0.21}$\\
\\[-1.5ex]

$\mathrm{FoG \; model} \rightarrow exp$&
$0.49^{+1.42}_{-1.56}$&
$-0.00^{+0.01}_{-0.01}$&
$-0.25^{+1.66}_{-1.80}$&
$-0.01^{+0.01}_{-0.01}$&
$1.06^{+4.03}_{-4.12}$&
$-0.02^{+0.02}_{-0.02}$&
$0.18^{+2.12}_{-2.03}$&
$0.01^{+0.01}_{-0.01}$\\
\\[-1.5ex]

$\mathrm{FoG \; model} \rightarrow gauss$&
$0.49^{+1.42}_{-1.56}$&
$0.00^{+0.00}_{-0.00}$&
$-0.27^{+1.66}_{-1.81}$&
$0.00^{+0.00}_{-0.00}$&
$1.07^{+4.04}_{-4.13}$&
$0.00^{+0.00}_{-0.00}$&
$0.18^{+2.13}_{-2.03}$&
$-0.00^{+0.00}_{-0.00}$\\
\\[-1.5ex]

\hline

\end{tabular}
\end{table*}


\begin{table*}
\caption{The fitting results from mocks from numerous variations of our fiducial fitting methodology for DR11 mock galaxy catalogs post-reconstruction without covariance corrections for the overlapping mock regions. Each methodology test have been described earlier in Section~\ref{sec:sysfit};  we clarify some of the less obvious ones here: ``$P_{pt}(k)$ floating'' refers to using RPT-inspired $P_{\rm pt}(k)$  template with $\beta$ floating and ``$P_{pt}$ fixed" with $\beta$ fixed to $0.4$ for pre-reconstructed and $\beta=0$ post-reconstruction, ``exp'' and ``Gaussian'' refer to various Finger-of-God models as described in Section~\ref{sec:streaming_sys}. The first line shows the median and 16th and 84th percentiles for $\alpha, \epsilon, \alpha_{||, \perp}$. The rest of the lines lists the median bias and median  variations $\Delta v=v_{i}-v_{f}$ with their corresponding 16th and 84th percentiles. The median bias $\widetilde{b}$, median variations $\widetilde{\Delta v}$, and percentiles are multiplied by 100.}
\label{tab:dr11rec}
\begin{tabular}{@{}lrrrrrrrr}

\hline
Model&
$\widetilde{\alpha}$&
-&
$\widetilde{\epsilon}$&
-&
$\widetilde{\alpha_{\parallel}}$&
-&
$\widetilde{\alpha_{\perp}}$&
-\\
\hline
\multicolumn{9}{c}{DR11 Post-Reconstruction}\\
\hline

Fiducial&
$0.9998^{+0.0094}_{-0.0084}$&-&
$0.0016^{+0.0124}_{-0.0116}$&-&
$1.0029^{+0.0272}_{-0.0247}$&-&
$0.9991^{+0.014}_{-0.016}$&-\\
\hline
Model&
$\widetilde{b_\alpha}$&
$\widetilde{\Delta\alpha}$&
$\widetilde{b_\epsilon}$&
$\widetilde{\Delta\epsilon}$&
$\widetilde{b_{\parallel}}$&
$\widetilde{\Delta\alpha_{\parallel}}$&
$\widetilde{b_{\perp}}$&
$\widetilde{\Delta\alpha_{\perp}}$\\
\hline
$r_{bin} \rightarrow 4\mathrm{Mpc}/h$&
$0.03^{+0.80}_{-0.84}$&
$0.03^{+0.31}_{-0.36}$&
$-0.09^{+1.22}_{-1.01}$&
$-0.03^{+0.37}_{-0.34}$&
$0.23^{+2.61}_{-2.36}$&
$-0.03^{+0.74}_{-0.71}$&
$-0.11^{+1.36}_{-1.36}$&
$0.05^{+0.48}_{-0.54}$\\
\\[-1.5ex]

$30<r<200\mathrm{Mpc}/h$&
$0.06^{+0.83}_{-0.85}$&
$0.06^{+0.14}_{-0.12}$&
$-0.27^{+1.25}_{-1.09}$&
$0.10^{+0.15}_{-0.11}$&
$0.60^{+2.62}_{-2.29}$&
$0.26^{+0.41}_{-0.32}$&
$-0.17^{+1.37}_{-1.48}$&
$-0.05^{+0.09}_{-0.09}$\\
\\[-1.5ex]

$\mathrm{2-term} \; A_\ell(r)$&
$0.07^{+0.87}_{-0.88}$&
$0.05^{+0.07}_{-0.07}$&
$-0.47^{+1.23}_{-1.03}$&
$0.31^{+0.16}_{-0.13}$&
$1.05^{+2.61}_{-2.32}$&
$0.68^{+0.35}_{-0.26}$&
$-0.39^{+1.29}_{-1.47}$&
$-0.26^{+0.13}_{-0.19}$\\
\\[-1.5ex]

$\mathrm{4-term} \; A_\ell(r)$&
$-0.08^{+0.92}_{-0.83}$&
$-0.05^{+0.06}_{-0.08}$&
$0.00^{+1.25}_{-1.17}$&
$-0.15^{+0.11}_{-0.13}$&
$-0.05^{+2.71}_{-2.48}$&
$-0.35^{+0.27}_{-0.32}$&
$0.02^{+1.33}_{-1.58}$&
$0.10^{+0.08}_{-0.07}$\\
\\[-1.5ex]

$\mathrm{Fixed \;} \beta=0.0$&
$-0.04^{+0.95}_{-0.84}$&
$-0.00^{+0.01}_{-0.03}$&
$-0.18^{+1.24}_{-1.14}$&
$0.02^{+0.09}_{-0.10}$&
$0.36^{+2.63}_{-2.52}$&
$0.04^{+0.16}_{-0.19}$&
$-0.15^{+1.41}_{-1.54}$&
$-0.02^{+0.10}_{-0.11}$\\
\\[-1.5ex]

$\spar=4,\sperp=2(\mathrm{Mpc/h})$&
$-0.02^{+0.93}_{-0.85}$&
$-0.00^{+0.03}_{-0.03}$&
$-0.21^{+1.21}_{-1.13}$&
$0.05^{+0.05}_{-0.05}$&
$0.37^{+2.66}_{-2.43}$&
$0.09^{+0.10}_{-0.10}$&
$-0.14^{+1.37}_{-1.56}$&
$-0.05^{+0.05}_{-0.06}$\\
\\[-1.5ex]

$\Sigma_s \rightarrow 3.0\mathrm{Mpc}/h$&
$0.00^{+0.92}_{-0.83}$&
$0.02^{+0.04}_{-0.04}$&
$-0.24^{+1.26}_{-1.15}$&
$0.09^{+0.08}_{-0.07}$&
$0.49^{+2.69}_{-2.50}$&
$0.20^{+0.19}_{-0.19}$&
$-0.15^{+1.39}_{-1.59}$&
$-0.06^{+0.05}_{-0.05}$\\
\\[-1.5ex]

$\mathrm{No \; prior}$&
$-0.01^{+0.95}_{-0.86}$&
$0.00^{+0.05}_{-0.02}$&
$-0.13^{+1.30}_{-1.16}$&
$-0.04^{+0.14}_{-0.15}$&
$0.28^{+2.72}_{-2.60}$&
$-0.08^{+0.28}_{-0.26}$&
$-0.05^{+1.41}_{-1.60}$&
$0.03^{+0.18}_{-0.14}$\\
\\[-1.5ex]

$\mathrm{Only} \; \log(B_0^2) \; \mathrm{prior}$&
$-0.01^{+0.94}_{-0.85}$&
$0.01^{+0.04}_{-0.01}$&
$-0.13^{+1.28}_{-1.14}$&
$-0.03^{+0.13}_{-0.13}$&
$0.32^{+2.67}_{-2.58}$&
$-0.05^{+0.26}_{-0.23}$&
$-0.04^{+1.39}_{-1.61}$&
$0.03^{+0.16}_{-0.12}$\\
\\[-1.5ex]

$\mathrm{Only} \; \beta \; \mathrm{prior}$&
$-0.03^{+0.94}_{-0.85}$&
$-0.01^{+0.01}_{-0.01}$&
$-0.15^{+1.25}_{-1.14}$&
$-0.00^{+0.02}_{-0.03}$&
$0.28^{+2.73}_{-2.49}$&
$-0.01^{+0.03}_{-0.07}$&
$-0.09^{+1.40}_{-1.60}$&
$0.00^{+0.02}_{-0.02}$\\
\\[-1.5ex]

$P_{pt}(k) \mathrm{\; with \; floating \;} \beta$&
$0.01^{+0.94}_{-0.87}$&
$0.03^{+0.03}_{-0.03}$&
$-0.11^{+1.26}_{-1.11}$&
$-0.03^{+0.05}_{-0.05}$&
$0.29^{+2.68}_{-2.47}$&
$-0.04^{+0.13}_{-0.11}$&
$-0.02^{+1.30}_{-1.55}$&
$0.06^{+0.05}_{-0.05}$\\
\\[-1.5ex]

$P_{pt}(k)\mathrm{\; with \;} \beta=0.0$&
$-0.00^{+0.94}_{-0.86}$&
$0.02^{+0.03}_{-0.03}$&
$-0.14^{+1.22}_{-1.14}$&
$-0.01^{+0.09}_{-0.10}$&
$0.33^{+2.58}_{-2.55}$&
$0.00^{+0.17}_{-0.22}$&
$-0.08^{+1.40}_{-1.50}$&
$0.04^{+0.10}_{-0.11}$\\
\\[-1.5ex]

\hline

\end{tabular}
\end{table*}

\begin{figure*}
   \centering   
 \includegraphics[width=1.0\columnwidth]{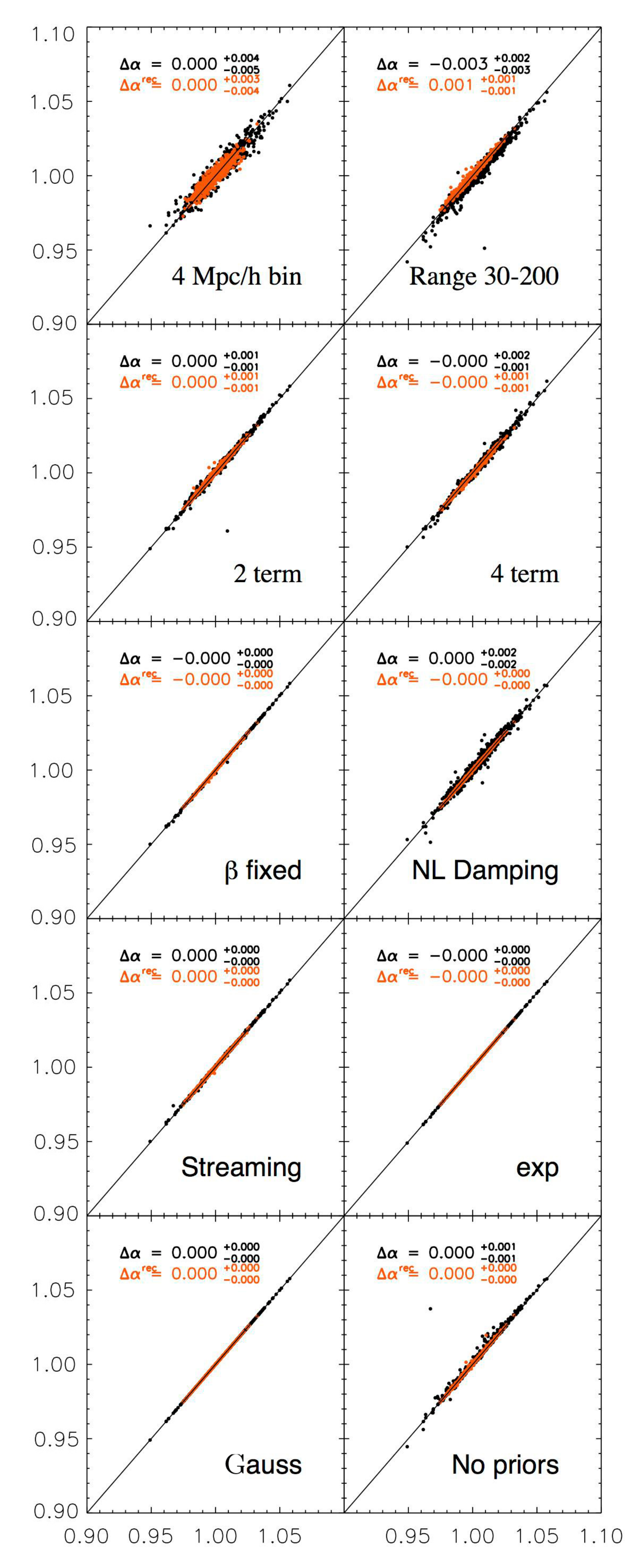}    
 \includegraphics[width=1.0\columnwidth]{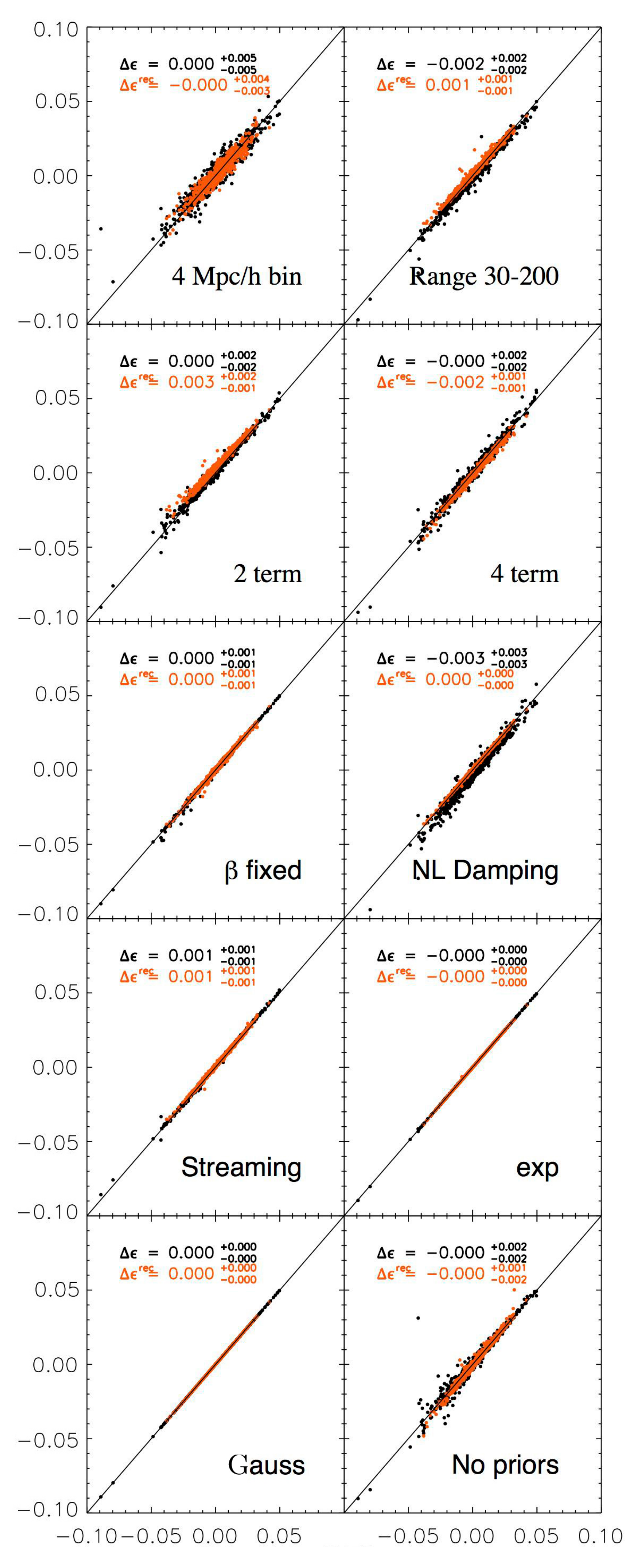} 
  \caption{Dispersion plots $\alpha$ [left] and $\epsilon$  [right] for the different cases enumerated in Tables.~\ref{tab:dr11norec} and~\ref{tab:dr11rec} for DR11 pre and post-reconstructed mocks. The dispersion is between the fiducial fitting methodology and particular methodology/fitting parameter change. 
For example, for the upper left-most plot, the x-axis is $\alpha$ using fiducial model, while is $\alpha$ the fiducial model except with the $\beta$ parameter fixed instead of varying $\beta$ with a prior of 0.2 centered around $0.35$.
``NL Damping'' denotes changing $\spar \; \& \; \sperp \rightarrow \spar=\sperp=8\hMpc$ for pre-reconstructed case and $\spar=\sperp \rightarrow \spar=4\hMpc \; \& \; \sperp=2\hMpc$ for post-reconstruction, ``$P_{pt}(k)$ floating'' refers to using RPT-inspired $P_{\rm pt}(k)$  template with $\beta$ floating and ``$P_{pt}(k)$ fixed" with $\beta$ fixed to $0.4$ for pre-reconstructed and $\beta=0$ post-reconstruction, ``exp'' and ``gaussian'' refers to various Finger-of-God models as described in Section~\ref{sec:streaming_sys}; ``Streaming" refers to changing the streaming value to $\Sigma_s =1.5\rightarrow 3.0 \hMpc$. Black denotes pre-reconstruction distributions, while red denotes post-reconstruction distributions. We can see that the dispersion is fairly small for nearly all cases shown here. }
   \label{fig:dr11v1ae}
\end{figure*}
\begin{figure*}
   \centering     
 \includegraphics[width=1.0\columnwidth]{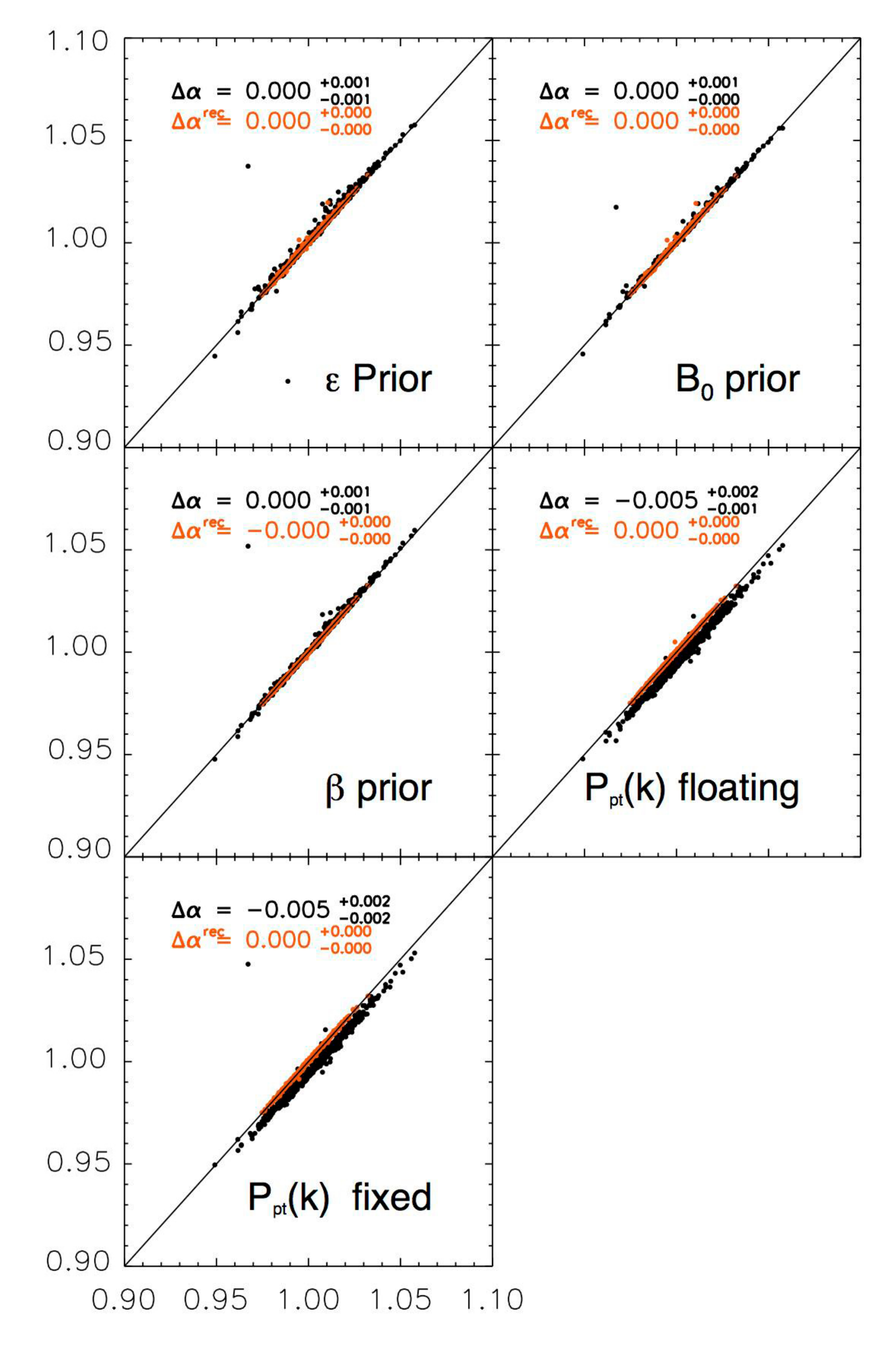}    
 \includegraphics[width=1.0\columnwidth]{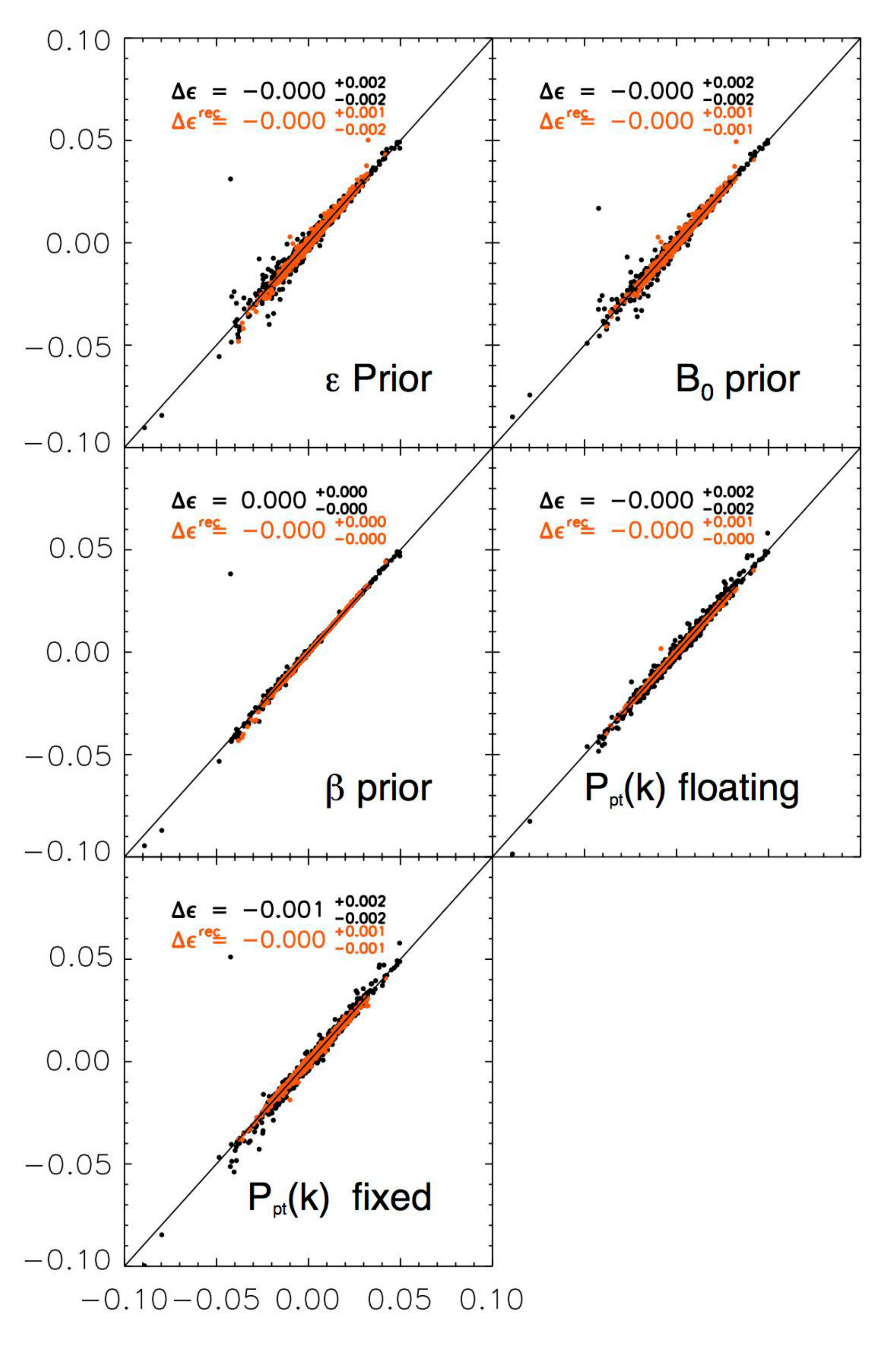} 
 \caption{Continuation of Fig.~\ref{fig:dr11v1ae}. Dispersion plots $\alpha$ and $\epsilon$ for methodology tests enumerated in Tables.~\ref{tab:dr11norec} and~\ref{tab:dr11rec}.  ``NL Damping'' denotes changing $\spar \; \& \; \sperp \rightarrow \spar=\sperp=8\hMpc$ for pre-reconstructed case and $\spar=\sperp \rightarrow \spar=4\hMpc \; and \; \sperp=2\hMpc$ for post-reconstruction, ``$P_{pt}(k)$ floating'' refers to using RPT-inspired $P_{\rm pt}(k)$  template with $\beta$ floating and ``$P_{pt}(k)$ fixed" with $\beta$ fixed to $0.4$ for pre-reconstructed and $\beta=0$ post-reconstruction, ``exp'' and ``Gaussian'' refer to various Finger-of-God models as described in Section~\ref{sec:streaming_sys}; ``Streaming" refers to changing the streaming value to $\Sigma_s =1.5\rightarrow 3.0 \hMpc$.  Black denotes pre-reconstruction distributions, while red denotes post-reconstruction distributions. We can see that the dispersion is fairly small for nearly all cases shown here. }
   \label{fig:dr11v2ae}
\end{figure*}

Table.~\ref{tab:dr11norec} and Table.~\ref{tab:dr11rec} presents the results of the anisotropic fit when we apply various changes in the methodology for DR11 pre- and post-reconstruction.  

The dispersion plots in $\alpha$ and $\epsilon$ for methodology tests listed in these tables are displayed in Fig.~\ref{fig:dr11v1ae} and Fig.~\ref{fig:dr11v2ae}. 
Each panel corresponds to the dispersion plot when we apply one of the methodology tests compared with the fiducial methodology. The legends of the plot indicate the median variation and the 16th and 84th percentiles of the median variation.
 We first consider the effects when we fit the pre- reconstructed mock catalogs.  We observe some dispersion in $\Delta \alpha$ in fourth cases: changing the bin size to smaller bins produces a 0.4 per cent dispersion, changing the fitting range to [30, 200] produces a 0.3 per cent dispersion and using higher order polynomials for modeling the broadband terms and when we use $P_{pt}$ templates shows both a  0.2\% dispersion. The remaining cases shows a small dispersion of $\le $0.1\%.

 In the case of $\Delta \epsilon$, the largest dispersions when we change the bin size showing a 0.5\% dispersion; when we use lower order polynomials for the broadband terms or when vary the NL damping parameters both produce a dispersion 0.3\%, and finally three cases show a 0.2\% dispersion: when using higher order polynomials for the broadband terms, the cases eliminating priors and using $P_{pt}$ templates. The remaining cases shows a small dispersion $\le 0.1\%$.
 
 In the post-reconstruction cases the dispersion is significantly reduced in all cases. With only one exception, the dispersion observed in  $\Delta \alpha$ is $\le 0.1\%$. The exceptional case is uses smaller bins, an has a dispersion of 0.3\%. The parameter $\Delta \epsilon$ also shows a small dispersion  of $\le 0.1\%$, with exception of  three cases: using smaller bins shows a dispersion of 0.4\% and using lower order polynomials for the broad band terms yields a 0.3\% dispersion and the fitting range case yields a 0.15\% dispersion.

Table.~\ref{tab:dr11norec} demonstrates biases in best-fit $\alpha$ values pre-reconstruction  are  $\le 0.5$\% for all robustness tests. 
Smaller biases in best fit $\alpha$ values are produced when we use $P_{\rm pt}$ template as the non-linear power-spectrum and when a larger fitting range is used. 
In the case of best fit $\epsilon$ values, systematic biases of $\sim$ 0.3\% are found in almost all cases. We also observe smaller biases in best fit $\epsilon$ values when the fitting range is increased, using the$P_{\rm pt}(k)$ template with fixing/floating $\beta$, or varying the $\Sigma_{||, \perp}$.
The fact that $P_{pt}(k)$ template produces a smaller bias pre reconstruction is not surprising because these templates includes a mode coupling term that is supposed to match the non-linear correlation function better than the de-wiggled template. This decrease in the bias associated a $P_{pt}(k)$ templates has been reported in previous works \citep{Kaz12,Sanchez12}. 
When we fit with reconstructed catalogs, observed biases  in best fit $\alpha$ values are all  $\le  $0.1\%
The fitting range results suggests that there is a trade off  between obtaining a smaller bias in pre- or post- reconstruction cases using same range. 
In the case of $\epsilon$, the only three cases that produce a larger bias than fiducial one ($2\%$) are those applying 2-term polynomials, enlarging the fitting range and the change in $\Sigma_s\rightarrow 3.0$Mpc$h^{-1}$.

We then turn to the differences each methodology change can cause when we compare them to  results using fiducial fitting methodology. 
Tables.~\ref{tab:dr11norec} and~\ref{tab:dr11rec} include  median variation in $\alpha$, $\widetilde{\Delta \alpha}$, and median variation in $\epsilon$, $\widetilde{\Delta \epsilon}$. The variation is defined as the difference of the test case compared to the fiducial case.
\begin{equation}
\Delta_{\rm var} =\alpha_{fid}-\alpha_{var}
\end{equation}
where \emph{fid} denotes the fiducial case and \emph{\rm var} refers to the different variations we apply to the fiducial methodology. 
The 16th, 86th percentiles of the distribution of $\Delta\alpha, \Delta\epsilon$ are also listed in the tables to quantify the dispersion of the best fit values.

With a large array of robustness tests, our fiducial methodology shows maximum differences of 0.5\%  in best fit  $\alpha$ value and 0.3\% in best fit $\epsilon$ value pre-reconstruction. Post reconstruction, the maximum differences are reduced to $\le$ 0.1\% in both $\alpha$ and $\epsilon$.
We list  methodology changes that produces the larger  variations pre-reconstruction: (i) changing the De-Wiggled Template to the $P_{\rm PT}$ template (0.5\% in best fit $\alpha$) ; (ii) changing the fitting range from  $[50,200]$ $h^{-1}$Mpc to $[30,200]$ $h^{-1}$Mpc  (0.3\% in best fit $\alpha$); (iii) Changing $\Sigma_{||,\perp}$ to  $\Sigma_{||}=\Sigma_\perp=8.0$Mpc$h^{-1}$ (0.3\% in best fit $\epsilon$).
Post-reconstruction, the variations in $\alpha$ and $\epsilon$ are impressively small, in all cases $\Delta \alpha,\Delta \epsilon \le 0.1\%$, except in the case of broad-band terms modeling, where a change to $2$ or $4$ term polynomials affects the $\epsilon$ post-reconstruction at still relatively modest level of $0.2-0.3\%$.  
We see a shift of +0.3\% with 2 terms and -0.2\% with 4 terms which suggests an interesting trend observed in the systematic polynomials: different order polynomials can cause a shift from positively biased $\epsilon$ to a  negatively biased in $\epsilon$.
These results also confirm the discussion in Section~\ref{sec:Fitting}, that changes in $\Sigma_{||}$, $\Sigma_\perp, \Sigma_s$  and also the change in the central value of the $\beta$ prior produces small  changes in the $\epsilon$.
After reconstruction, the variations in $\alpha$ and $\epsilon$ are  $\Delta \alpha < 0.1$\% except for the case of using polynomials of lower or higher order than the fiducial case to describe the broad band contribution of the correlation function, where we observe a  0.1-0.3\% fluctuation.

\subsubsection{Effects of methodology choices on best fit uncertainties} 

\begin{figure}
   \centering     
   \includegraphics[width=1.0\columnwidth]{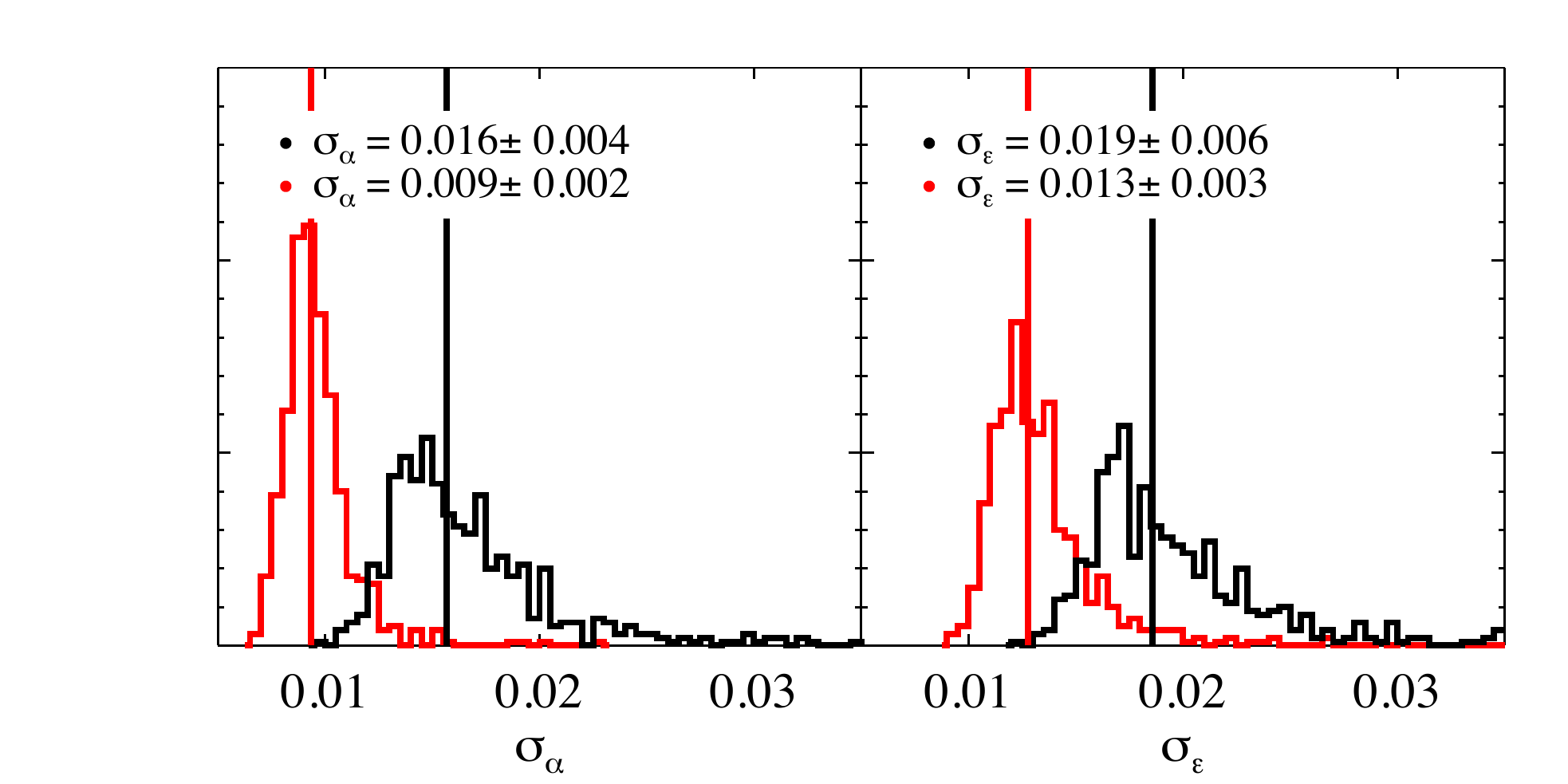}
   \caption{Histograms of $\sigma_\alpha$[left]  , $\sigma_\epsilon$[right] for fiducial fitting methodology pre-reconstruction [black] and post-reconstruction [red]. The legends indicate the median and RMS of the distributions. We can see that reconstruction clearly reduces the errors on fitted parameters.}
   \label{fig:sigfid}
\end{figure}

Fig.~\ref{fig:sigfid} shows the distribution of $\sigma_{\alpha}$ and $\sigma_\epsilon$ from fitting the DR11 mock galaxy catalogs using fiducial fitting methodology. 
The black [red]  solid lines indicates the pre- (post-) reconstruction results, and legends list the mean and the standard deviation on the mean. 
The median errors (and their quartiles)  are $\widetilde{\sigma_\alpha}=0.016^{+0.004}_{-0.002}$ and $\widetilde{\sigma_\epsilon}=0.019^{+0.005}_{-0.002}$. 
When we apply the fiducial fitting methodology to reconstructed DR11 mock galaxy catalogs, 
 $\widetilde{\sigma_\alpha}=0.009^{+0.001}_{-0.001}$ and  $\widetilde{\sigma_\epsilon}=0.013^{+0.002}_{-0.001}$, respectively.
 The distributions of $\sigma$ are highly skewed, the large tails extending to larger values of $\sigma$ in the pre-reconstruction are  significantly reduced post reconstruction.
 From the 16th and 86th percentiles of the $\sigma_\alpha, \sigma_\epsilon$ distributions we can observe the $\sigma_\epsilon$ distribution appears to be more skewed. 

\begin{figure*}
   \centering     
	   \includegraphics[width=1.0\columnwidth]{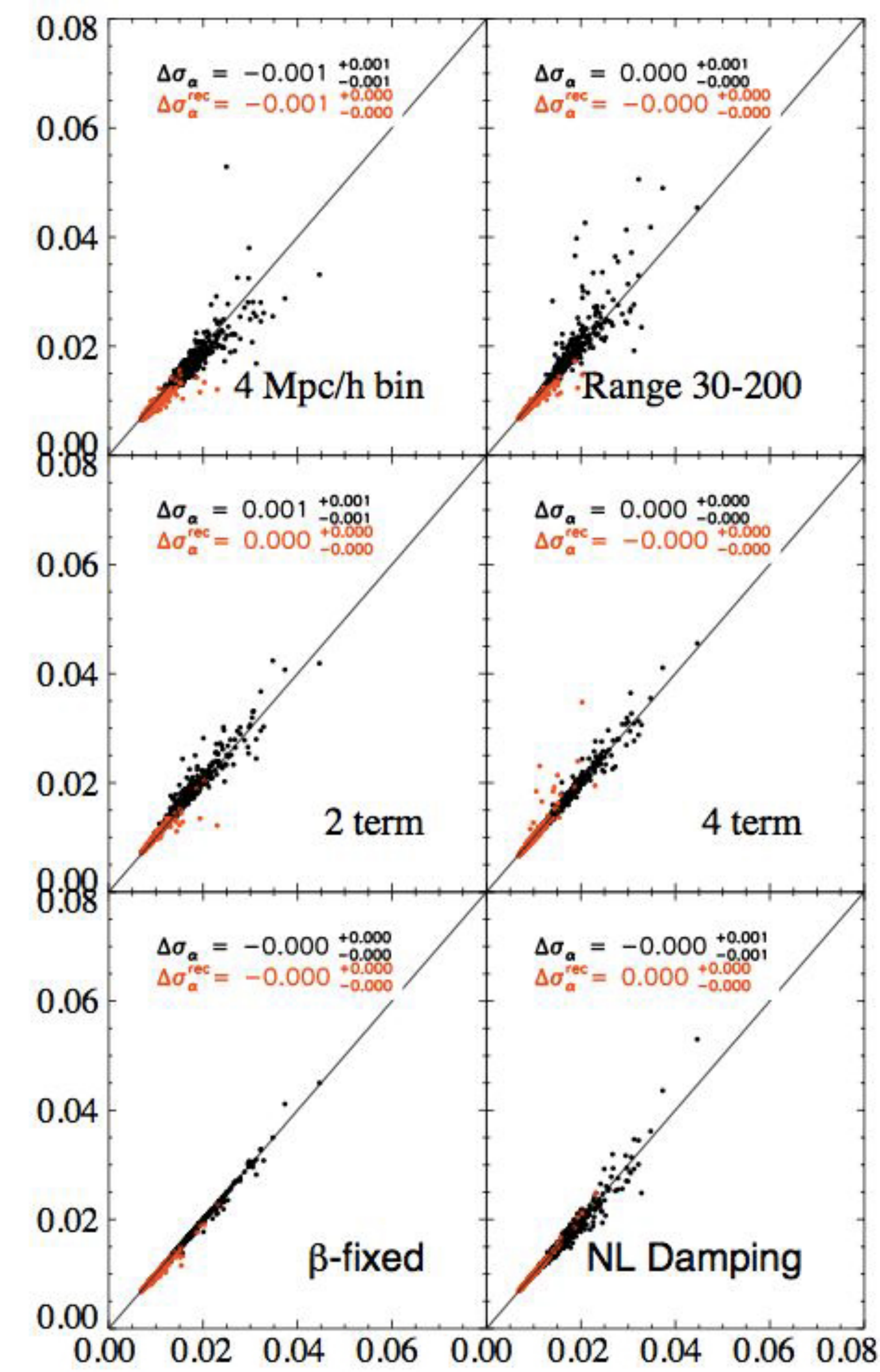} 
   \includegraphics[width=1.0\columnwidth]{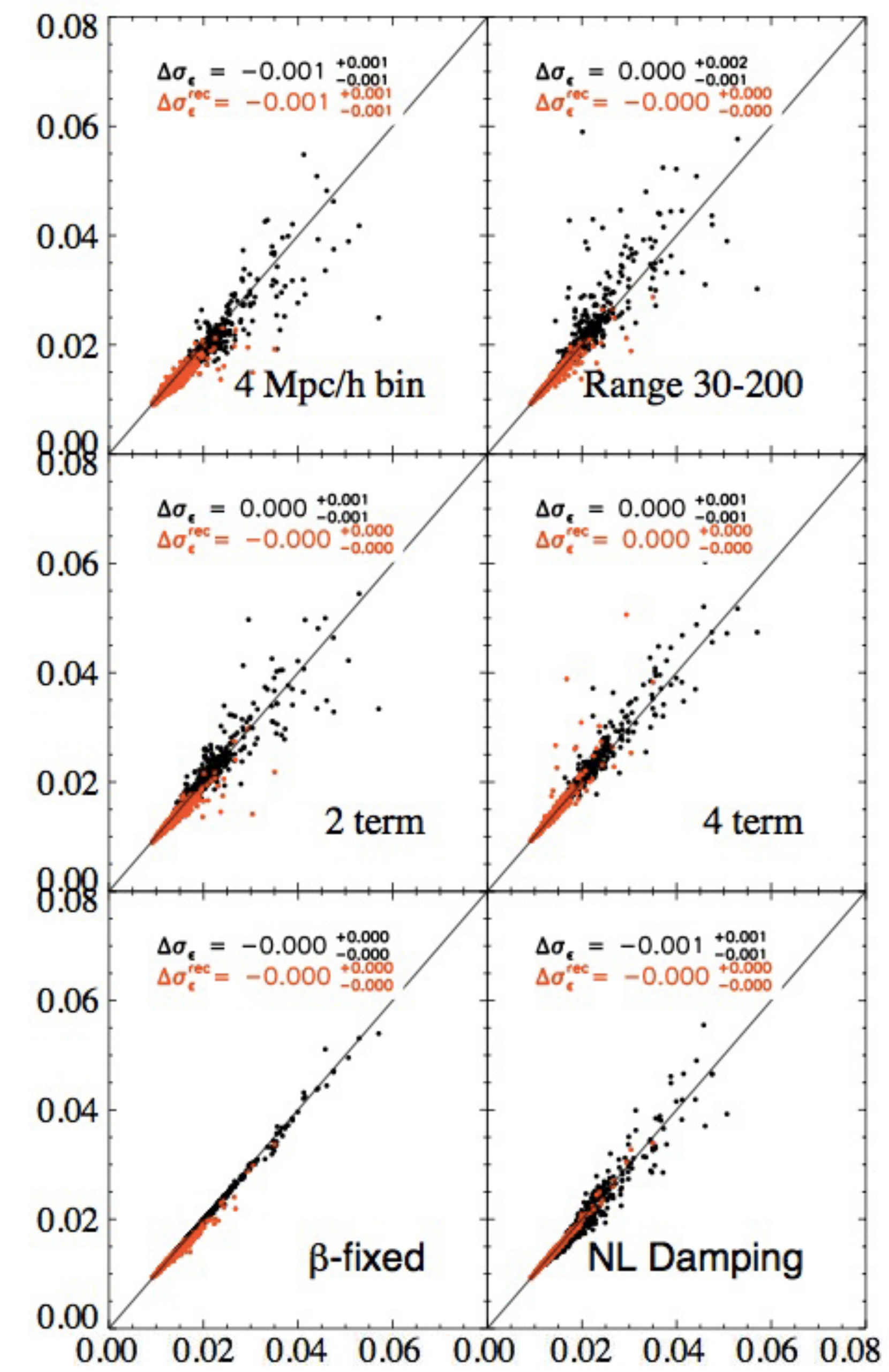}
 \caption{
 Dispersion plots of  $\sigma_{\alpha}$[left] and $\sigma_{\epsilon}$ [right] from various fitting methodologies to DR11 pre- and post reconstructed mock galaxy catalogs. The fitting methodologies tested are listed in Tables.~\ref{tab:sigmadr11norec} and~\ref{tab:sigmadr11rec}. 
The dispersion is between the fiducial fitting methodology  and the modified methodology. For example, for the upper left-most plot, the x-axis is $\sigma_{\alpha}$ using fiducial model, while the y-axis is $\sigma_{\alpha}$  applying fiducial methodology but with $\beta$ parameter fixed instead of varying $\beta$ with a prior. Black denotes pre-reconstruction distributions, while red denotes post-reconstruction distributions. We can see that the dispersion is fairly small (though not as small as the dispersion observed for the fitted values of the measured parameters) for nearly all cases shown here.}
      \label{fig:sigmadr11v1ae}
\end{figure*}

\begin{figure*}
   \centering     
   \includegraphics[width=1.0\columnwidth]{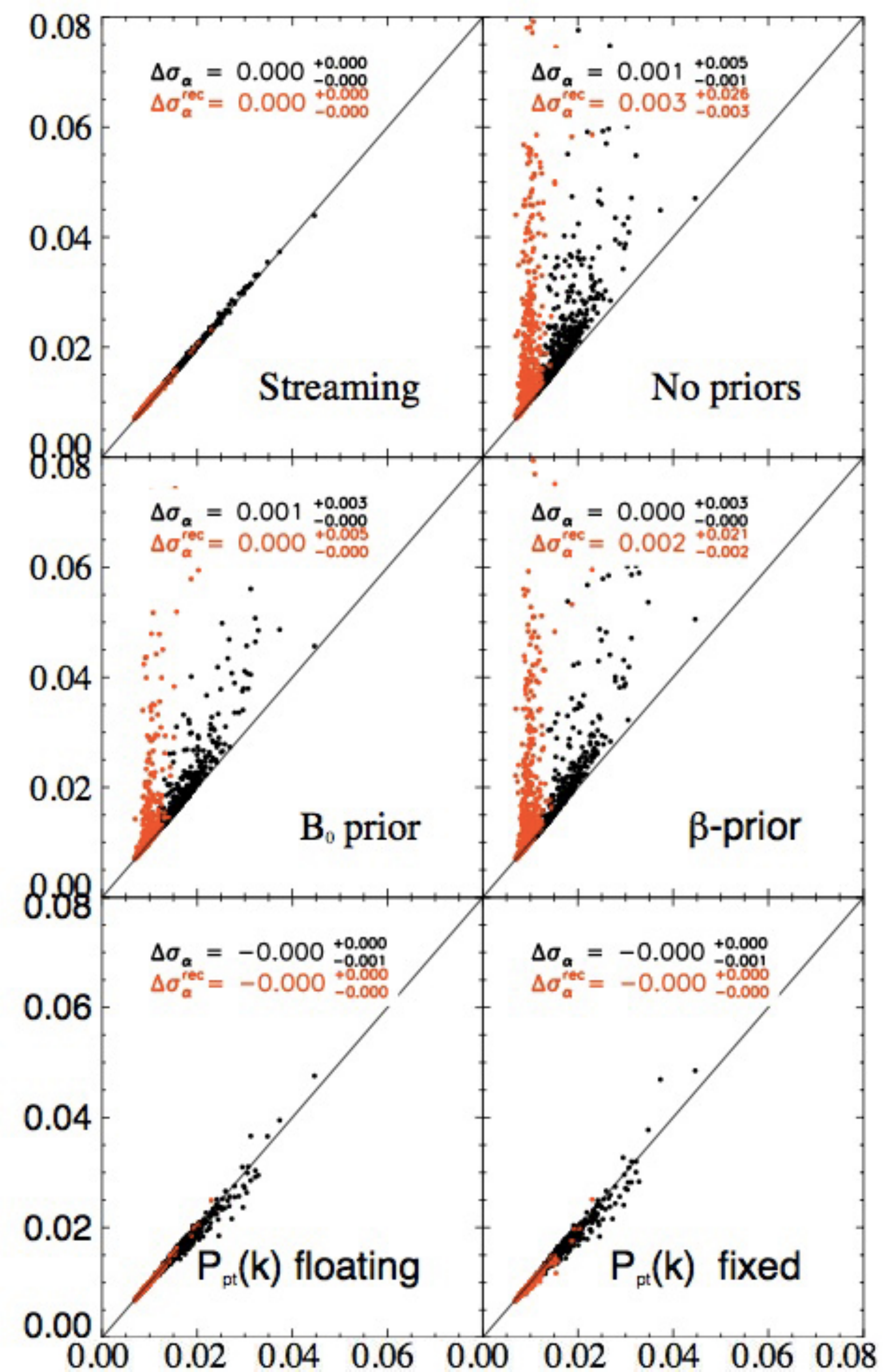} 
   \includegraphics[width=1.0\columnwidth]{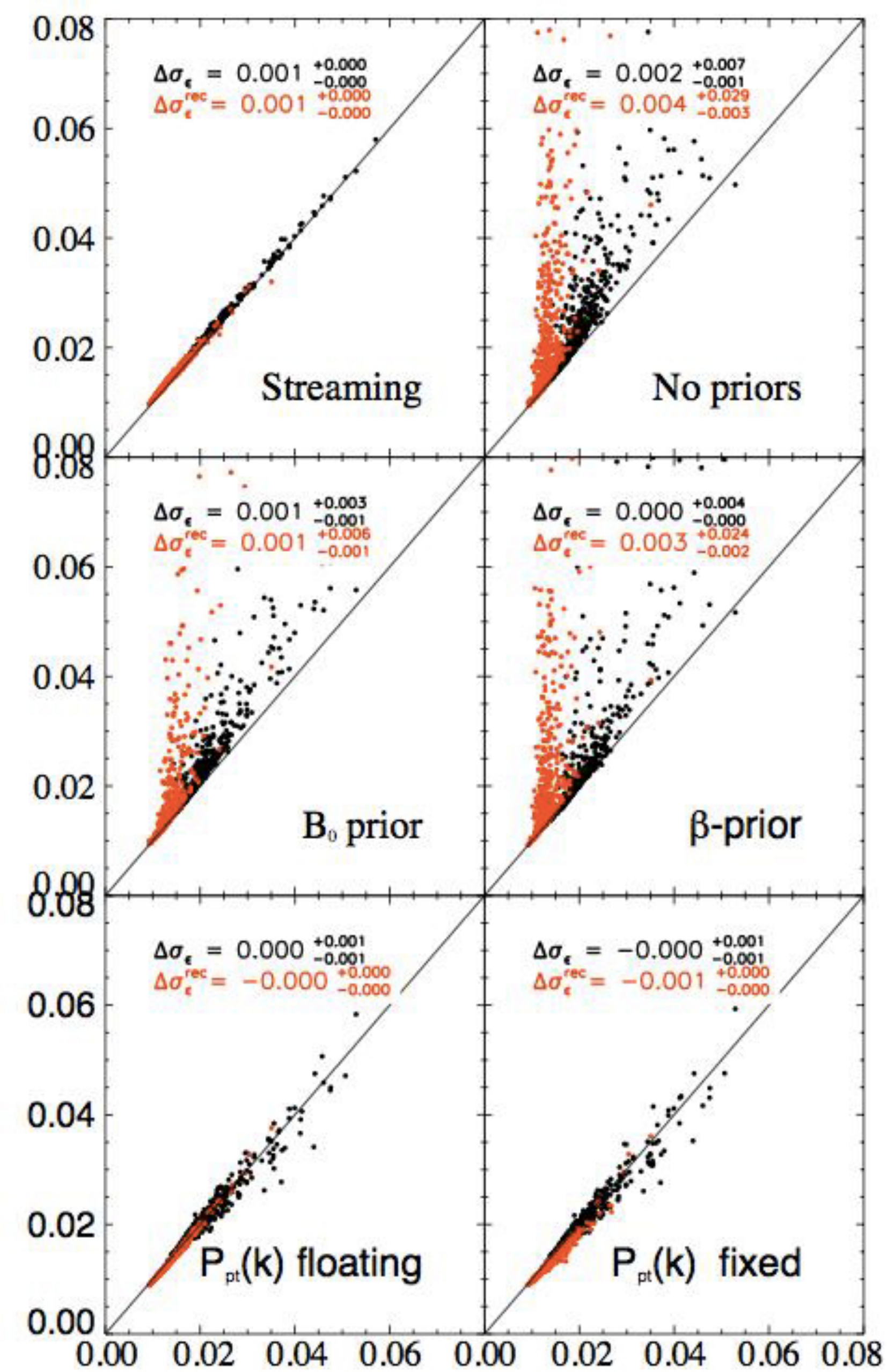}
 \caption{A continuation of Fig.~\ref{fig:sigmadr11v1ae}. Dispersion plots of $\sigma_{\alpha}$[left] and $\sigma_{\epsilon}$ [right] when we apply various fitting methodologies to DR11 pre- and post reconstructed mock galaxy catalogs. The fitting methodologies tested are listed in Tables.~\ref{tab:sigmadr11norec} and~\ref{tab:sigmadr11rec}.  Black denotes pre-reconstruction distributions, while red denotes post-reconstruction distributions. We can see that the dispersion is fairly small (though not as small as the dispersion observed for the fitted values of the measured parameters) for cases not involving changing priors. These prior related cases tested here do not limit the integration range for the errors to physically meaningful range, and will be discussed further in the text.}
   \label{fig:sigmadr11v2ae}
\end{figure*}

We now move to analyze the dispersion and variations in the uncertainties of the best fit parameter when  we modify the fiducial fitting methodology. 
We concentrate on Fig.~\ref{fig:sigmadr11v1ae} and Fig.~\ref{fig:sigmadr11v2ae} for the following discussion. 
\comment{DISCUSS DISPERSION PLOTS PLSSSS -- Grammatically corrected and added full sentences !! } 
To summarize, we observe small dispersions in the uncertainties of the best fit parameters (both  $\Delta \sigma_\alpha, \Delta \sigma_\epsilon$), at $\le 0.001$ ($\sim 6\%$), except when we change the fitting range or the priors we apply in the methodology. 
A larger fitting range produces a dispersion of $\Delta \sigma_\epsilon$ at 0.002 ($\sim$ 11\%) when the fitting methodology is applied to pre-reconstructed mock catalogs. 

We observe relatively large dispersions in $\Delta \sigma_\alpha, \Delta \sigma_\epsilon$ when certain priors are removed when fitting both pre- and post-reconstructed mock catalogs.  
We first consider the effects when fitting the pre-reconstructed mock catalogs (Table.~\ref{tab:sigmadr11norec}). 
The dispersion in $\Delta \sigma_{\alpha}$is 0.005 (31\%) when no priors are applied. 
When only $\beta, B_0$ priors are applied,  $\Delta \sigma_{\alpha}$  shows a 0.003 scatter ($\sim$ 19\%). 
The quantity $\Delta \sigma_\epsilon$ also possesses large dispersions when we eliminate certain priors. For example, applying no priors at all, $\Delta \sigma_\epsilon$ shows a dispersion of  0.007 ($\sim$ 36\%); when we apply only beta prior, a dispersion of  0.004 ($\sim$ 21\%) is produced; applying only $B_0$ prior produces a dispersion of 0.003 ($\sim$ 15\%). 

We now describe the fitting results of the post-reconstructed mock catalogs (Table.~\ref{tab:sigmadr11rec}). 
Applying no priors in the fitting methodology, $\Delta \sigma_\alpha$ yields a dispersion of 0.025 ($\sim$ 277\%);  applying only $\beta$ prior, produces a dispersion of 0.020 ($\sim$ 222\%) and when we apply only $B_0$ prior produces a dispersion of 0.005 ($\sim$ 55\%). 
The dispersion in $\Delta \sigma_\epsilon$,  also displays similarly large dispersions. 
For example, when no priors are set, the dispersion is 0.029 ($\sim$ 223\%); the $\beta$-only prior produces a dispersion of 0.023 ($\sim$ 176\%). If we only apply  a $B_0$ prior, the dispersion is 0.006 ($\sim$ 46\%). 

\comment{CHECK THE WAY TO QUOTE THE DISPERSION (0.025/0.009 $\sim$ 277\%), (0.020/0.009 $\sim$ 222\%) and (0.005/0.009 $\sim$ 55\%), for $\Delta \sigma_\epsilon$, 0.029/0.013$\sim$ 223\%), (0.023/0.013 $\sim$176\%) and (0.006/0.013$\sim$ 46\%)
CHECK ALSO THIS COMPUTATION no priors (0.007/0.019 $\sim$ 36\%), only beta (0.004/0.019 $\sim$  21\%) and only $B_0$ (0.003/0.019$\sim$15\%);}

\begin{table*}
\caption{
The best fit errors with numerous variations of our fiducial fitting methodology for DR11 mock galaxy catalogs pre-reconstruction without covariance corrections corrections for the overlapping mock regions. Each methodology test has been described earlier in Section~\ref{sec:sysfit};  we clarify some of the less obvious ones here: ``$P_{pt}(k)$ floating'' refers to using RPT-inspired $P_{\rm pt}(k)$  template with $\beta$ floating and ``$P_{pt}(k)$ fixed" with $\beta$ fixed to $0.4$ for pre-reconstructed and $\beta=0$ post-reconstruction, ``exp'' and ``Gaussian'' refers to various Finger-of-God models as described in Section~\ref{sec:streaming_sys}. The columns list the median and 16th and 84th percentiles of the $\sigma_{\alpha,\epsilon, ||, \perp}$ and variations $\Delta v=v_{i}-v_{f}$. Except for the fiducial case all quantities are multiplied by 100.}
\label{tab:sigmadr11norec}
\begin{tabular}{@{}lrrrrrrrr}

\hline
Model&
$\widetilde{\sigma_{\alpha}}$&
$\widetilde{\Delta \sigma_{\alpha}}$&
$\widetilde{\sigma_{\epsilon}}$&
$\widetilde{\Delta \sigma_{\epsilon}}$&
$\widetilde{\sigma_{\alpha_{\parallel}}}$&
$\widetilde{\Delta \sigma_{\alpha_{\parallel}}}$&
$\widetilde{\sigma_{\alpha_{\perp}}}$&
$\widetilde{\Delta \sigma_{\alpha_{\perp}}}$\\

\hline
\multicolumn{9}{c}{DR11 Pre-Reconstruction}\\
\hline
\\[-1.5ex]

Fiducial&
$0.0157^{+0.0037}_{-0.0023}$&-&
$0.0186^{+0.0051}_{-0.0024}$&-&
$0.0438^{+0.0133}_{-0.0069}$&-&
$0.0211^{+0.0038}_{-0.0022}$&-\\
\\[-1.5ex]

$r_{bin}\rightarrow 4\mathrm{Mpc}/h$&
$1.49^{+0.36}_{-0.20}$&
$-0.08^{+0.08}_{-0.09}$&
$1.76^{+0.48}_{-0.21}$&
$-0.09^{+0.10}_{-0.11}$&
$4.23^{+1.27}_{-0.69}$&
$-0.16^{+0.27}_{-0.33}$&
$1.99^{+0.29}_{-0.21}$&
$-0.13^{+0.08}_{-0.11}$\\
\\[-1.5ex]

$30<r<200\mathrm{Mpc}/h$&
$1.59^{+0.45}_{-0.24}$&
$0.02^{+0.08}_{-0.05}$&
$1.92^{+0.63}_{-0.26}$&
$0.04^{+0.16}_{-0.06}$&
$4.57^{+1.68}_{-0.78}$&
$0.13^{+0.40}_{-0.21}$&
$2.11^{+0.40}_{-0.24}$&
$-0.01^{+0.05}_{-0.04}$\\
\\[-1.5ex]

$\mathrm{2-term} \; A_\ell(r)$&
$1.64^{+0.39}_{-0.24}$&
$0.06^{+0.07}_{-0.06}$&
$1.87^{+0.59}_{-0.25}$&
$0.00^{+0.12}_{-0.08}$&
$4.43^{+1.63}_{-0.89}$&
$-0.04^{+0.40}_{-0.19}$&
$2.15^{+0.48}_{-0.29}$&
$0.02^{+0.19}_{-0.11}$\\
\\[-1.5ex]

$\mathrm{4-term} \; A_\ell(r)$&
$1.57^{+0.41}_{-0.23}$&
$0.01^{+0.04}_{-0.04}$&
$1.89^{+0.57}_{-0.24}$&
$0.04^{+0.08}_{-0.05}$&
$4.45^{+1.52}_{-0.71}$&
$0.08^{+0.18}_{-0.11}$&
$2.14^{+0.37}_{-0.24}$&
$0.02^{+0.04}_{-0.03}$\\
\\[-1.5ex]

$\mathrm{Fixed \;} \beta=0.4$&
$1.56^{+0.37}_{-0.22}$&
$-0.01^{+0.01}_{-0.02}$&
$1.84^{+0.48}_{-0.24}$&
$-0.02^{+0.01}_{-0.02}$&
$4.35^{+1.34}_{-0.66}$&
$-0.03^{+0.07}_{-0.09}$&
$2.10^{+0.37}_{-0.22}$&
$-0.02^{+0.04}_{-0.04}$\\
\\[-1.5ex]

$\spar=\sperp=8\mathrm{Mpc}/h$&
$1.54^{+0.37}_{-0.21}$&
$-0.02^{+0.05}_{-0.06}$&
$1.79^{+0.59}_{-0.25}$&
$-0.06^{+0.10}_{-0.09}$&
$3.99^{+1.47}_{-0.73}$&
$-0.38^{+0.25}_{-0.22}$&
$2.30^{+0.39}_{-0.26}$&
$0.17^{+0.09}_{-0.08}$\\
\\[-1.5ex]

$\Sigma_s \rightarrow 3.0\mathrm{Mpc}/h$&
$1.59^{+0.38}_{-0.23}$&
$0.02^{+0.02}_{-0.01}$&
$1.92^{+0.52}_{-0.24}$&
$0.06^{+0.03}_{-0.02}$&
$4.57^{+1.38}_{-0.73}$&
$0.17^{+0.07}_{-0.05}$&
$2.13^{+0.39}_{-0.22}$&
$0.02^{+0.01}_{-0.01}$\\
\\[-1.5ex]

$\mathrm{No \; Priors}$&
$1.77^{+0.84}_{-0.37}$&
$0.14^{+0.51}_{-0.10}$&
$2.14^{+1.16}_{-0.42}$&
$0.20^{+0.67}_{-0.13}$&
$4.90^{+3.04}_{-1.18}$&
$0.43^{+1.74}_{-0.41}$&
$2.29^{+0.75}_{-0.36}$&
$0.16^{+0.42}_{-0.15}$\\
\\[-1.5ex]

$\mathrm{Only} \; \log (B_0^2) \; \mathrm{prior}$&
$1.66^{+0.59}_{-0.27}$&
$0.07^{+0.26}_{-0.05}$&
$2.01^{+0.82}_{-0.31}$&
$0.14^{+0.34}_{-0.09}$&
$4.66^{+2.12}_{-1.02}$&
$0.21^{+0.97}_{-0.25}$&
$2.23^{+0.60}_{-0.33}$&
$0.10^{+0.28}_{-0.13}$\\
\\[-1.5ex]

$\mathrm{Only} \; \beta \; \mathrm{prior}$&
$1.64^{+0.68}_{-0.29}$&
$0.05^{+0.29}_{-0.04}$&
$1.96^{+0.93}_{-0.32}$&
$0.05^{+0.41}_{-0.05}$&
$4.68^{+2.52}_{-0.94}$&
$0.14^{+1.24}_{-0.12}$&
$2.16^{+0.46}_{-0.25}$&
$0.03^{+0.08}_{-0.02}$\\
\\[-1.5ex]

$\mathrm{P_{pt}(k) \;with \; floating \;} \beta$&
$1.53^{+0.37}_{-0.21}$&
$-0.03^{+0.04}_{-0.05}$&
$1.86^{+0.54}_{-0.24}$&
$0.00^{+0.06}_{-0.07}$&
$4.20^{+1.39}_{-0.70}$&
$-0.15^{+0.16}_{-0.18}$&
$2.23^{+0.37}_{-0.25}$&
$0.10^{+0.07}_{-0.07}$\\
\\[-1.5ex]

$\mathrm{P_{pt}(k)\; with \;} \beta=0.0$&
$1.51^{+0.37}_{-0.20}$&
$-0.04^{+0.05}_{-0.06}$&
$1.84^{+0.52}_{-0.24}$&
$-0.02^{+0.07}_{-0.08}$&
$4.15^{+1.30}_{-0.64}$&
$-0.20^{+0.15}_{-0.23}$&
$2.23^{+0.35}_{-0.25}$&
$0.10^{+0.07}_{-0.08}$\\
\\[-1.5ex]

$\mathrm{FoG \; model} \rightarrow exp$&
$1.57^{+0.37}_{-0.23}$&
$-0.00^{+0.00}_{-0.00}$&
$1.85^{+0.52}_{-0.24}$&
$-0.01^{+0.00}_{-0.00}$&
$4.35^{+1.33}_{-0.69}$&
$-0.02^{+0.01}_{-0.01}$&
$2.11^{+0.38}_{-0.22}$&
$-0.00^{+0.00}_{-0.00}$\\
\\[-1.5ex]

$\mathrm{FoG \; model} \rightarrow gauss$&
$1.57^{+0.37}_{-0.23}$&
$0.00^{+0.00}_{-0.00}$&
$1.86^{+0.51}_{-0.24}$&
$0.00^{+0.00}_{-0.00}$&
$4.38^{+1.33}_{-0.69}$&
$0.00^{+0.00}_{-0.00}$&
$2.11^{+0.38}_{-0.22}$&
$0.00^{+0.00}_{-0.00}$\\
\\[-1.5ex]

\hline

\end{tabular}
\end{table*}

 \begin{table*}
\caption{The best fit errors with numerous variations of our fiducial fitting methodology for DR11 mock galaxy catalogs  post-reconstruction without covariance corrections corrections for the overlapping mock regions. Each methodology tests has been described earlier in Section~\ref{sec:sysfit};  we clarify some of the less obvious ones here: ``$P_{pt}(k)$ floating'' refers to using RPT-inspired $P_{\rm pt}(k)$  template with $\beta$ floating and ``$P_{pt}(k)$ fixed" with $\beta$ fixed to $0.4$ for pre-reconstructed and $\beta=0$ post-reconstruction, ``exp'' and ``Gaussian'' refers to various Finger-of-God models as described in Section~\ref{sec:streaming_sys}. The columns lis the median and 16th and 84th percentiles of the $\sigma_{\alpha,\epsilon, ||, \perp}$ and variations $\Delta v=v_{i}-v_{f}$. Except for the fiducial case all quantities are multiplied by 100.}
\label{tab:sigmadr11rec} 
\begin{tabular}{@{}lrrrrrrrr}

\hline
Model&
$\widetilde{\sigma_{\alpha}}$&
$\widetilde{\Delta \sigma_{\alpha}}$&
$\widetilde{\sigma_{\epsilon}}$&
$\widetilde{\Delta \sigma_{\epsilon}}$&
$\widetilde{\sigma_{\alpha_{\parallel}}}$&
$\widetilde{\Delta \sigma_{\alpha_{\parallel}}}$&
$\widetilde{\sigma_{\alpha_{\perp}}}$&
$\widetilde{\Delta \sigma_{\alpha_{\perp}}}$\\

\hline
\multicolumn{9}{c}{DR11 Post-Reconstruction}\\
\hline
\\[-1.5ex]

Fiducial&
$0.0093^{+0.0013}_{-0.0011}$&-&
$0.0128^{+0.0024}_{-0.0014}$&-&
$0.0280^{+0.0064}_{-0.0041}$&-&
$0.0151^{+0.0021}_{-0.0015}$&-\\
\\[-1.5ex]

$r_{bin} \rightarrow 4\mathrm{Mpc}/h$&
$0.87^{+0.12}_{-0.10}$&
$-0.07^{+0.04}_{-0.05}$&
$1.22^{+0.19}_{-0.13}$&
$-0.06^{+0.06}_{-0.07}$&
$2.67^{+0.57}_{-0.34}$&
$-0.13^{+0.17}_{-0.21}$&
$1.41^{+0.15}_{-0.12}$&
$-0.10^{+0.06}_{-0.07}$\\
\\[-1.5ex]

$30<r<200\mathrm{Mpc}/h$&
$0.91^{+0.12}_{-0.09}$&
$-0.03^{+0.02}_{-0.03}$&
$1.26^{+0.21}_{-0.13}$&
$-0.01^{+0.03}_{-0.04}$&
$2.79^{+0.57}_{-0.37}$&
$-0.00^{+0.11}_{-0.13}$&
$1.46^{+0.19}_{-0.14}$&
$-0.04^{+0.03}_{-0.03}$\\
\\[-1.5ex]

$\mathrm{2-term} \; A_\ell(r)$&
$0.96^{+0.13}_{-0.11}$&
$0.02^{+0.02}_{-0.02}$&
$1.25^{+0.20}_{-0.14}$&
$-0.04^{+0.03}_{-0.05}$&
$2.80^{+0.58}_{-0.41}$&
$-0.00^{+0.08}_{-0.12}$&
$1.44^{+0.18}_{-0.13}$&
$-0.06^{+0.02}_{-0.04}$\\
\\[-1.5ex]

$\mathrm{4-term} \; A_\ell(r)$&
$0.92^{+0.14}_{-0.10}$&
$-0.01^{+0.02}_{-0.02}$&
$1.27^{+0.26}_{-0.14}$&
$0.00^{+0.04}_{-0.03}$&
$2.77^{+0.69}_{-0.39}$&
$-0.02^{+0.09}_{-0.09}$&
$1.52^{+0.20}_{-0.15}$&
$0.00^{+0.02}_{-0.01}$\\
\\[-1.5ex]

$\mathrm{Fixed \;} \beta=0.0$&
$0.92^{+0.13}_{-0.10}$&
$-0.01^{+0.00}_{-0.02}$&
$1.26^{+0.20}_{-0.13}$&
$-0.02^{+0.02}_{-0.04}$&
$2.74^{+0.54}_{-0.32}$&
$-0.04^{+0.09}_{-0.15}$&
$1.48^{+0.19}_{-0.14}$&
$-0.03^{+0.03}_{-0.03}$\\
\\[-1.5ex]

$\spar=4\; \& \; \sperp=2$&
$0.93^{+0.14}_{-0.10}$&
$0.00^{+0.01}_{-0.01}$&
$1.28^{+0.22}_{-0.14}$&
$-0.00^{+0.01}_{-0.02}$&
$2.84^{+0.62}_{-0.40}$&
$0.03^{+0.03}_{-0.04}$&
$1.48^{+0.21}_{-0.15}$&
$-0.03^{+0.01}_{-0.01}$\\
\\[-1.5ex]

$\Sigma_s \rightarrow 3.0\mathrm{Mpc}/h$&
$0.95^{+0.13}_{-0.11}$&
$0.02^{+0.01}_{-0.01}$&
$1.35^{+0.23}_{-0.15}$&
$0.06^{+0.02}_{-0.02}$&
$2.97^{+0.67}_{-0.42}$&
$0.16^{+0.06}_{-0.05}$&
$1.52^{+0.20}_{-0.15}$&
$0.02^{+0.01}_{-0.01}$\\
\\[-1.5ex]

$\mathrm{No \; prior}$&
$1.30^{+2.52}_{-0.39}$&
$0.31^{+2.46}_{-0.27}$&
$1.73^{+3.02}_{-0.49}$&
$0.37^{+2.90}_{-0.32}$&
$4.21^{+8.99}_{-1.65}$&
$1.15^{+8.55}_{-1.09}$&
$1.64^{+0.36}_{-0.23}$&
$0.11^{+0.19}_{-0.06}$\\
\\[-1.5ex]

$\mathrm{Only} \; \log(B_0^2) \; \mathrm{prior}$&
$1.00^{+0.52}_{-0.14}$&
$0.04^{+0.47}_{-0.03}$&
$1.38^{+0.77}_{-0.21}$&
$0.07^{+0.65}_{-0.05}$&
$3.04^{+2.35}_{-0.66}$&
$0.15^{+2.02}_{-0.19}$&
$1.59^{+0.26}_{-0.19}$&
$0.06^{+0.09}_{-0.04}$\\
\\[-1.5ex]

$\mathrm{Only} \; \beta \; \mathrm{prior}$&
$1.21^{+2.10}_{-0.33}$&
$0.22^{+2.02}_{-0.21}$&
$1.62^{+2.49}_{-0.41}$&
$0.25^{+2.33}_{-0.23}$&
$3.86^{+7.64}_{-1.28}$&
$0.86^{+6.95}_{-0.79}$&
$1.55^{+0.28}_{-0.18}$&
$0.02^{+0.11}_{-0.01}$\\
\\[-1.5ex]

$P_{pt}(k) \mathrm{\; with \; floating \;} \beta$&
$0.86^{+0.12}_{-0.09}$&
$-0.08^{+0.04}_{-0.04}$&
$1.17^{+0.20}_{-0.12}$&
$-0.11^{+0.07}_{-0.08}$&
$2.57^{+0.57}_{-0.33}$&
$-0.22^{+0.17}_{-0.21}$&
$1.38^{+0.16}_{-0.11}$&
$-0.13^{+0.06}_{-0.07}$\\
\\[-1.5ex]

$P_{pt}(k)\mathrm{\; with \;} \beta=0.0$&
$0.91^{+0.13}_{-0.10}$&
$-0.02^{+0.01}_{-0.02}$&
$1.20^{+0.20}_{-0.12}$&
$-0.07^{+0.03}_{-0.05}$&
$2.65^{+0.53}_{-0.33}$&
$-0.14^{+0.10}_{-0.16}$&
$1.45^{+0.20}_{-0.14}$&
$-0.05^{+0.03}_{-0.03}$\\
\\[-1.5ex]

\hline

\end{tabular}
\end{table*}

We summarize the results on the best fit uncertainties in Table.~\ref{tab:sigmadr11norec} and Table.~\ref{tab:sigmadr11rec}. In particular, we concentrate on the results from testing the methodological changes on DR11 pre-and post-reconstruction mock galaxy catalogs. 
Table.~\ref{tab:sigmadr11norec} and Table.~\ref{tab:sigmadr11rec} demonstrates that only a few cases show variations in $\sigma_\alpha$; the effects of different robustness tests affect mostly the $\sigma_\epsilon$.  
In particular, there are large variations in the cases where we change our prior assumptions; 
these large variations will be discussed in the following section devoted to the priors (Section~\ref{sec:priors_errors}). Here we present only cases not related to priors assumptions in the fitting methodology.

When  fitting using DR11 pre-reconstructed mock galaxy catalogs,  
$\sigma_\alpha$ is only affected when we apply lower order polynomials for broad-band terms, and the bin sizes are changed. In both cases, $\sigma_\alpha$ shows a variation  of  $0.001/0.016 \sim 6\% $.  
The quantity $\sigma_\epsilon$  displays ($\Delta \sigma_\epsilon=\pm0.001$) 5\% variation in three different cases: 
using smaller bins (when we bin monopoles and quadrupoles), changing the value of streaming parameter or changing the non linear damping parameters.  
The variations when fitting using DR11 pre-reconstructed mock galaxy catalogs,  we do not produce variations in $\sigma_\alpha$, and only in a few cases, we find a small variation of $\sigma_\epsilon$  at $\sim 7\%$ level ($\Delta \sigma_\epsilon=0.001/0.013$) when we switch to using smaller bin size in binning the correlation function, change $\Sigma_s$  and when we apply the $P_{\rm pt}(k)$ non-linear power-spectrum template. 


\subsection{Discussion of Individual Robustness Tests} 

We will now turn to the discussion of individual robustness tests to examine the effects of these methodological changes. These tests are listed in Section~\ref{sec:sysfit} in the same order as the following discussion.  We will quote all results from DR11, as DR10 behaves similarly, and the discussion of DR10 results will be relegated to the Appendix. 

\subsubsection{Model Templates} \label{sec:template_results} 

Table~\ref{tab:dr11norec} shows that  changing from the ``De-Wiggled" template to the RPT-inspired $P_{\rm pt}(k)$ template, the best fit values  using pre-reconstructed DR11 mock galaxy catalogs are slightly affected, on the order of $0.5\%$ on $\alpha$, and $\le 0.1\%$ on $\epsilon$, while the best fit values are changed by $\le 0.1\%$ once we use post-reconstructed mock catalogs (Table~\ref{tab:dr11rec}).  The fitted uncertainties in both pre- and post-reconstructed catalogs are well within $0.1\%$. 

In addition, the best fit results are slightly biased (we compare measured $\alpha$ to 1 and $\epsilon$ to 0 as the input cosmology of the mock catalogs is known), and the``De-Wiggled" template produces biased best fit values in the opposite direction ($\widetilde \alpha = 1.0049$, $\widetilde \epsilon = 0.0027$) when compared with bias in best fit values ($\widetilde \alpha = 0.996 $, $\widetilde \epsilon = -0.002$) using $P_{\rm pt}(k)$ template. 
This inverted trend generates quite different results in term of $\alpha_{||}-\alpha_{\perp}$ parametrization. For the ``De-Wiggled" template, the bias in best fit $\alphapar$ reaches 1.1\% but only 0.2\% shift in $\alphaper$, while using $P_{\rm pt}(k)$ template produces 0.6 \% in $\alphapar$ and $-0.3 \%$ in $\alphaper$. 
After reconstruction, the templates have consistent results, the bias on the best fit $\alpha$ reduces to $<0.1\%$ and $\epsilon$ has a slightly larger bias with ``De-Wiggled" template of 0.2\% compared to 0.1\% for the $P_{\rm pt}(k)$  template. 

The information in Table~\ref{tab:sigmadr11norec} and Table~\ref{tab:sigmadr11rec} demonstrates  that the change of non-linear power-spectrum template does not significantly affect the uncertainties on the best fit parameters for both pre- and post-reconstruction; the changes are well within $\le 0.1\%$ for all cases. 

\subsubsection{Fitting Range and Bin Sizes}   \label{sec:fitting_range_results}

\cite{And13}, found that the optimal fitting range for anisotropic clustering in DR9 mock galaxy catalog is $[50,200]$ $h^{-1}$Mpc, and that using $[30,200]$ $h^{-1}$Mpc produces a more biased measurement of $\alpha$ and $\epsilon$. 
However, when find that fitting the DR11 pre-reconstructed mock galaxy catalogs using $[30,200]$ $h^{-1}$Mpc yields less biased best fit values. 
This result is unexpected since it is the opposite sense of the DR9 findings (Table~\ref{tab:dr11norec}). 
On the other hand, the differences between using one fitting range and another are consistently small in both the DR9 \citep{And13} and the DR11 pre-reconstructed galaxy mock catalogs.  Furthermore, once we apply reconstruction to the galaxy catalogs, the differences in best fit values between using different fitting ranges are well within $0.1\%$ ($0.3\%$) for DR11 (DR9)  (see Table~\ref{tab:dr11rec} for more details). 
Applying the larger fitting ranges to the post-reconstructed galaxy catalogs, produces a slight increase in the bias in best fit values. In particular, the bias on the best fit $\alpha$ ($\epsilon$)  is 0.06\% ($0.1\%$). 
Table~\ref{tab:sigmadr11norec} and Table~\ref{tab:sigmadr11rec}, shows that changing the fitting range does not affect the estimated uncertainties of any of the fitted parameters both pre- and post-reconstruction. 
Table~\ref{tab:dr11norec} and Table~\ref{tab:dr11rec} demonstrate that  using smaller bins has a negligible effect in the best fit values. The variations on the errors are also small; identical results are obtained when fitting both pre- and post-reconstructed mock catalogs.

\subsubsection{Nuisance Terms Model} \label{sec:nuisance_terms_results} 

\comment{This section can be updated with Ross polygons perhaps within the 3-weeks period not before}

The effects of the broadband modeling are most prominent when one examines the best fit values on post-reconstructed mock galaxy catalogs as shown in Tables.~\ref{tab:dr11norec} and~\ref{tab:dr11rec}. 
The fiducial fitting methodology uses three terms, providing relatively unbiased best fit values of $\alpha$ and $\epsilon$.  For the pre-reconstructed galaxy catalogs, biases of best fit values are $\le 0.5\%$. For post-reconstructed galaxy catalogs, the best fit values are only biased by $\le 0.2\%$.  
Varying the number of terms included in the broad-band modeling, produces little effect on the best fit values using pre-reconstructed mock galaxy catalogs. 
However, when we use post-reconstructed mock galaxy catalogs, increasing the number of terms removes the bias on best fit $\epsilon$ completely, while decreasing the number of terms in the broad-band modeling to 2-terms increases the bias of the best fit $\epsilon$ by $0.3\%$. 

\subsubsection{Priors}   \label{sec:priors_errors}

Table.~\ref{tab:dr11norec} and Table.~\ref{tab:dr11rec}  show that the application of different priors have a relatively large effect on uncertainties of the best fit values, especially on  the uncertainties in best fit $\epsilon$. 
For example, applying the fiducial methodology to DR11 pre-reconstructed mock galaxy catalogs, producesmedian variations of $\Delta \alpha\sim 0.001$, $\Delta \epsilon \le 0.002$. These increase to as large as $\Delta \alpha \sim 0.002$ and $\Delta \epsilon \le $0.004 in $\epsilon$ when we apply the same methodology to post-reconstructed mock catalogs.  
In addition, a large dispersion is observed in $\sigma_\alpha$ and $\sigma_\epsilon$ among the results where we apply fitting to reconstructed mock galaxy catalogs. The large dispersion observed  is also quite obvious in the dispersion plots  in Fig.~\ref{fig:sigmadr11v1ae}  and Fig.~\ref{fig:sigmadr11v2ae} for DR11. The presence of``column'' structures in the dispersion plot indicating the relatively large difference between some of the mocks.

\begin{table*}
\caption{\comment{MISSING cases with $B_0$ and $\beta$ prior, with no RL!!  AND $B_0$, $\beta$ + $\alpha$ prior, but not epsilon prior!} 
The median variations of the fitted values and the fitted errors from DR11 reconstructed mocks for different changes on the fiducial fitting methodology.  The variation is defined as  $\Delta v=v^{i}-v^{fid}$, where $v=\alpha$, $\epsilon$,$\alpha_{\parallel}$, $\alpha_{\perp}$,$ \sigma_{||}$, $\sigma_{\perp}$ and $i$, denotes the modification in the methodology being tested. 
The median variations $\widetilde{\Delta v}$, and percentiles are multiplied by 100. Note that (RL) stands for calculating the errors by integrating over specific intervals in the likelihood surfaces in 
$\alpha$-$\epsilon$,  $\alpha=[0.8,1.2]$ and $\epsilon=[-0.15,0.15]$.  
\comment{and maybe there are other cases to be included?? we only copied Aardwolf, but Aardwolf said that we will discuss more cases in THIS paper! We should take the table that you already sent to Daniel a while ago (Crosschecking.pdf) and put in some of the cases there}}
\label{tab:priorsdeW}

\begin{tabular}{@{}lrrrrrrrr}

\hline
%
%
Model&
$\widetilde{\Delta \alpha}$&
$\widetilde{\Delta \sigma_{\alpha}}$&
$\widetilde{\Delta \epsilon}$&
$\widetilde{\Delta \sigma_{\epsilon}}$&
$\widetilde{\Delta \alpha_{\parallel}}$&
$\widetilde{\Delta \sigma_{\alpha_{\parallel}}}$&
$\widetilde{\Delta \alpha_{\perp}}$&
$\widetilde{\Delta \sigma_{\alpha_{\perp}}}$\\
\\[-1.5ex]
\hline

$\mathrm{Only}\; \alpha\; \mathrm{prior}$&
$0.00^{+0.05}_{-0.02}$&
$0.31^{+2.46}_{-0.27}$&
$-0.04^{+0.14}_{-0.15}$&
$0.37^{+2.90}_{-0.32}$&
$-0.08^{+0.28}_{-0.26}$&
$1.15^{+8.55}_{-1.09}$&
$0.03^{+0.18}_{-0.14}$&
$0.11^{+0.19}_{-0.06}$\\
\\[-1.5ex]

$\mathrm{Only}\;B_0\; \alpha\;\mathrm{priors}$&
$0.01^{+0.04}_{-0.01}$&
$0.04^{+0.47}_{-0.03}$&
$-0.03^{+0.13}_{-0.13}$&
$0.07^{+0.65}_{-0.05}$&
$-0.05^{+0.26}_{-0.23}$&
$0.15^{+2.02}_{-0.19}$&
$0.03^{+0.16}_{-0.12}$&
$0.06^{+0.09}_{-0.04}$\\
\\[-1.5ex]

$\mathrm{Only}\;\beta \;\alpha\; \mathrm{priors}$&
$-0.01^{+0.01}_{-0.01}$&
$0.22^{+2.02}_{-0.21}$&
$-0.00^{+0.02}_{-0.03}$&
$0.25^{+2.33}_{-0.23}$&
$-0.01^{+0.03}_{-0.07}$&
$0.86^{+6.95}_{-0.79}$&
$0.00^{+0.02}_{-0.02}$&
$0.02^{+0.11}_{-0.01}$\\
\\[-1.5ex]
\hline
$\mathrm{No \; priors}$&
$0.00^{+0.05}_{-0.02}$&
$0.95^{+4.61}_{-0.87}$&
$-0.04^{+0.14}_{-0.15}$&
$1.05^{+5.46}_{-0.95}$&
$-0.08^{+0.28}_{-0.26}$&
$3.28^{+15.77}_{-3.05}$&
$0.03^{+0.18}_{-0.14}$&
$0.13^{+0.37}_{-0.08}$\\
\\[-1.5ex]

$\mathrm{Only} \;B_0\; \mathrm{prior}$&
$0.01^{+0.04}_{-0.01}$&
$0.07^{+0.91}_{-0.05}$&
$-0.03^{+0.13}_{-0.13}$&
$0.09^{+1.11}_{-0.08}$&
$-0.05^{+0.26}_{-0.23}$&
$0.24^{+3.48}_{-0.28}$&
$0.03^{+0.16}_{-0.12}$&
$0.07^{+0.12}_{-0.04}$\\
\\[-1.5ex]

$\mathrm{Only} \; \beta \; \mathrm{prior}$&
$-0.01^{+0.01}_{-0.01}$&
$0.73^{+4.53}_{-0.68}$&
$-0.00^{+0.02}_{-0.03}$&
$0.84^{+5.15}_{-0.77}$&
$-0.01^{+0.03}_{-0.07}$&
$2.67^{+15.15}_{-2.45}$&
$0.00^{+0.02}_{-0.02}$&
$0.03^{+0.29}_{-0.02}$\\
\\[-1.5ex]
\hline
$\mathrm{No \; priors(RL)}$&
$0.00^{+0.05}_{-0.02}$&
$0.03^{+0.10}_{-0.02}$&
$-0.04^{+0.14}_{-0.15}$&
$0.07^{+0.22}_{-0.05}$&
$-0.08^{+0.28}_{-0.26}$&
$0.11^{+0.72}_{-0.14}$&
$0.03^{+0.18}_{-0.14}$&
$0.07^{+0.10}_{-0.04}$\\
\\[-1.5ex]

$\mathrm{Only} \;B_0\; \mathrm{prior(RL)}$&
$0.01^{+0.04}_{-0.01}$&
$0.03^{+0.08}_{-0.02}$&
$-0.03^{+0.13}_{-0.13}$&
$0.05^{+0.16}_{-0.03}$&
$-0.05^{+0.26}_{-0.23}$&
$0.07^{+0.51}_{-0.13}$&
$0.03^{+0.16}_{-0.12}$&
$0.05^{+0.08}_{-0.03}$\\
\\[-1.5ex]

$\mathrm{Only} \; \beta \; \mathrm{prior(RL)}$&
$-0.01^{+0.01}_{-0.01}$&
$0.00^{+0.01}_{-0.00}$&
$-0.00^{+0.02}_{-0.03}$&
$0.02^{+0.03}_{-0.01}$&
$-0.01^{+0.03}_{-0.07}$&
$0.03^{+0.07}_{-0.02}$&
$0.00^{+0.02}_{-0.02}$&
$0.01^{+0.02}_{-0.01}$\\
\\[-1.5ex]
\hline
$\beta\;\;B_0$&
$0.00^{+0.00}_{0.00}$&
$0.01^{+0.20}_{-0.01}$&
$0.00^{+0.00}_{0.00}$&
$0.01^{+0.25}_{-0.01}$&
$0.00^{+0.00}_{0.00}$&
$0.03^{+0.85}_{-0.03}$&
$0.00^{+0.00}_{0.00}$&
$0.00^{+0.01}_{-0.00}$\\
\\[-1.5ex]
\hline
$\beta\;\epsilon\;B_0$&
$0.00^{+0.00}_{0.00}$&
$0.00^{+0.00}_{-0.00}$&
$0.00^{+0.00}_{0.00}$&
$0.00^{+0.00}_{-0.00}$&
$0.00^{+0.00}_{0.00}$&
$0.00^{+0.01}_{-0.00}$&
$0.00^{+0.00}_{0.00}$&
$0.00^{+0.00}_{-0.00}$\\
\\[-1.5ex]

$\mathrm{Only}\;\epsilon$&
$0.00^{+0.05}_{-0.02}$&
$0.03^{+0.13}_{-0.02}$&
$-0.04^{+0.14}_{-0.15}$&
$0.07^{+0.24}_{-0.05}$&
$-0.08^{+0.28}_{-0.26}$&
$0.12^{+0.77}_{-0.15}$&
$0.03^{+0.18}_{-0.14}$&
$0.07^{+0.10}_{-0.04}$\\
\\[-1.5ex]

$\epsilon\;B_0$&
$0.01^{+0.04}_{-0.01}$&
$0.03^{+0.09}_{-0.02}$&
$-0.03^{+0.13}_{-0.13}$&
$0.05^{+0.17}_{-0.03}$&
$-0.05^{+0.26}_{-0.23}$&
$0.08^{+0.58}_{-0.13}$&
$0.03^{+0.16}_{-0.12}$&
$0.05^{+0.08}_{-0.03}$\\
\\[-1.5ex]

$\epsilon\;\beta$&
$-0.01^{+0.01}_{-0.01}$&
$0.00^{+0.02}_{-0.00}$&
$-0.00^{+0.02}_{-0.03}$&
$0.02^{+0.04}_{-0.01}$&
$-0.01^{+0.03}_{-0.07}$&
$0.03^{+0.09}_{-0.02}$&
$0.00^{+0.02}_{-0.02}$&
$0.01^{+0.02}_{-0.01}$\\
\\[-1.5ex]

\hline

\end{tabular}
\end{table*}


In order to explore the origin of these relatively large variations and dispersions we expand our investigation into  the priors-related cases using  DR11 post-reconstructed mock catalogs. The results are listed in Table.~\ref{tab:priorsdeW}. 
In addition to test cases shown in previous Tables (No priors, only $\beta$ and only $B_0$), we add similar cases switching on $\alpha$ prior (discussed in Section~\ref{sec:sysfit} and Section~\ref{sec:prior_alpha}) and $\epsilon$ prior (Section~\ref{sec:sysfit} and Section~\ref{sec:prior_epsilon}). 
We also include the same test cases where a large fluctuation is observed,  but we restrict our integration intervals in the likelihood surfaces when the uncertainties are calculated for the best fit parameters. We choose the range for our integration intervals by restricting ourselves only to the fitting ranges, limiting $\alpha$ and $\epsilon$ to ranges which would not lead us outside our fitting range of [50,200] $h^{-1}$Mpc. These cases are denoted as ``Range Limited'' (RL).  The reason for these Range Limited cases will be discussed later. 
To summarize from Table.~\ref{tab:priorsdeW}, when we apply  $\beta$ and $B_0$ prior  without any $\alpha$ or $\epsilon$ prior, the fitting of DR11 post-reconstructed data produces 1.4\%\; rms for $\epsilon$, compared to $1.3\%$ when we include in addition the  $\alpha$ and $\epsilon$ prior are included. 

\comment{need to add the case of $B_0$, beta prior only (NO RL) for this discussion, and why are we missing all the $\epsilon$ prior case? In general, we need a more complete table for this discussion}

\begin{figure*}
   \centering     
   \includegraphics[width=5.0in]{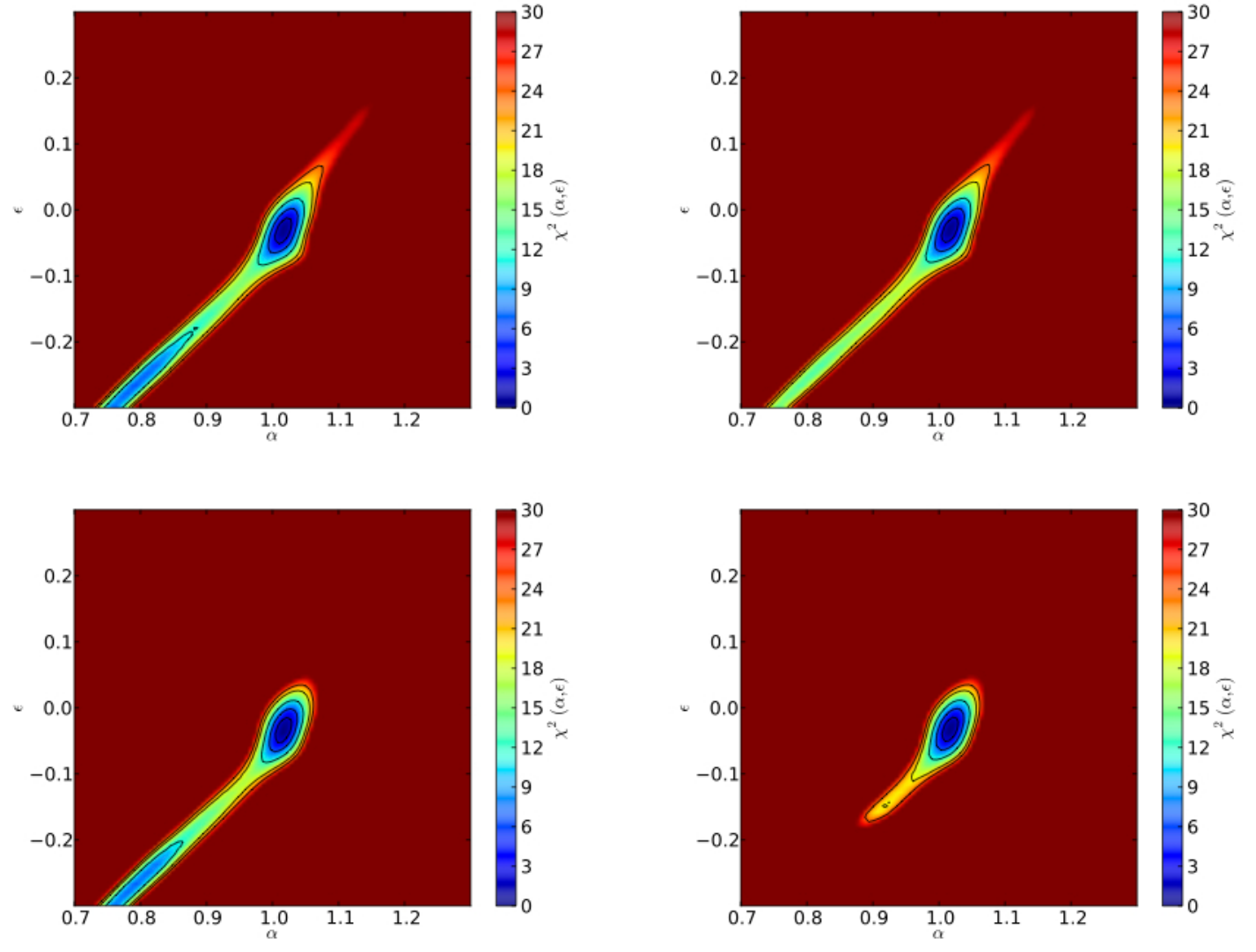} 
   \caption{Contours plots of $\Delta \chi^2(\alpha,\epsilon)$.  Top left panel shows the $\Delta \chi^2(\alpha,\epsilon)$  when no priors are applied  in the fitting methodology on DR11 post-reconstructed CMASS dataset,  the top right panel applies only $B_0$ prior, the bottom left panel applies only $\beta$ prior, and the bottom right utilizes all priors.
  \comment{Shirley:  ADDED the 2D plot, because it shows the degeneracy direction for alpha,epsilon and it explains why we restrict things to +/- 0.15 for epsilon!!  And please make the 2D plot's axes a LOT larger, it is impossible to see the alpha-epsilon label as it is right now.} }
   \label{fig:chisq_ae}
\end{figure*}

\comment{ Shirley: ADDED  discussion and 2D plot here, basically we don' t pull the +/- 0.15 (or +/-0.2)  out of a hat, we have to explain why this is the case }  

Given the seemingly strong dependence of our fit on the various priors, we examine the 2D $\Delta \chi^2$ surfaces of $\alpha$ and $\epsilon$. 
Fig.~\ref{fig:chisq_ae} shows that the 2D likelihood surfaces are highly degenerate along $\alpha=1+\epsilon$ direction. 
The long tail when we do not apply all of the priors  corresponds to large variations in $\alphapar$, but only small
variations in $\alphaper$.  The variations in $\alphapar$  for the extreme cases are of order $1.25^3$, which is $\approx$2.  
In other words, these cases correspond to places where the acoustic peak along the line of
sight has been shifted out of our [50-200] $h^{-1}$Mpc fitting range.
The asymmetry toward small $\alpha,\epsilon$ refers to the case when the peak shifts to larger apparent
scale where the errors are larger. One therefore should not be surprised that when the data lacks a good
acoustic peak along the line of sight, that the fitter can place one at
huge scale, beyond our fitted range of correlation function scales.
This feature motivates to placing of a bound on $\epsilon$, which is not a cosmological prior, but rather, it is  a 
statement of where we have ``searched''.   
Therefore, we propose to examine the effects of integrating over a range (with flat priors) in both $\alpha$ and $\epsilon$ to calculate uncertainties on these parameters. 
We adopt an integration interval of $\alpha=[0.8,1.2]$ and $\epsilon=[-0.15,0.15]$ \comment{CHECK interval}, which  corresponds to a maximum dilation of $\sim 1.7$, which should force the peak to be contained within our fitted domain. 
These cases are presented as ``RL ''  in Table.~\ref{tab:priorsdeW}. 
With the limited range of integration, the fitted uncertainties are extremely stable with or without the application of any other priors. 
Thus we decide to adopt these intervals as the standard integration intervals in all uncertainties quoted in  \cite{Aar13}. 
This integration interval, however, is merely used to decide a quoted error; but it does not change the likelihood surface and its down-stream cosmological analysis in \cite{Aar13} as the full likelihood surface for anisotropic fitting is used for all cosmological analysis in \cite{Aar13}.

\subsubsection{Interdependency between $\Sigma_{||, \perp}$, $\Sigma_s$ and $\epsilon$}  \label{sigma_param_results}

We observe a variation  of 0.3\% in the median of the $\epsilon$ when the values of $\Sigma_{||\perp}$ and $\Sigma_s$
are modified, and measure a variation in $\epsilon$ around 0.1\% when we change $\Sigma_s =1.5\rightarrow 3.0$. 
Changes in $\Sigma_{NL}$ values affect only the quadrupole because we are not changing the overall value of $\Sigma_{NL}$ but only the relative contribution of the components ($\Sigma_{||, \perp}$). In particular, 
we change  anisotropic values, $\Sigma_{||}=11$ and $\Sigma_{\perp}=6$Mpc$h^{-1}$ to isotropic values and set $\Sigma_{||}=\Sigma_{\perp}=8$Mpc$h^{-1}$, post-reconstruction. 
Thus, the effect is best described as lowering the contrast in the crest-trough structure.
 In the case of $\Sigma_s$ , we increase the streaming  value that corresponds to an enhancement of this structure. 
\comment{What does this mean??  --- We do expect that the modifications applied affects in different directions. } 
These parameters are degenerate with each other; the 
results in Table.~\ref{tab:sigmadr11norec} suggest  that reducing this contrast in the structure of the quadrupole when fitting the pre-reconstructed mock catalogs decreases the observed biases in best fit parameters. 
 
Results in Table.~\ref{tab:sigmadr11rec} show that   changes in values of $\Sigma_{||, \perp}$ do not have any effect on best fit values  or fitted errors when for the post-reconstructed mock catalogs analysis. 
The negligible effect  post-reconstruction is not surprising as reconstruction is supposed to eliminate most of the quadrupole at larges scales, thus residual differences between our model of  the non linear correlation function that includes the redshift distortion should in principle has a better match to the reconstructed correlation functions. 
The effect of a large $\Sigma_s$ is not surprising neither, as this large value of the streaming is unrealistic for reconstructed mocks, thus an artificial enhancement of the crest-thought structure should give a poorer match to the multipoles measured. 
The effects of the $\Sigma_{||, \perp, s}$ on the best fitting values indicate that the calibration of these parameters is important for improving performance of the fitting methodology.  The residual mismatch is compensated by systematic polynomials. However, as we explain in Section~\ref{sec:nuisance_terms_results},   the order of the polynomials introduces also extra variations in  best fit  values pre- and post-reconstruction.  We  leave the interplay between the broadband terms and the non-linear damping and streaming parameters to a future study.

\subsubsection{Covariance matrix corrections} \label{sec:covar_corr_results}

\begin{table*}
\caption{A comparison of fitting results in DR11 mock galaxy catalogs including both covariance corrections and the correction for overlapping regions of mocks (CC) versus not including any of those corrections.
The median bias $\widetilde{b}$, median errors $\widetilde{\Delta \sigma}$, and percentiles are multiplied by 100.}
\label{tab:sigmadr11CC}
\begin{tabular}{@{}lrrrrrrrr}

\hline
Model&
$\widetilde{b_\alpha}$&
$\widetilde{\sigma_{\alpha}}$&
$\widetilde{b_\epsilon}$&
$\widetilde{\sigma_{\epsilon}}$&
$\widetilde{b_{\parallel}}$&
$\widetilde{\sigma_{\alpha_{\parallel}}}$&
$\widetilde{b_{\perp}}$&
$\widetilde{\sigma_{\alpha_{\perp}}}$\\

\hline
%

$\mathrm{No \;Recon. \;No CC}$&
$0.49^{+1.42}_{-1.56}$&
$1.57^{+0.37}_{-0.23}$&
$0.27^{+1.66}_{-1.81}$&
$1.86^{+0.51}_{-0.24}$&
$1.07^{+4.04}_{-4.13}$&
$4.38^{+1.33}_{-0.69}$&
$0.18^{+2.13}_{-2.03}$&
$2.11^{+0.38}_{-0.22}$\\
\\[-1.5ex]

$\mathrm{No \;Recon\; CC}$&
$0.51^{+1.41}_{-1.59}$&
$1.44^{+0.34}_{-0.21}$&
$0.26^{+1.65}_{-1.82}$&
$1.69^{+0.44}_{-0.21}$&
$1.08^{+4.02}_{-4.17}$&
$3.97^{+1.19}_{-0.62}$&
$0.17^{+2.13}_{-2.03}$&
$1.93^{+0.33}_{-0.20}$\\
\\[-1.5ex]

%

$\mathrm{Recon.\; CC}$&
$-0.02^{+0.94}_{-0.84}$&
$0.86^{+0.12}_{-0.10}$&
$0.15^{+1.25}_{-1.13}$&
$1.17^{+0.22}_{-0.13}$&
$0.29^{+2.72}_{-2.48}$&
$2.57^{+0.58}_{-0.37}$&
$-0.08^{+1.36}_{-1.59}$&
$1.39^{+0.19}_{-0.14}$\\
\\[-1.5ex]

$\mathrm{Recon.\; No \;CC}$&
$-0.02^{+0.94}_{-0.84}$&
$0.93^{+0.13}_{-0.11}$&
$0.16^{+1.24}_{-1.16}$&
$1.28^{+0.24}_{-0.14}$&
$0.29^{+2.72}_{-2.47}$&
$2.80^{+0.64}_{-0.41}$&
$-0.09^{+1.36}_{-1.59}$&
$1.51^{+0.21}_{-0.15}$\\
\\[-1.5ex]

%
%

\hline

 \end{tabular}

\end{table*}

This section presents the results including all covariance corrections described Section~\ref{sec:sysfit}. 
We measure the systematic error introduced in the results by not applying the correction for the overlapping region in the mock generation. We quantify the effect in DR10 and DR11 to integrate this error with our previous results that do not consider this defect of the mocks. Table.~\ref{tab:sigmadr11CC} presents the median and rms observed when including the correction factors A, B, C and r described in Section~\ref{sec:sysfit} pre- and post-reconstruction for the two templates, ``De-Wiggled"  and $P_{\rm pt}(k)$ templates. 
We do not include the $m_1$ factor. The final result should be still rescaled by $\sqrt{m_1}$ given by equation~\ref{eqn:m1}. This factor has a value of 1.0198 for DR10 and 1.0221 for DR11.

We do not find any variation in the best fit values of $\alpha$ and $\epsilon$ in all cases, which is to be expected, since these corrections only change the covariance matrices.  We do (as expected) observe variations on the uncertainties of best-fit $\epsilon$ when we apply the covariance corrections to pre-reconstructed mock galaxy catalogs.

\comment{commented out the following sentences since they are for DR10 ...----- 
the variations for DR10 on the uncertainties are in $\Delta \sigma_{\alpha}=0.002$ and $\Delta \sigma_\epsilon$=0.002-0.003 in the pre-reconstruction case when including the covariance matrix errors from overlapping regions, this correction reduces the errors. 
Post-reconstruction we found a variation in the uncertainties of $\Delta \sigma_{\alpha},\Delta \sigma_\epsilon=0.000-0.001$. 
These correction should be included in addition to the others systematic errors for all previous test.}

\subsubsection{Effect of Grids Sizes in the Likelihood Surface} \label{sec:grid_results} 

Table.~\ref{tab:grids} we summarizes the results for tests performed for the four different data sets varying the grid-size and fixing the range for the $\alpha$ and $\epsilon$ grid to [0.7, 1.3] and [-0.3, 0.3]  \comment{The interval is different from the TABLE's caption!}, respectively. 
Table.~\ref{tab:grids} shows the $\sigma_\alpha$ and $\sigma_\epsilon$ as well as the mean $\alpha$  ($\left < \alpha\right >$) and mean $\epsilon$ ($\left< \epsilon \right >$) when varying the number of grid points in $\alpha$  ($n_\alpha$), and,  in $\epsilon$ ($n_\epsilon$).
 We test  three  different  $n_{\alpha}=$(401,  241 and 121), which correspond to $\Delta_\alpha=$ 0.0015 ,0.0025 and  0.005,  and test three different   $n_{\epsilon}$ = 61, 121, 241, that correspond to $\Delta_\epsilon=$ 0.01, 0.005, 0.0025. 
The results show that variations on these parameter do not have any effect in the estimation of the best fit values nor their respective uncertainties from the likelihood surfaces. 

\begin{table}
\caption{Uncertainties on the best fit values when varying  the bin size of the grids for a fixed interval of $\alpha$=[0.8,1.2] .  \comment{HERE IT says 0.8,1.2, but in the section it said 0.7 to 1.3 so which one is which? that also affect the $\Delta_\alpha$ you say in the section text ..., and what is the grid size for epsilon?} }
\label{tab:grids}
\begin{tabular}{@{}lccccc}

\hline
$n_\alpha/n_\epsilon$
&$\sigma_\alpha$
&$\sigma_\epsilon$
&$\sigma_{\alpha,\epsilon}$
&$\left < \alpha\right >$
&$\left< \epsilon \right >$\\
\hline
DR10 Pre-Recon\\
121/61
&0.0229
&0.0441
&0.00062
&1.006
&-0.027\\
241/121
&0.0229
&0.0440
&0.00061
&1.006
&-0.027\\
401/201
&0.0229
&0.0441
&0.00061
&1.006
&-0.027\\
DR10 Post-Recon\\
121/61
&0.0112
&0.0233
&0.00054
&1.018
&-0.012\\
241/121
&0.0112
&0.0233
&0.00054
&1.018
&-0.012\\
401/201
&0.0112
&0.0233
&0.00054
&1.018
&-0.012\\
DR11 Pre-Recon\\
121/61
&0.0192
&0.0360
&0.00047
&1.021
&-0.015\\
241/121
&0.0192
&0.0375
&0.00047
&1.021
&-0.015\\
401/201
&0.0192
&0.0376
&0.00047
&1.021
&-0.015\\
DR11 Post-Recon\\
121/61
&0.0108
&0.0186
&0.00013
&1.011
&-0.036\\
241/121
&0.0108
&0.0186
&0.00013
&1.011
&-0.036\\
401/201
&0.0108
&0.0186
&0.00013
&1.011
&-0.036\\

\hline

\end{tabular}

\end{table}


\section{Comparison with other Methodologies}\label{sec:other_method}
We compare the methodology  followed in this work with other methodologies. We describe first the methodologies explored in this section and then  enumerate the main differences. 
\begin{itemize}

\item {\bf Multipoles-Gridded (hereafter Multip. Grid)}: 
This is our fiducial fitting methodology, which is described extensively in Section~\ref{sec:fiducial}. 
We consider the non-linear template RPT-inspired  $P_{\rm pt}(k)$ described in Section~\ref{sec:rpt}, instead of the fiducial de-wiggled template, to make a fair comparison with the others methodologies. 

\item {\bf Multipoles-MCMC (hereafter Multip. MCMC)}:  
This approach follows the multipoles-fitting methodology combined with a Monte Carlo Markov Chains sampling of the posterior in $\alpha_{||}, \alpha_\perp$, while marginalizing over all the remaining parameters. The template for the nonlinear power spectrum is  $P_{\rm pt}(k)$ as described in Section~\ref{sec:rpt}. 

\item {\bf Clustering Wedges (hereafter Wedges)}:  Described previously in Section~\ref{sec:wedges_description}, the clustering wedges analysis consists of an alternative set of moments, for a detailed description, please refer to \citet{Kaz12}, \citet{Kaz13}. This methodology also samples the posterior using Multip. MCMC in the $\alpha_{||}-\alpha_{\perp}$ parametrization. The template for the nonlinear power spectrum is also the RPT-inspired template, $P_{\rm pt}(k)$, described in Section~\ref{sec:rpt}. 

\end{itemize}
The main differences between the methodologies listed above are: i) The Multip. MCMC and the Wedges approaches have in common that the parametrization is in $\alpha_{||}$-$\alpha_{\perp}$ instead of  $\alpha$-$\epsilon$ of the Multip. Grid method. ii) The sampling of the posterior is generated with a MCMC instead of our grid approach. iii) Both Multip. MCMC and Wedges  apply flat priors on $\alpha_{||}$ and $\alpha_{\perp}$, compared with a Gaussian prior on $\epsilon$ and $\alpha$.  iv) There is a difference in the model, Wedges and Mutlip. MCMC methodologies fix $\beta$ and include a normalization factor in quadrupole, $rB_\perp$. 
\begin{equation}
\begin{array}{l}
\xi_\perp (r)=B_\perp \xi_{0,\perp}(r) + A_\perp(r)\\
\xi_{||}(r)=rB_{\perp}\xi_{0,||}(r) + A_{||}(r)
\end{array}
\end{equation}
v)There is a difference in the quoted values. We use the best fitting values for Multip. Grid method whereas the mean values are adopted for Multip. MCMC and Wedges. \footnote{For Multip. MCMC, the mean values and the best fitting values are similar but not identical, between them, the mean values are the most robust estimator. The contrary happens for the Multip. Grid method; the best values are more robust and the mean values are poor estimates of the parameters for low signal to noise ratio BAO features.} Regarding uncertainties, the values of $\sigma$ for Multip. MCMC correspond to the symmetrized percentiles and for Multip.Grid method uses the rms from the likelihoods.

We fitted the 600 mocks using these three different methodologies and we compare the fitted parameters and fitted errors. For this test we consider the full covariance matrix corrections. The tests were performed  for DR10/DR11 in the pre- and post-reconstructed mocks, but for clarity we only quoted DR11 results. Also in this section we use the $\alpha_{||, \perp}$ parametrization to compare the techniques, as the other two methodologies results use this parametrization.

\subsection{Results from the mocks}
 
Table.~\ref{tab:methodology_mocks} shows the median and standard deviation of the best fits values and errors estimated with the results of the mocks for the three methodological choices. 
\begin{table*}
\caption{ The fitting results on $\alpha_{||}$ and $\alpha_{\perp}$ and their respective errors when different anisotropic clustering fitting methodologies are used on different mock galaxy catalogs (DR10, DR11, pre- and post-reconstruction). We present median values since median values are more representative of the skewed $\sigma$ distributions and slightly asymmetric $\alpha_\parallel$ pre-reconstruction distributions.The columns are the median values of the bias $b_{\alpha, \epsilon, ||,\perp}$, median values of errors on $\alpha_{||},\alpha_{\perp}$ ($\widetilde{ \sigma_{\alpha, \epsilon, \alpha_{||},\alpha_{\perp}}}$), the standard deviation of $\alpha_{||},\alpha_{\perp}$, ($S_{\alpha, \epsilon, \alpha_{||}, \alpha_{\perp}}$), and finally the standard deviation of the errors  $S_{\sigma_{\alpha, \epsilon, \alpha_{||},\alpha_{\perp}}}$. The median bias, the median values of the errors $\widetilde{\Delta v}$, and the standard deviations are multiplied by 100.
}
\label{tab:methodology_mocks}
\begin{tabular}{@{}lcccccccccccccccc}

\hline
Method&
$\widetilde{\alpha}$&
$S_\alpha$&
$\widetilde{\epsilon}$&
$S_{\epsilon}$&
$\widetilde{\alpha_{\parallel}}$&
$S_{\alpha_{\parallel}}$&
$\widetilde{\sigma_{\alpha_{\parallel}}}$&
$S_{\sigma_{\alpha_{\parallel}}}$&
$\widetilde{\alpha_{\perp}}$&
$S_{\alpha_{\perp}}$&
$\widetilde{\sigma_{\alpha_{\perp}}}$&
$S_{\sigma_{\alpha_{\perp}}}$\\
\hline
DR11 Post-recon\\
Multip. Grid&
0.02&0.92
&0.11&1.22
&0.32&2.66
&2.48&0.72
&-0.01&1.49
&1.37&0.18\\
Multip. MCMC&
0.00&0.89
&0.15&1.12
&0.19&2.47
&2.73&0.55
&-0.11&1.39
&1.48&0.22\\
Wedges&
0.03&0.90
&0.05&1.24
&0.06&2.64
&2.96&0.52
&-0.07&1.53
&1.61&0.26\\

\hline
Multip. Grid&
-0.05&1.55
&0.22&1.89
&0.58&4.43
&3.84&1.50
&-0.35&2.10
&2.04&0.32\\
Multip. MCMC&
-0.08&1.54
&0.03&2.06
&0.17&4.82
&4.29&2.26
&-0.18&2.15
&2.22&0.47\\
Wedges&
-0.09&1.52
&-0.11&2.07
&-0.35&4.75
&4.66&1.37
&0.07&2.22
&2.30&0.86\\

\hline

DR10 Post-recon\\
Multip. Grid&
0.15&1.26
&0.04&1.53
&0.28&3.38
&3.17&1.25
&0.04&1.92
&1.73&0.28\\
Multip. MCMC&
0.20&1.27
&-0.00&1.50
&0.05&3.46
&3.30&2.21
&0.13&1.82
&1.85&0.38\\
Wedges&
0.21&1.19
&-0.15&1.53
&-0.03&3.35
&3.62&0.88
&0.37&1.88
&1.93&0.42\\
\hline
DR10 Pre-recon\\
Multip. Grid&
-0.25&1.92
&0.25&2.54
&0.27&5.86
&4.98&2.40
&-0.31&2.69
&2.60&0.73\\
Multip. MCMC&
-0.10&2.07
&-0.02&3.11
&-0.25&7.52
&5.04&4.09
&-0.07&2.98
&2.72&1.37\\
Wedges&
-0.17&1.99
&-0.23&2.94
&-0.46&6.66
&5.50&3.47
&0.05&3.07
&2.83&1.88\\


\hline

\end{tabular}
\end{table*}
In the pre-reconstruction mocks, we found small variations in the bias between methodologies but there is no indication that a methodology is more biased with respect to the others. For DR11, the bias are  $|b_{||}| \le 0.6\%$ and $|b_\perp| \le 0.4\%$. 
 For the post reconstruction mocks, the three variations in methodology produce consistent results; the bias is less than 0.3\% for $\alpha_{||, \perp}$.
\begin{figure}
   \centering       
   \includegraphics[width=1.0\columnwidth]{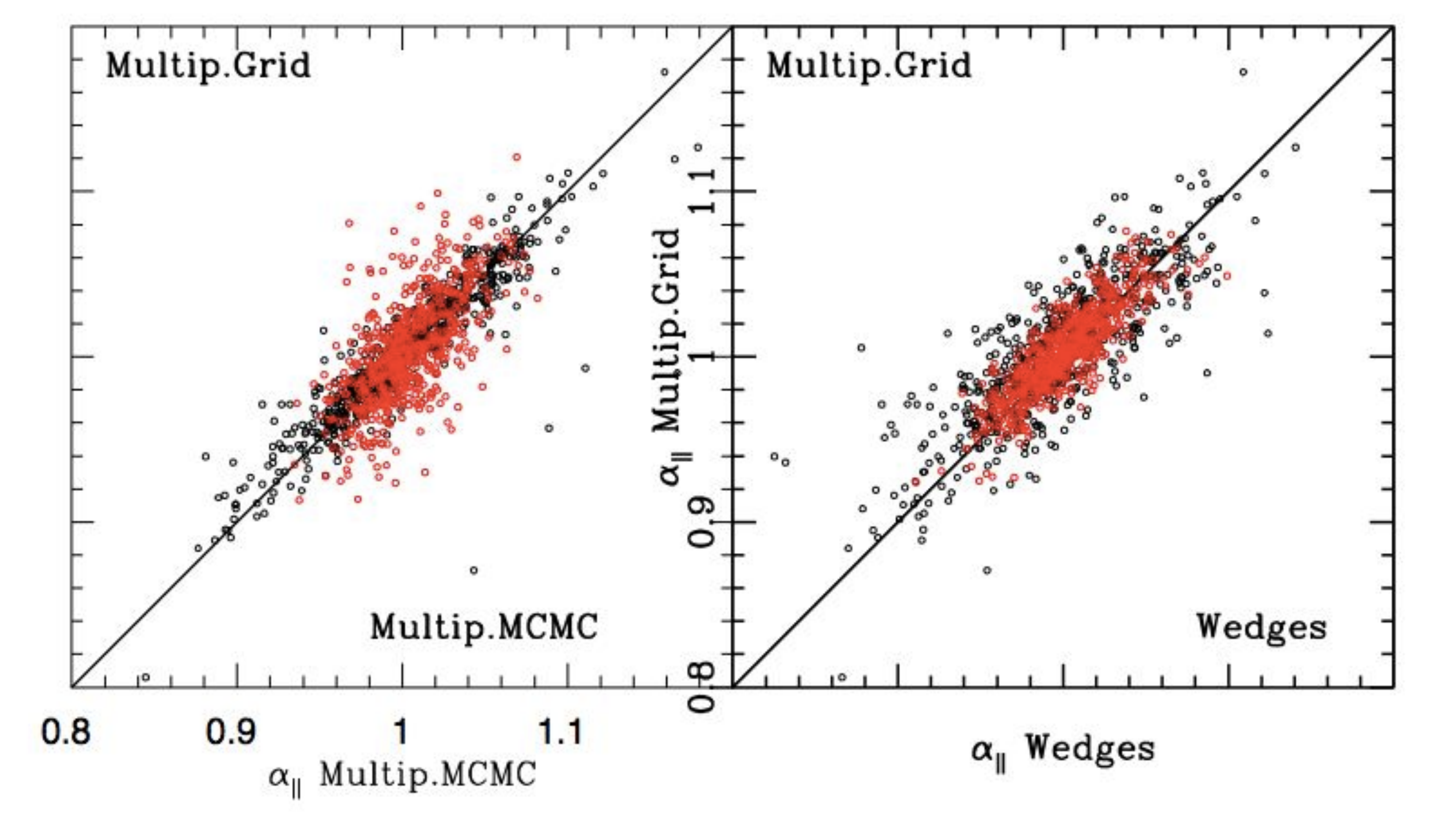} 
   \caption{Dispersion plots of $\alpha_{||}$, pre [black] and post-reconstruction [red] using different methodologies. The left panel compares multipoles using Multip. Grid and Multip. MCMC approaches, and right panel compares Gid and Wedges.}
   \label{fig:alphasmetho1}
\end{figure}
\begin{figure}
   \centering       
   \includegraphics[width=1.0\columnwidth]{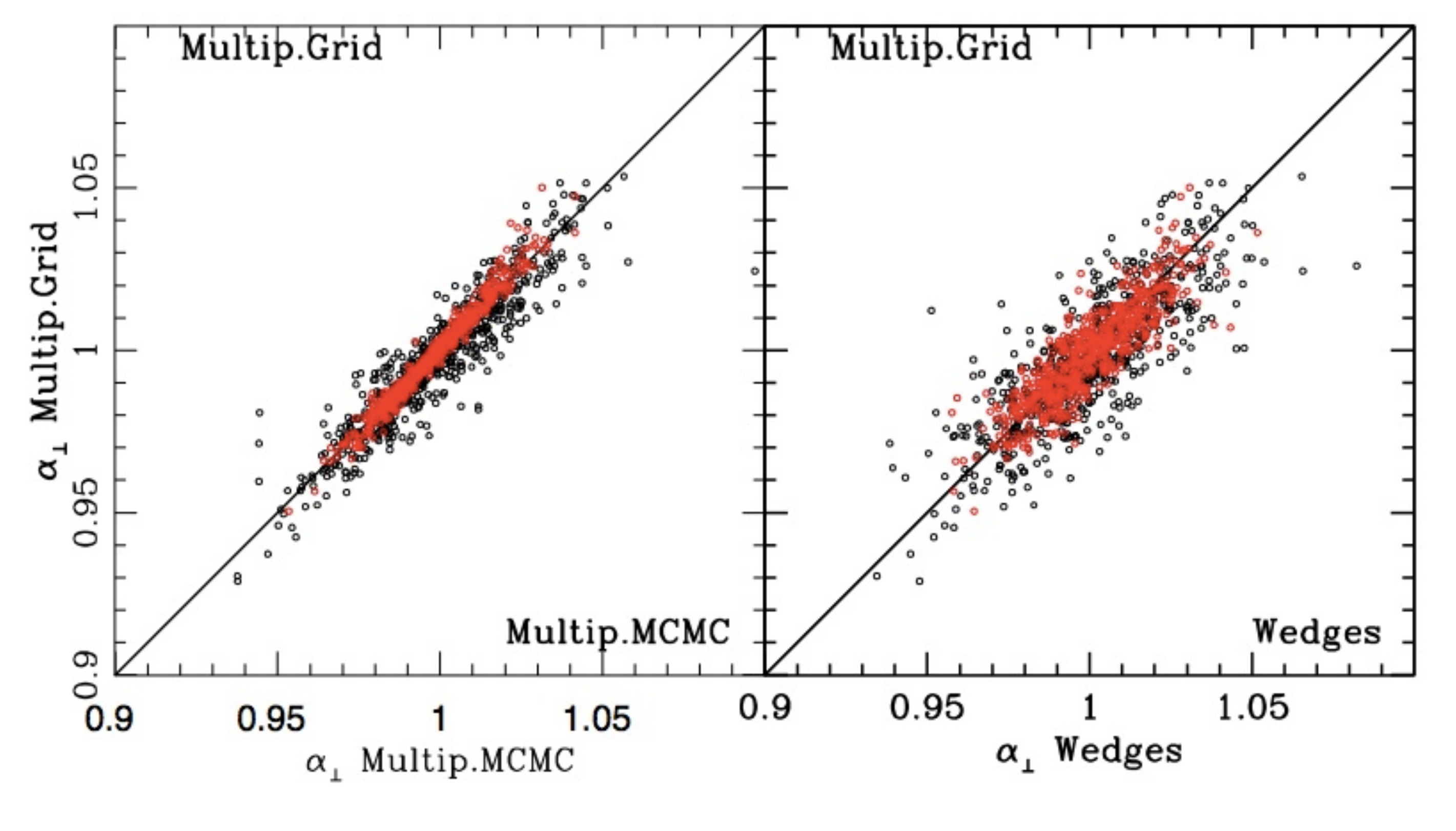} 
   \caption{Dispersion plots of $\alpha_{\perp}$, pre [black] and post-reconstruction [red] using different methodologies. The left panel compares multipoles using Multip. Grid and Multip. MCMC, and right panel compares Multip. Grid and Wedges.}
   \label{fig:alphasmetho2}
\end{figure}

The second observation is that the RMS of the $\alpha$, as well as the median and RMS of the $\sigma$ of the three methods are consistent. The agreement is even better post-reconstruction, however, they are small differences in their values. In this section we explore the small discrepancies observed between the three methodological choices. 

Figs.~\ref{fig:alphasmetho1} and~\ref{fig:alphasmetho2} show the dispersion plots of $\alpha_{||}$ and  $\alpha_{\perp}$, pre- and post-reconstruction using different methodologies for DR11. 
The left panel compares Multip. MCMC approach and Wedges, and right panel compare Multip. MCMC and Multip. Grid method.
In general the two dispersion plots show a good correspondence between different methodologies. There is, however, some dispersion in the parallel direction that decreases post reconstruction.
\begin{table*}
\caption{Dispersion observed between the different methodologies in the fitted parameters and fitted errors. We define the variation $\Delta \alpha_{||, \perp, \alpha,\epsilon}=\alpha^{Method \; a}_{||, \perp, \alpha, \epsilon}-\alpha^{Method\;b}_{||, \perp, \alpha, \epsilon}$. The columns show the median variation and the 16th and 84th percentiles. The median values $\widetilde{\Delta v}$, and percentiles are multiplied by 100}
\label{tab:dispersion}

\begin{tabular}{@{}lrrrrrr}

\hline
Model&
$\widetilde{\Delta \alpha}$&
$\widetilde{\Delta \epsilon}$&
$\widetilde{\Delta \alpha_{\parallel}}$&
$\widetilde{\Delta \sigma_{\alpha_{\parallel}}}$&
$\widetilde{\Delta \alpha_{\perp}}$&
$\widetilde{\Delta \sigma_{\alpha_{\perp}}}$\\
\hline
DR11 Post-Recon.\\
Multip. Grid-Multip.MCMC&
$0.02_{-0.07}^{+0.08}$&
$-0.02_{-0.19}^{+0.21}$&
$-0.02_{-0.38}^{+0.44}$&
$-0.22_{-0.36}^{+0.35}$&
$0.04_{-0.20}^{+0.21}$&
$-0.11_{-0.13}^{+0.12}$\\
Multip. Grid-Wedges&
$0.02_{-0.21}^{+0.18}$&
$0.14_{-0.71}^{+0.69}$&
$0.27_{-1.37}^{+1.32}$&
$-0.46_{-0.39}^{+0.46}$&
$-0.11_{-0.73}^{+0.75}$&
$-0.24_{-0.18}^{+0.14}$\\
Multip. MCMC-Wedges&
$-0.01_{-0.18}^{+0.16}$&
$0.12_{-0.64}^{+0.70}$&
$0.23_{-1.27}^{+1.31}$&
$-0.24_{-0.17}^{+0.19}$&
$-0.15_{-0.72}^{+0.69}$&
$-0.14_{-0.10}^{+0.09}$\\
\hline
DR11 Pre-Recon\\
Multip. Grid-Multip.MCMC&
$-0.02_{-0.47}^{+0.40}$&
$0.16_{-0.59}^{+0.60}$&
$0.32_{-1.29}^{+1.17}$&
$-0.39_{-0.59}^{+0.58}$&
$-0.17_{-0.75}^{+0.70}$&
$-0.21_{-0.24}^{+0.23}$\\
Multip. Grid-Wedges&
$-0.02_{-0.47}^{+0.57}$&
$0.37_{-1.31}^{+1.04}$&
$0.72_{-2.61}^{+2.26}$&
$-0.77_{-0.78}^{+0.75}$&
$-0.40_{-1.13}^{+1.40}$&
$-0.28_{-0.33}^{+0.26}$\\
Multip. MCMC-Wedges&
$0.06_{-0.31}^{+0.28}$&
$0.18_{-1.01}^{+1.00}$&
$0.38_{-2.09}^{+1.99}$&
$-0.40_{-0.35}^{+0.33}$&
$-0.14_{-0.97}^{+0.98}$&
$-0.08_{-0.20}^{+0.17}$\\

\hline
DR10 Post-Recon\\
Multip. Grid-Multip.MCMC&
$-0.03_{-0.15}^{+0.11}$&
$0.06_{-0.26}^{+0.30}$&
$0.08_{-0.52}^{+0.65}$&
$-0.10_{-0.49}^{+0.65}$&
$-0.13_{-0.25}^{+0.27}$&
$-0.12_{-0.19}^{+0.18}$\\
Multip. Grid-Wedges&
$-0.06_{-0.25}^{+0.26}$&
$0.20_{-0.88}^{+0.87}$&
$0.32_{-1.81}^{+1.70}$&
$-0.37_{-0.66}^{+0.65}$&
$-0.21_{-0.97}^{+0.84}$&
$-0.20_{-0.26}^{+0.21}$\\
Multip. MCMC-Wedges&
$-0.01_{-0.21}^{+0.21}$&
$0.15_{-0.89}^{+0.79}$&
$0.27_{-1.74}^{+1.56}$&
$-0.31_{-0.28}^{+0.24}$&
$-0.16_{-0.85}^{+0.88}$&
$-0.08_{-0.15}^{+0.13}$\\

\hline
DR10 Pre-Recon\\
Multip. Grid-Multip.MCMC&
$-0.07_{-0.74}^{+0.79}$&
$0.18_{-1.01}^{+1.00}$&
$0.28_{-2.07}^{+2.30}$&
$-0.10_{-0.89}^{+1.00}$&
$-0.32_{-1.04}^{+1.25}$&
$-0.13_{-0.40}^{+0.35}$\\
Multip. Grid-Wedges&
$0.02_{-0.82}^{+0.81}$&
$0.35_{-1.50}^{+1.44}$&
$0.74_{-3.30}^{+2.92}$&
$-0.48_{-1.08}^{+1.31}$&
$-0.37_{-1.57}^{+1.56}$&
$-0.18_{-0.50}^{+0.39}$\\
Multip. MCMC-Wedges&
$0.06_{-0.34}^{+0.41}$&
$0.23_{-1.25}^{+1.20}$&
$0.45_{-2.41}^{+2.48}$&
$-0.38_{-0.56}^{+0.54}$&
$-0.13_{-1.20}^{+1.25}$&
$-0.06_{-0.26}^{+0.26}$\\
\hline

\end{tabular}
\end{table*}
To perform a quantitative comparison of the differences, we estimate the median variation, where we define the variation as $\Delta \alpha_{||, \perp}=\alpha^{Method \; a}_{||, \perp}-\alpha^{Method\;b}_{||, \perp}$. To quantify the dispersion we show the 16th and 84th percentiles of the variation. The different comparisons permit to discern the different contributions of the discrepancy observed between the three methodological choices. Table.~\ref{tab:dispersion} summarizes the median difference and dispersion observed between the different methodologies in the fitted parameters and fitted errors.

\subsubsection{Comparison of Multip. MCMC and Multip. Grid Approaches.}
 We first consider the effects when fitting to the pre- reconstructed mock catalogs. Table.~\ref{tab:dispersion}, demonstrates the median difference is small, $|\widetilde{\alpha_{||,\perp}}|\le 0.003$. 
The observed dispersion is $\sim 0.012$ in the parallel direction and 0.007 in the perpendicular direction. 
The dispersion of the Multip. MCMC-Multip.Grid is illustrated in right panels of Figs.~\ref{fig:alphasmetho1} and~\ref{fig:alphasmetho2}  for the parallel and perpendicular directions. 
\comment{Shirley, I just think that if you you really think is better to eliminate the dispersion plots in this section it is ok, these dispersion plots in section 7 are not my favorites :)}

\begin{figure}
   \centering      
   \includegraphics[width=1.0\columnwidth]{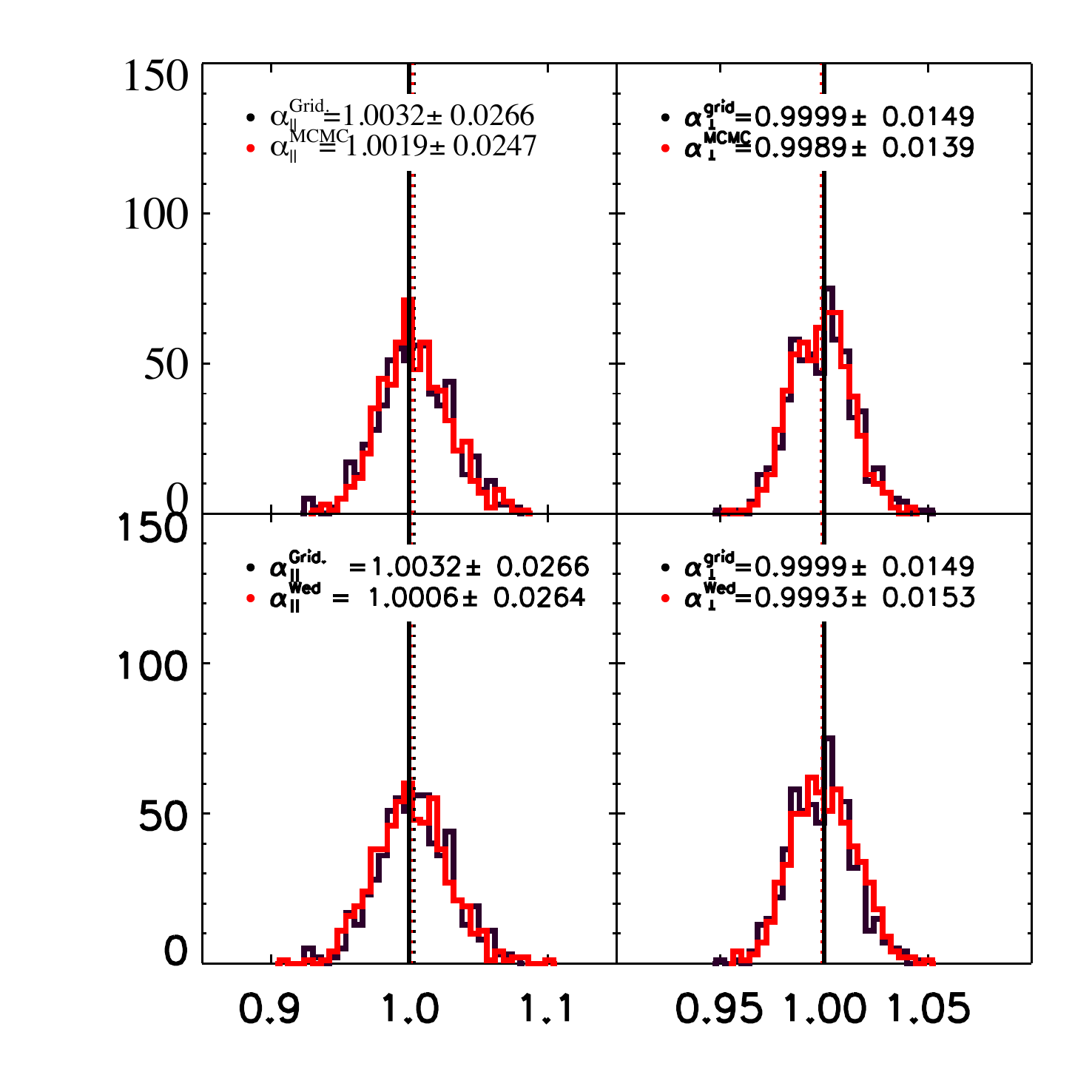}
   \caption{Top panels: comparisons between the Multip. Grid and Multip. MCMC  methodologies. Bottom panels: comparison between the Wedges and Multip. Grid. The panels show the $\alpha_{||}$ and $\alpha_{\perp}$ distributions of best fit values of  PTHALOS DR11 post-reconstruction mocks. }
   \label{fig:wedgesmcmcgrid}
\end{figure}
The post-reconstruction mocks possesses smaller dispersion and median differences. The median difference is zero for both $\alpha_{||, \perp}$ and the dispersion is almost half of the value pre-reconstruction. The dispersion in $\alpha_{||}$ is only 0.004 and in $\alpha_{\perp}$ is 0.002. The smaller dispersion post-reconstruction is also clearly  observed in the dispersion plots (see Figs.~\ref{fig:alphasmetho1} and~\ref{fig:alphasmetho2}).
Fig.~\ref{fig:wedgesmcmcgrid} shows the distributions of $\alpha_{||, \perp}$ post-reconstruction, the distributions in $\alpha_\perp$ are almost identical.
\begin{figure}
   \centering
    \includegraphics[width=1.0\columnwidth]{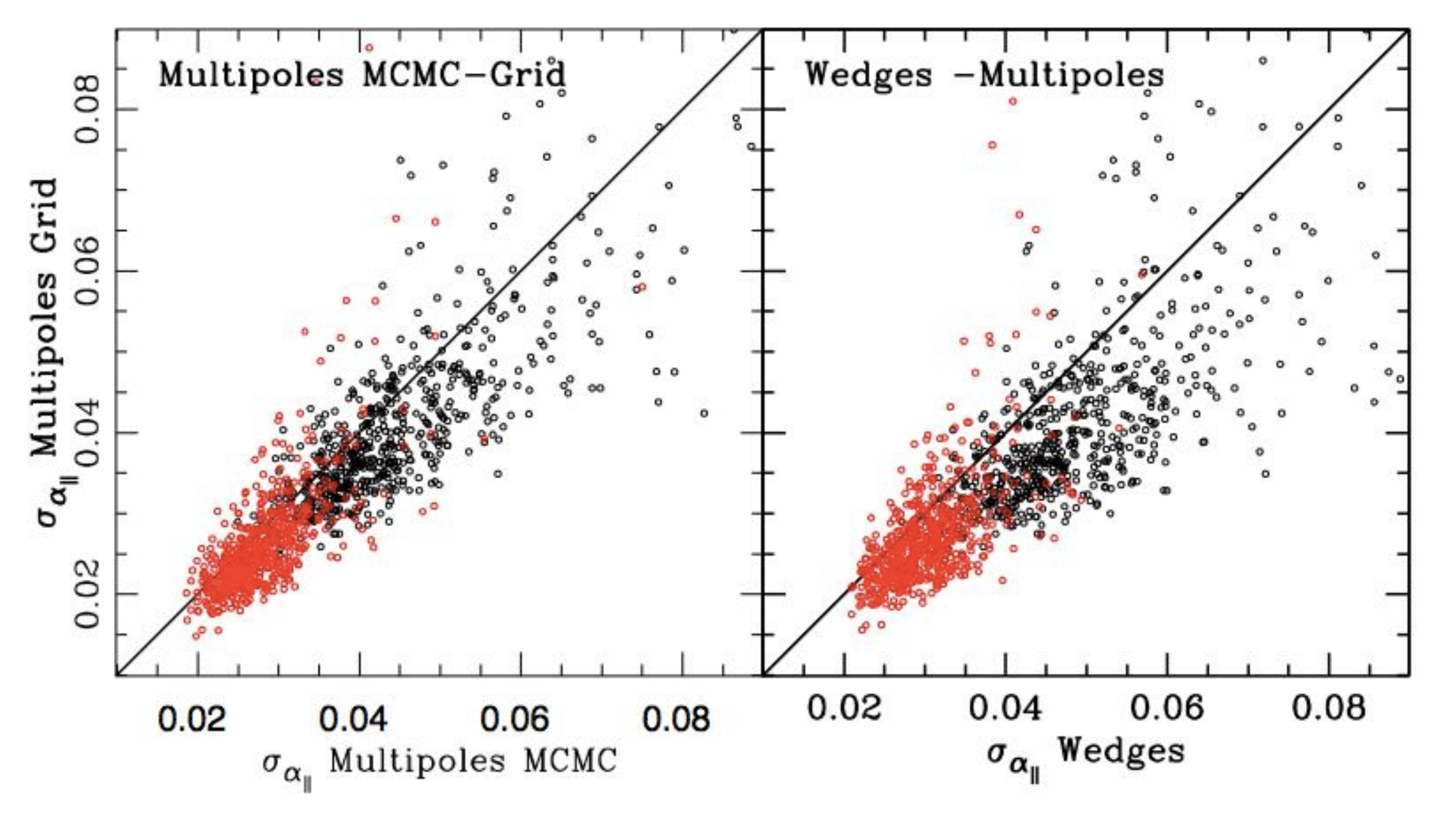}    
       \caption{Dispersion plots of $\sigma_{||}$ pre [black] and post-reconstruction [red] using different methodologies. Left panel compares multipoles using Multip. MCMC approach and Multip. Grid, and  right panel compares Multip. Grid and Wedges.}
   \label{fig:sigmasmetho1}
\end{figure}
\begin{figure}
   \centering
     \includegraphics[width=1.0\columnwidth]{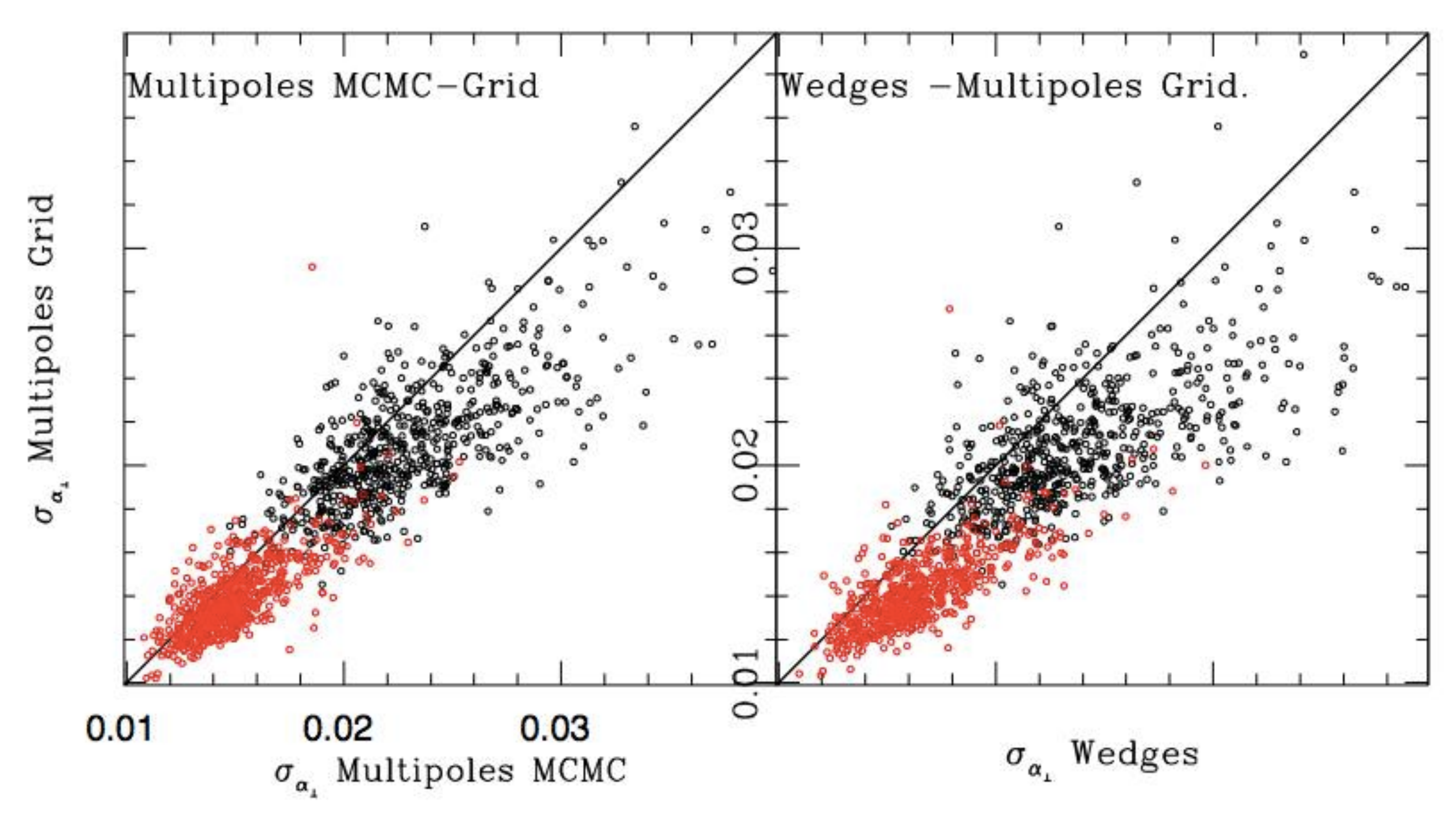} 
       \caption{Dispersion plots of $\sigma_{\perp}$,  pre [black] and post-reconstruction [red] using different methodologies. Left panel compares multipoles using Multip. MCMC approach and Multip. Grid, and right panel compare Multip. Grid and Wedges.}
   \label{fig:sigmasmetho2}
\end{figure}

Concerning the fitted errors, Figs.~\ref{fig:sigmasmetho1}  and~\ref{fig:sigmasmetho2} show the dispersion plots for $\sigma_{||}$[left], and $\sigma_\perp$[right] comparing Multip. MCMC and Multip.Grid results pre- and post reconstruction for DR11.  The figures demonstrate that the errors are well correlated for two methodologies, however, there is some level of dispersion. The dispersion is mostly observed in the parallel direction, and the dispersion decreases significantly post-reconstruction

\begin{figure}
   \centering     
   \includegraphics[width=1.0\columnwidth]{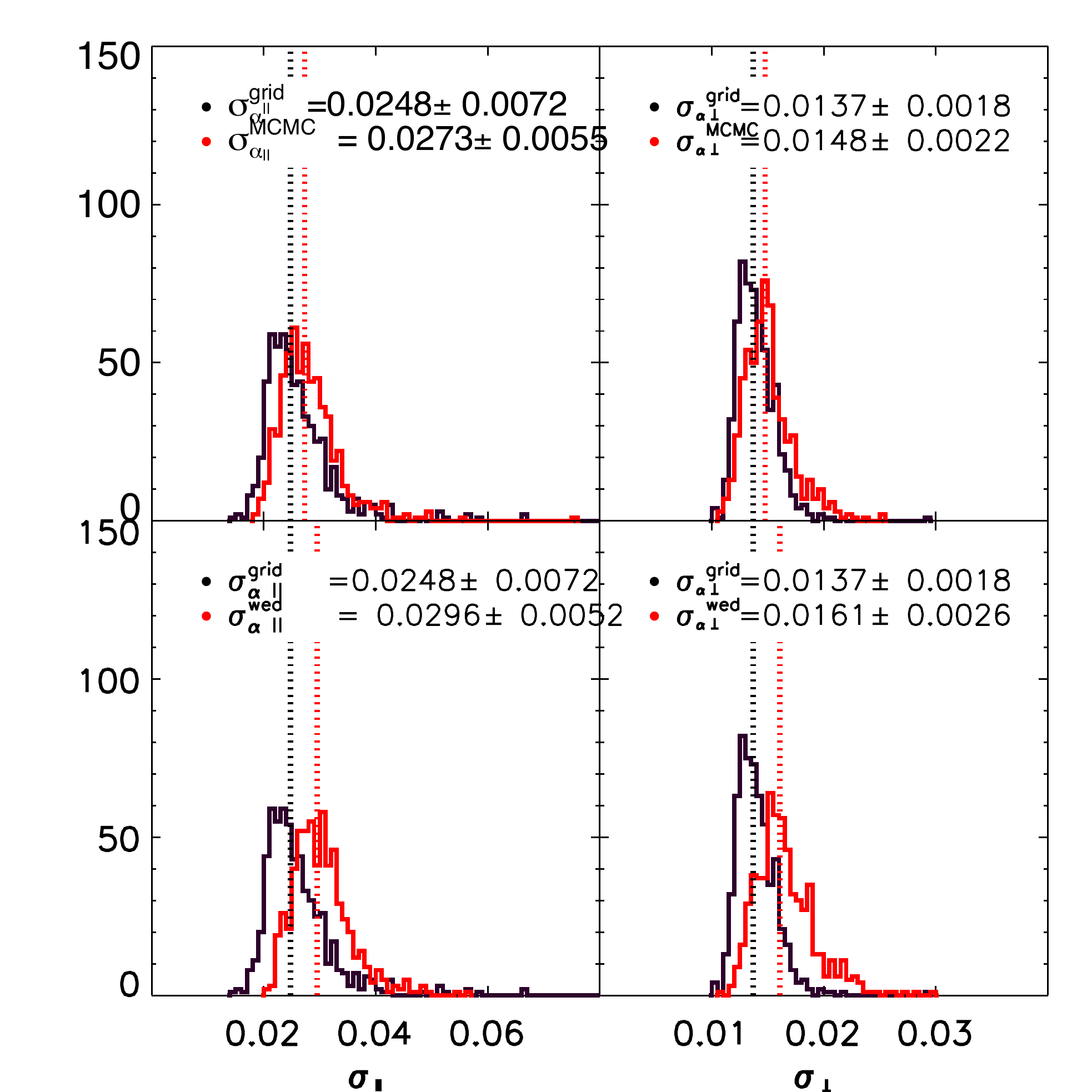}
   \caption{Histograms of $\sigma_{\alpha_{||}}$[left]  , $\sigma_{\alpha_\perp}$[right] from 600 DR10 simulations, comparing Multip. MCMC and Multip. Grid approaches in top panels and Multip. Grid vs Wedges.}
   \label{fig:mcmcgrid2}
\end{figure}
The actual values we obtained from the pre-reconstructed mocks are for the median variation $\widetilde{\Delta \sigma_{||}}=-0.004$ with 0.006 dispersion and $\widetilde{\Delta \sigma_{\perp}}=-0.002$ with $\sim0.003$ dispersion. Post-reconstruction, mocks have smaller variations and dispersion levels. The median differences and dispersions are approximately half of pre-reconstruction values,  $\widetilde{\sigma_{||}}=-0.002$  with 0.004 dispersion  and $\widetilde{\sigma_{\perp}}=-0.001$ with 0.001 dispersion. Fig.~\ref{fig:mcmcgrid2} shows the $\sigma$ distribution from the mocks for DR11 post-reconstructed. The figures show similar $\sigma$ distributions with a small shift between the peak of the distributions. 

\comment{In principle, these two methods, the Multip. MCMC and Multip. Grid approaches, should give the same results if the same region of the parameter space is explored, independently of the parametrization considered in the sampling, however the difference of priors could make possible a difference in the region of parameters explored. Also the slight difference in the way to include de normalization terms in the modeling could generate some differences.
Another source of dispersion is the different quoted values we are comparing. }
To conclude, there is a good agreement between the Multip. MCMC and Multip. Grid methdologies, although remains a small level of scatter.  The discrepancies could be explained by the slight differences in the implementations (and codes).

\subsubsection{Comparing Wedges and Multipoles Methodologies}

We analyze two cases: Multip. MCMC/Wedges and Multip. Grid/Wedges. This comparison enables to separate the sources of dispersion between the results. Comparing the Multip. MCMC and Wedges will account for errors associated with different estimators without any methodological modification, and comparing the Multip. Grid-Wedges quantifies the extra systematic error/ dispersion coming from slightly different methodological choices (in addition to the fact of using different implementations codes).

Table.~\ref{tab:dispersion} shows that for DR11 pre-reconstruction the median difference is $\Delta \alpha_{||}=-0.007_{-0.026}^{+0.023}$  and $\Delta \alpha_\perp=+0.004_{-0.011}^{+0.014}$ for Multip. Grid-Wedges, compared to $\Delta \alpha_{||}=-0.004_{-0.021}^{+0.020}$ and $\Delta \alpha_\perp=+0.001_{-0.010}^{+0.010}$ from Wedges-Multip. MCMC. These results indicate that the majority of the scatter is produced by from the different estimators; however, the difference in the median difference indicates that the remaining methodological differences are increasing the median discrepancy in the values. \comment{If we analyze the $\alpha-\epsilon$ parametrization, these discrepancies corresponds to median variations in $\epsilon$ at the level of 0.004.}
In mocks post-reconstruction, the different multipoles implementations do not produces any difference in the best fitting values results, thus the median differences and scatter are only determined by the different estimator. For the DR11 results  for Wedges-Multip.Grid the  median variations are $\alpha_{\parallel, \perp}=+0.003$ with $\sim 0.014$ scatter,  compared to with $+0.002$ with 0.013 scatter from Wedges-Mutlip. MCMC. 

To understand the small differences, in bottom panels of Fig.~\ref{fig:wedgesmcmcgrid} we display the distribution of fitting parameters obtained from Wedges compared with those issued from Multip. Grid methodology post-reconstruction for DR11 for $\alpha_{\parallel, \perp}$. The distributions are quite similar, but the Multip. Grid  approach shows a slightly higher bias in the $\alpha_{||}$ and Wedges in the $\alpha_\perp$.

 For the uncertainties the difference in estimator and the remaining methodological differences are contributing at the same level to the dispersion, but the median of the differences have closer values indicating that the difference in errors is primarily related to the difference in estimator.  For Wedges-Multip.Grid we get $\Delta \sigma_{||}=+0.008_{-0.008}^{+0.007}$, and $\Delta \sigma_{\perp}=+0.003_{-0.003}^{+0.003}$, to be compare with $\Delta \sigma_{||}=+0.004_{-0.004}^{+0.003}$ and $\Delta \sigma_{\perp}=+0.001_{-0.002}^{+0.002}$ for Wedges-Multip.MCMC. 
 
The post-reconstruction mocks have the two contributions in the median difference and dispersion of the errors. The median variations of the errors in Wedges-Multip. Grid is two times higher compared with Wedges-Multip. MCMC, and also the dispersion in Wedges-Multip. Grid is  two twice the scatter in Wedges-Multip. MCMC. 

Finally, the bottom panels in Fig.~\ref{fig:mcmcgrid2} present the $\sigma$ distribution from the mocks for DR11 post-reconstructed for $\sigma_{\alpha_{||}}$ and $\sigma_{\alpha_\perp}$ for Wedges and Multip. Grid methods. We observe from the plots that we get some differences in the error distributions. The distribution peaks are slightly shifted, and the Wedges distributions have slightly larger tails pushing the median to a lower value. The differences in the medians are less significant in the perpendicular direction. \comment{These small differences in the errors are not related with the different definition for estimating the errors, as the distributions comparing the Multip. MCMC and Multip. Grid do not show this behavior. Further exploration need to be done to determine the origin of this differences.}

Summarizing, the median differences in the best fit values and in their errors between Wedges and Multip. Grid methods are small pre-reconstruction. In the post-reconstruction mocks, the discrepancies are even smaller, $\Delta \alpha_{||, \perp} \le 0.003$ for the fitted $\alpha_{||,\perp}$ and $\Delta \sigma_{||, \perp}\le 0.005$ for the errors. 
The results clearly indicate two contributions in the median variation and the dispersion. The contribution arising from the estimator itself (Wedges vs Multipoles), and the contribution coming from the different implementations. In principle, we should not expect discrepancies between multipoles and wedges analysis, as  the Wedges is only a basis rotation, however, the way  the priors impact the different estimators could generate some small dispersions. 

\comment{Shirley you can also eliminate this.The analyze performed in this work confirm previous studies comparing multipoles and wedges methodology with hight signal to noise \citep{And13,Kaz12}. We can compare directly the results reported on Table 3. of Anderson et al 2013 to the values of Table.~\ref{tab:dispersion}. The DR9 results show a 2 times larger median difference in $\alpha_{||}$ (identical $\alpha_\perp$) and 1.5-2 times larger scatter probably related with weaker signal to noise results.
  Also for the case of the $\sigma_{||}$ we observed similar median difference, however in $\sigma_{\perp}$ the median difference is slightly larger in DR11 compared with DR9. For post-reconstruction results, DR9 results show a $\sim1.7$ larger median difference in parallel direction and similar in perpendicular but $\sim$2 times larger scatter 
 compared with DR10 results.  And  for $\sigma$ in DR10 we get similar median differences but the scatter in DR10 is reduced to $\sim$1/4 from those of DR9. }

%
\section{Results with the data}\label{sec:data}
\subsection{Robustness Test on Data}
\label{sec:fitdata}

\begin{figure}
\resizebox{84mm}{!}{\includegraphics{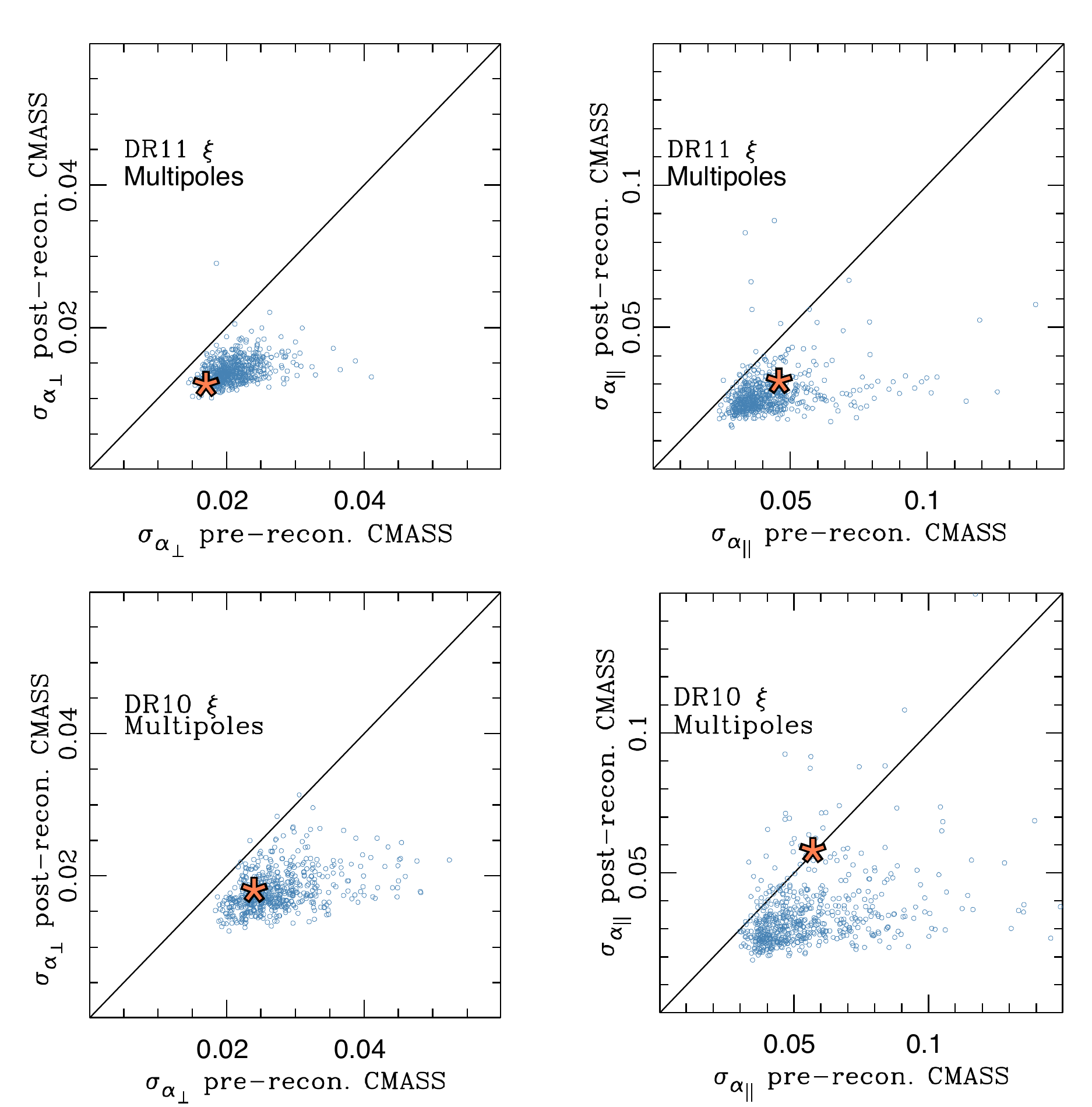}}
\caption{Plots of $\sigma_\alpha$ pre- and post-reconstruction:
mocks (circles) + data (star) for $\sigma_{\alpha_{\perp}}$ and $\sigma_{\alpha_{||}}$ for  CMASS DR10 and DR11. The reconstruction improves the precision in nearly all  of the 600 mock galaxy samples for both DR10 and DR11. The fits here uses RPT-inspired $P_{\rm PT}(k)$ template instead of the De-Wiggled template. This approach is adopted since the $P_{\rm PT}$ template produces less biased results pre-reconstruction when everything else is held fixed, and also the choice made for \protect \cite{Aar13}. }
\label{fig:sigdr11_multip}
\end{figure}

In this section we present the results of applying the same robustness tests applied to  the mocks to the DR11 CMASS data described in Section~\ref{sec:cmass}.
For these tests we apply all covariance corrections described in Section~\ref{sec:cov_corr}. 
Fig.~\ref{fig:sigdr11_multip} shows the data compares with the results observed on the mocks for our fiducial case using RPT-inspired $P_{\rm PT}(k)$ template. This approach is adopted since $P_{\rm PT}$ template produces less biased results pre-reconstruction when compared to De-Wiggled Template and also the choice made for \cite{Aar13}. Fig.~\ref{fig:sigdr11_multip} illustrates that the uncertainties  recovered from the data (orange stars)  on the anisotropic BAO measurements  for our fiducial case (with $P_{\rm PT}(k)$ template)  are typical of those found in the mock samples. The data point lies within the locus of those recovered from mock samples (blue points).

\begin{table*}
\caption{The results of the robustness test on CMASS post-reconstruction DR11 data.
We show the best fit $\alpha$ and $\epsilon$ values as well as the corresponding $\alphapar$ and $\alphaper$ with their respective errors and the $\chi^2/d.o.f$ of the fit for the fiducial case (with De-Wiggled Template). The remaining lines of the table show the difference between the fitted value or error with respect to the fiducial case $\Delta v=v-v^{fid}$.  (RL) stands for calculating the errors by integrating over specific intervals in the likelihood surfaces in 
$\alpha$-$\epsilon$ or $\alphapar$-$\alphaper$. The median variations $\widetilde{\Delta v}$ are multiplied by 100. The definitions of various rows are defined similarly as in earlier tables such as Table~\ref{tab:dr11norec}.}
\label{tab:datatest}
\begin{tabular}{@{}lrrrrrrrrr}
\hline
Model&$ \alpha$&$\sigma_\alpha$&$ \epsilon$&$ \sigma_\epsilon$&$ \alpha_{||}$& $ \sigma_{\alpha_{||}}$&$\alpha_\perp$&$  \sigma_{\alpha_\perp}$&$\chi^2/d.o.f$\\
\hline
Fiducial
&$ 1.0166$
&$ 0.0089$
&$-0.0324$
&$ 0.0133$
&$ 0.9518$
&$ 0.0314$
&$ 1.0507$
&$ 0.0127$
&    21./30\\
\hline
Model&$\Delta \alpha$&$\Delta \sigma_\alpha$&$\Delta \epsilon$&$\Delta \sigma_\epsilon$&$\Delta \alpha_{||}$& $\Delta \sigma_{\alpha_{||}}$&$\Delta \alpha_\perp$&$ \Delta \sigma_{\alpha_\perp}$&$\chi^2/d.o.f$\\
\hline
$\mathrm{Fitting} \; 30<r<200\hMpc$
&$   0.23$
&$  -0.09$
&$   0.30$
&$  -0.07$
&$   0.80$
&$  -0.26$
&$  -0.08$
&$   0.03$
&    36./30\\
$\mathrm{2-term} \; A_0(r) \; \& \; A_2(r)$
&$   0.09$
&$   0.04$
&$   0.82$
&$  -0.03$
&$   1.70$
&$   0.04$
&$  -0.79$
&$  -0.09$
&    38./30\\
$\mathrm{4-term} \; A_0(r) \; \& \; A_2(r)$
&$  -0.07$
&$  -0.02$
&$  -0.16$
&$  -0.05$
&$  -0.39$
&$  -0.16$
&$   0.10$
&$   0.02$
&    16./30\\
$\mathrm{Fixed \;} \beta=0.0$
&$   0.04$
&$  -0.05$
&$  -0.15$
&$  -0.08$
&$  -0.26$
&$  -0.22$
&$   0.20$
&$  -0.02$
&    21./30\\

$\spar=\sperp=8\mathrm{Mpc/h}$
&$   0.01$
&$   0.02$
&$   0.06$
&$   0.02$
&$   0.12$
&$   0.08$
&$  -0.06$
&$  -0.03$
&    20./30\\
$\Sigma_s \rightarrow 3.0\mathrm{Mpc/h}$
&$   0.02$
&$   0.05$
&$   0.12$
&$   0.05$
&$   0.26$
&$   0.17$
&$  -0.11$
&$   0.00$
&    21./30\\
$\mathrm{FoG \; model} \rightarrow exp$
&$  -0.00$
&$  -0.00$
&$  -0.02$
&$  -0.01$
&$  -0.05$
&$  -0.02$
&$   0.02$
&$   0.00$
&    21./30\\
$\mathrm{FoG \; model} \rightarrow gauss$
&$   0.00$
&$   0.00$
&$   0.00$
&$   0.00$
&$   0.00$
&$   0.00$
&$  -0.00$
&$   0.00$
&    21./30\\


$\mathrm{No \; prior}(RL)$
&$  -0.08$
&$   0.14$
&$   0.05$
&$   0.20$

&$   0.03$
&$   0.56$
&$  -0.14$
&$   0.04$
&     20./30\\
$\mathrm{Only} \; \beta \; \mathrm{prior}(RL)$
&$  -0.05$
&$   0.03$
&$  -0.12$
&$   0.06$
&$  -0.29$
&$   0.15$
&$   0.08$
&$   0.02$
&    21./30\\
$\mathrm{Only} \; \log(B_0^2) \; \mathrm{prior}(RL)$
&$  -0.03$
&$   0.09$
&$   0.17$
&$   0.12$
&$   0.30$
&$   0.35$
&$  -0.22$
&$   0.01$
&    20./30\\
$P_{pt}(k)\;\mathrm{floating}\; \beta$
&$   0.02$
&$  -0.01$
&$  -0.01$
&$  -0.02$
&$  -0.01$
&$  -0.05$
&$   0.04$
&$  -0.01$
&    20./30\\
$P_{pt}(k)\;\mathrm{Fitting} \; 30<r<200\hMpc$
&$   0.24$
&$  -0.10$
&$   0.26$
&$  -0.09$
&$   0.73$
&$  -0.30$
&$  -0.03$
&$   0.02$
&    36./30\\
$P_{pt}(k)\;\mathrm{2-term} \; A_\ell(r)$
&$   0.08$
&$   0.02$
&$   0.72$
&$  -0.05$
&$   1.49$
&$  -0.02$
&$  -0.69$
&$  -0.11$
&    37./30\\
$P_{pt}(k)\;\mathrm{4-term} \; A_\ell(r)$
&$  -0.06$
&$  -0.04$
&$  -0.18$
&$  -0.08$
&$  -0.40$
&$  -0.23$
&$   0.14$
&$   0.00$
&    16./30\\
$P_{pt}(k)\mathrm{\;Fixed\;} \beta$
&$   0.06$
&$  -0.07$
&$  -0.18$
&$  -0.10$
&$  -0.30$
&$  -0.28$
&$   0.26$
&$  -0.03$
&    21./30\\
$P_{pt}(k)\mathrm{\;Only} \; \beta  \;  \mathrm{ prior}(RL)$
&$  -0.02$
&$   0.01$
&$  -0.12$
&$   0.03$
&$  -0.25$
&$   0.08$
&$   0.11$
&$   0.01$
&    20./30\\
$P_{pt}(k)\mathrm{\;Only} \; B_0 \; \mathrm{ prior}(RL)$
&$  -0.00$
&$   0.06$
&$   0.17$
&$   0.08$
&$   0.33$
&$   0.25$
&$  -0.18$
&$  -0.01$
&    20./30\\


$P_{pt}(k)\mathrm{\;No \; priors}(RL)$
&$  -0.04$
&$   0.11$
&$   0.07$
&$   0.15$
&$   0.09$
&$   0.44$
&$  -0.12$
&$   0.02$
&    20./30\\
\hline

\end{tabular}
\end{table*}
The results of the robustness test on data are summarized in Table.~\ref{tab:datatest} for DR11 post-reconstruction. The first line lists the best $\alpha$ and $\epsilon$ values as well as the corresponding $\alphapar$ and $\alphaper$ with their respective errors and the $\chi^2/d.o.f$ of the fit for the fiducial case. 
The remaining lines of Table.~\ref{tab:datatest} present the difference between fitted values or errors with respect to the fiducial case $\Delta v=v-v^{fid}$. 

First we consider variations in best fit $\alpha$ due to changes in the fitting methodology; the largest variations observed  are at 0.2\% level. When the range of fitting is changed regardless of template used. Small variations at 0.1\% level are observed in a few cases, such as when the order of the systematic polynomials is changed, as well as eliminating all priors and when fixing the $\beta$ for $P_{\rm PT}(k)$ template uniquely.
All these cases agree with the behavior observed when we apply the robustness tests are applied to mock galaxy catalogs as discussed in Section~\ref{sec:resultsmocks}. 
The variations in best fit value of $\epsilon$ are all $ \le 0.3\%$ with two exceptions. 
The changes are in agreement with the results of the mocks as described in Section~\ref{sec:resultsmocks}. 
The two exceptions show a variation of  $\sim0.7-0.8\%$, these occur when the lower order polynomials for the broad band terms is used, and this is a feature of both the De-Wiggled and $P_{\rm PT}(k)$ templates. 
Higher order polynomials, produce a variation at 0.2\% level.

Since the best fit values of $\alphaper$ and $\alphapar$ are generated from $\alpha$ and $\epsilon$, the variations observed in $\alpha$ and $\epsilon$ 
are also observed in $\alphaper$ and $\alphapar$ . 
In particular, $\Delta \alpha_{||} \le 0.4$\%   and $\Delta \alpha_{\perp} \le 0.3$\%, except for a few cases:
(a) Changing fitting range produces variations at 0.7-1.8\% 
(b) Using the lower orders polynomials produces variations at $1.5-1.7\%$ in $\alphapar$.
(c) For lower order polynomials, $\alphaper$ shows variations at 0.7-0.8\% level. 
(d) Fit  without priors produces  a 0.6\% variation in $\alphaper$. 

Now we turn our attention to the errors on the various fitted values. For all of the robustness tests, $\Delta \sigma_\alpha =\Delta \sigma_\epsilon \le0.002$. 
For errors on  $\alphapar$ and $\alphaper$, the variations are all within $0.3\%$
All of these cases agreed with the behavior observed when the robustness tests are applied to mock galaxy catalogs as discussed in Section~\ref{sec:resultsmocks}.

Finally, a fixed $\beta$ parameter does not lead to much smaller error  nor does it change our central values of fitted parameters when compared to the fiducial case.

\subsection{Data results with different methodologies} 
\label{sec:methodology_data} 

We apply different methodologies to the data and  summarize the results in Table.~\ref{tab:datametho}. 
We test four different methodologies, namely Multipoles-Gridded with two templates (De-Wiggled and $P_{\rm PT}$ templates), Multipoles-MCMC  and Clustering Wedges. We also include the results of the isotropic fitting in order to compare the isotropic to the anisotropic fitting  results. 
The isotropic results are described in \cite{Aar13}, and we refer the reader to this reference for more details.

Table.~\ref{tab:datametho} shows that for DR11 post-reconstruction the variations $\Delta {\alpha}\le 0.002$ and $\Delta {\epsilon}\le 0.004$  between variations of Multipoles fitting method (including using De-Wiggled Template, $P_{\rm PT}$ Template, 
and MCMC). 

However, Table.~\ref{tab:datametho} reveals that the DR11 post-reconstruction Multipoles (RPT) and Wedges results disagree by nearly to 1$\sigma$: $\alpha_{\parallel,{\rm Mult}}= 0.952 \pm 0.031$, $\alpha_{\parallel,{\rm Wedges}}= 0.982 \pm 0.031$; $\alpha_{\perp,{\rm Mult}}= 1.051 \pm 0.012$, $\alpha_{\perp,{\rm Wedges}}= 1.038 \pm 0.012$ .
The difference in $\alpha_\parallel$ is $0.030$. We then turn to the galaxy mock catalogs to see whether this behavior is common. In 39 cases out of 600 that show the same or larger differences are produced between the two methods. The mean difference is $0.005$ with a RMS of $0.016$, implying that this difference is at 1.9 $\sigma$ level.
The difference in $\alpha_\perp$ is  $0.013$; 45 out of 600  cases show the same or larger differences between the two methods.
The mean difference found in the mocks is $0.001$ and the RMS is $0.008$, suggesting that the difference in data is a  1.6$\sigma$ event.
Table.~\ref{tab:datametho}, indicates that this difference is primarily driven by their differences in $\epsilon$ fitted results, as the $\alpha$'s from both methodologies only differ by $0.2\%$, while $\epsilon$ changes   by $1.5\%$, which is comparable to the 1-$\sigma$ error on $\epsilon$.
We thus turn to a discussion using $\alpha$-$\epsilon$ parametrization for ease of discussion.

Pre-reconstruction, the Wedges and Multipoles measurements in $\alpha$ and $\epsilon$ differ by less than 0.25$\sigma$ as shown in Table.~\ref{tab:datametho}.
Fig.~\ref{fig:DaH} shows that, as reconstruction tightens the constraints from both methods, the central values have shifted slightly along the axis of constant $\alpha$. The discrepancy in the measurements is best quantified as a $1.5\%$ difference in $\epsilon$.
When we examine at this comparison in our mocks, we find a root mean square difference in $\epsilon$ fits of $0.007$, indicating that the data are a $2.004 \sigma$
outlier.  $27$ of 600 mocks have differences more extreme than $\pm 0.015$.  A total ofThe other 3 cases (DR10 and DR11 pre-reconstruction) show smaller variations.  We conclude that this event is consistent with normal scatter of the two estimators.

We will only briefly discuss the DR10 fitting results, as the differences between all methodologies are fairly small compared with the errors, and the best fit results in all parameterizations are all relatively consistent with each other. 
The only point of interest is in DR10 post-reconstruction,  $\Delta {\epsilon}  \approx 0.014$ in the Wedges and Multipoles (for all three variations) comparisons. 
This variation level is also observed in 81 out of 600 mocks, and thus indicating that this is a 1.4-$\sigma$ event. 
These differences are in general agreement  with the differences  obtained with the mocks otherwise.

\subsection{Combining Results from different methodologies}
 
The tests on our fitting methodology, presented for the mock samples in Section~\ref{sec:resultsmocks} and on the DR11 data in Section~\ref{sec:fitdata}, suggest that no systematic issue is causing the observed discrepancies in data. Thus, we combine results produced from the Multipoles Grid and Wedges measurements to create our consensus anisotropic BAO measurement shown in Table.~\ref{tab:datametho}.
To combine results from clustering wedges with 
multipoles, we average the likelihood distributions recovered from the Multipoles and Wedges . We addd the systematic errors to the consensus
anisotropic BAO results. 
Although the marginalized constraints on $\alpha_\perp$ and
$\alpha_{||}$ quoted in the paper include covariance correction factors to
account for the noise in our covariance matrix, this correction was not
included in the full 2D posteriors that we combine to get the consensus
result. We include an extra factor of $\sqrt{(m1)}$ centered
at the corresponding best fit point. These dilations were applied to the
posteriors from wedges and multipoles before averaging them to get the
consensus constraints.

We include additional 
uncorrelated systematic error terms on $\alpha$ and $\epsilon$
of $\sigma_{\alpha_{sys}} = \sqrt{0.003^2 + 0.003^2}$ and $
\sigma_{\epsilon_{sys}}  = 0.003$. Translated into systematic errors on $\alpha_\perp$ and $\alpha_{||}$, these
values correspond to $\sigma_{\perp_{sys}} = 0.005$ and $\sigma_{{||}_{sys}}  = 0.007$ with a small correlation factor. 
To include these systematics in the
consensus posterior a convolution with the Gaussian describing the
systematic errors was performed.

\begin{figure*}
{\includegraphics[width=3.in]{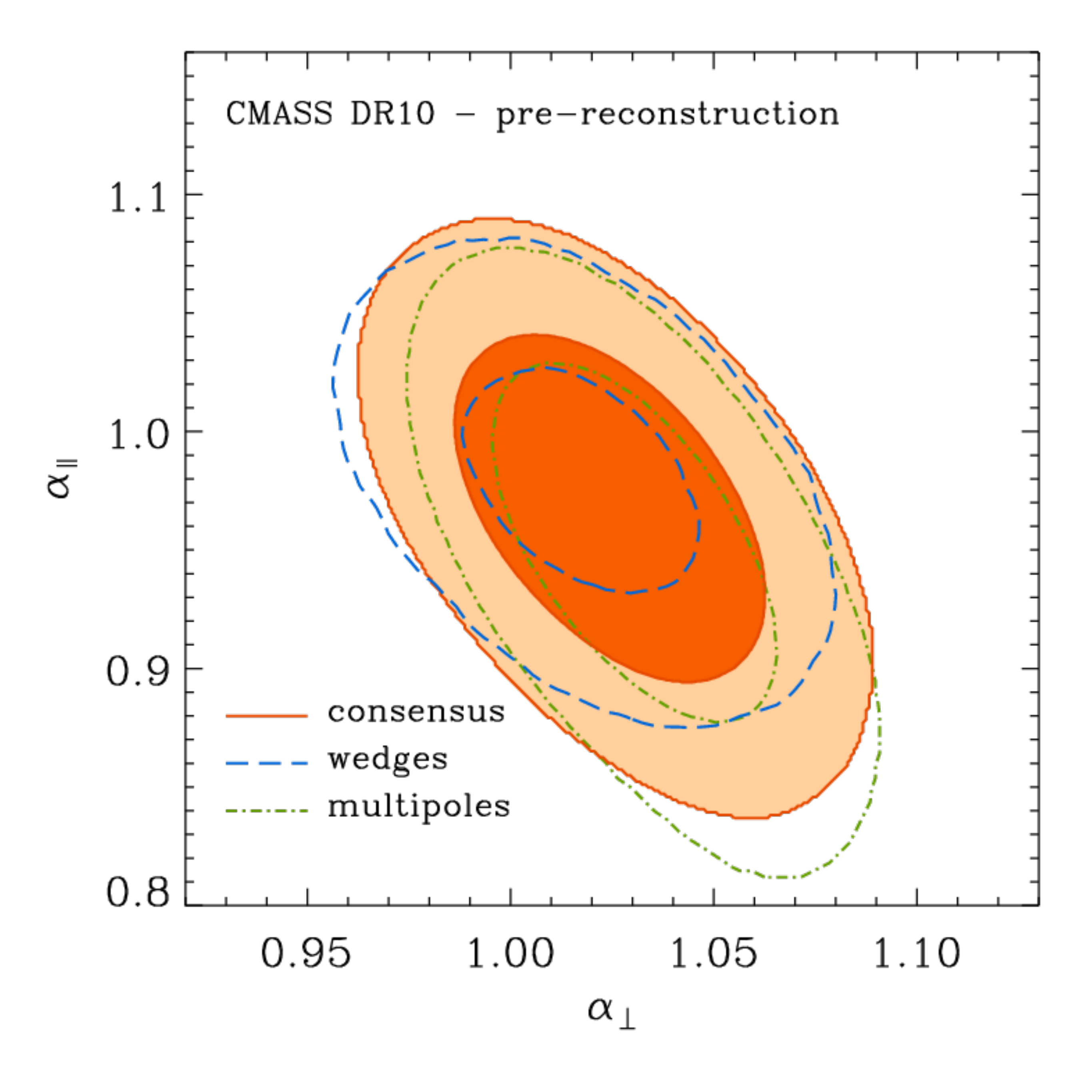}}
{\includegraphics[width=3.in]{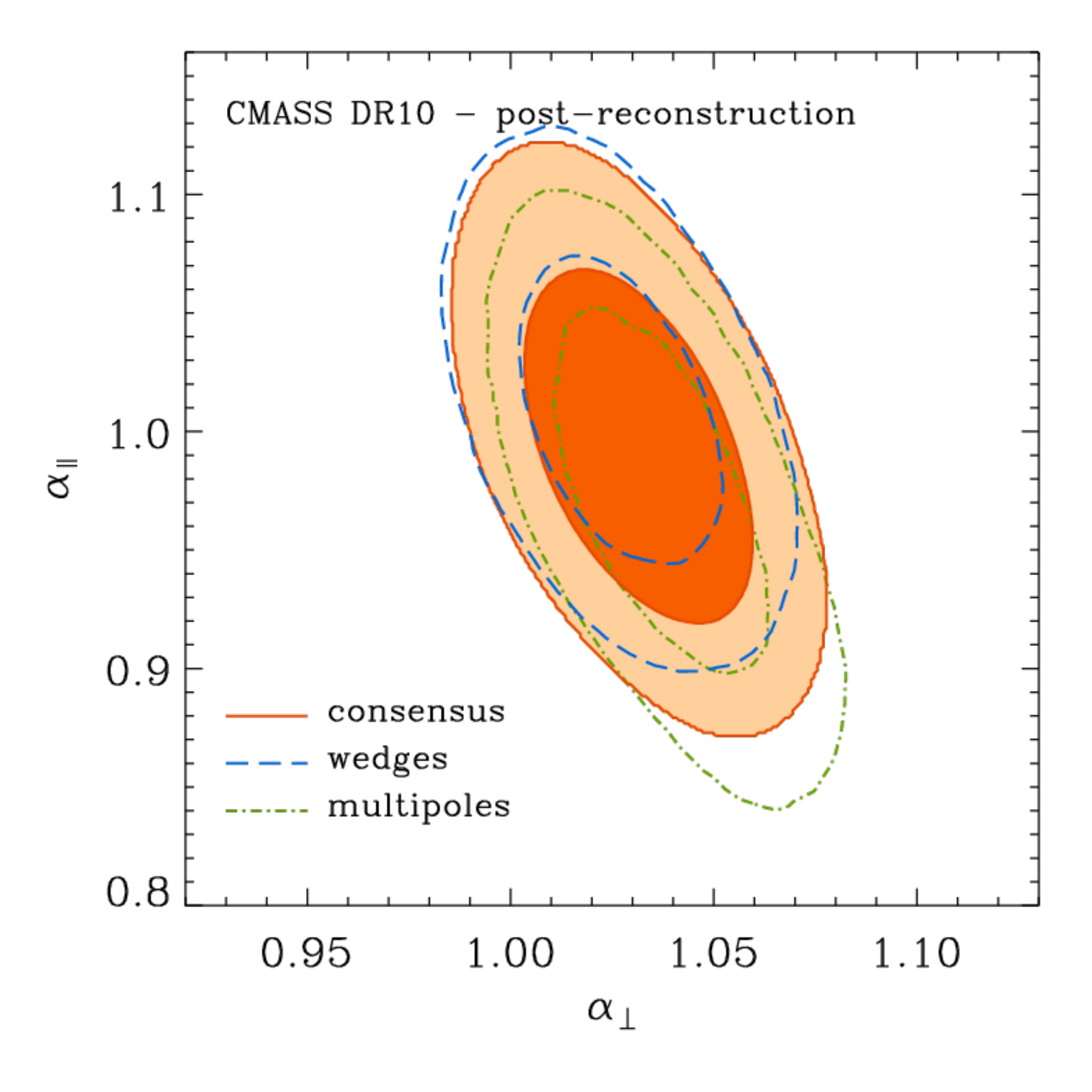}}
{\includegraphics[width=3.in]{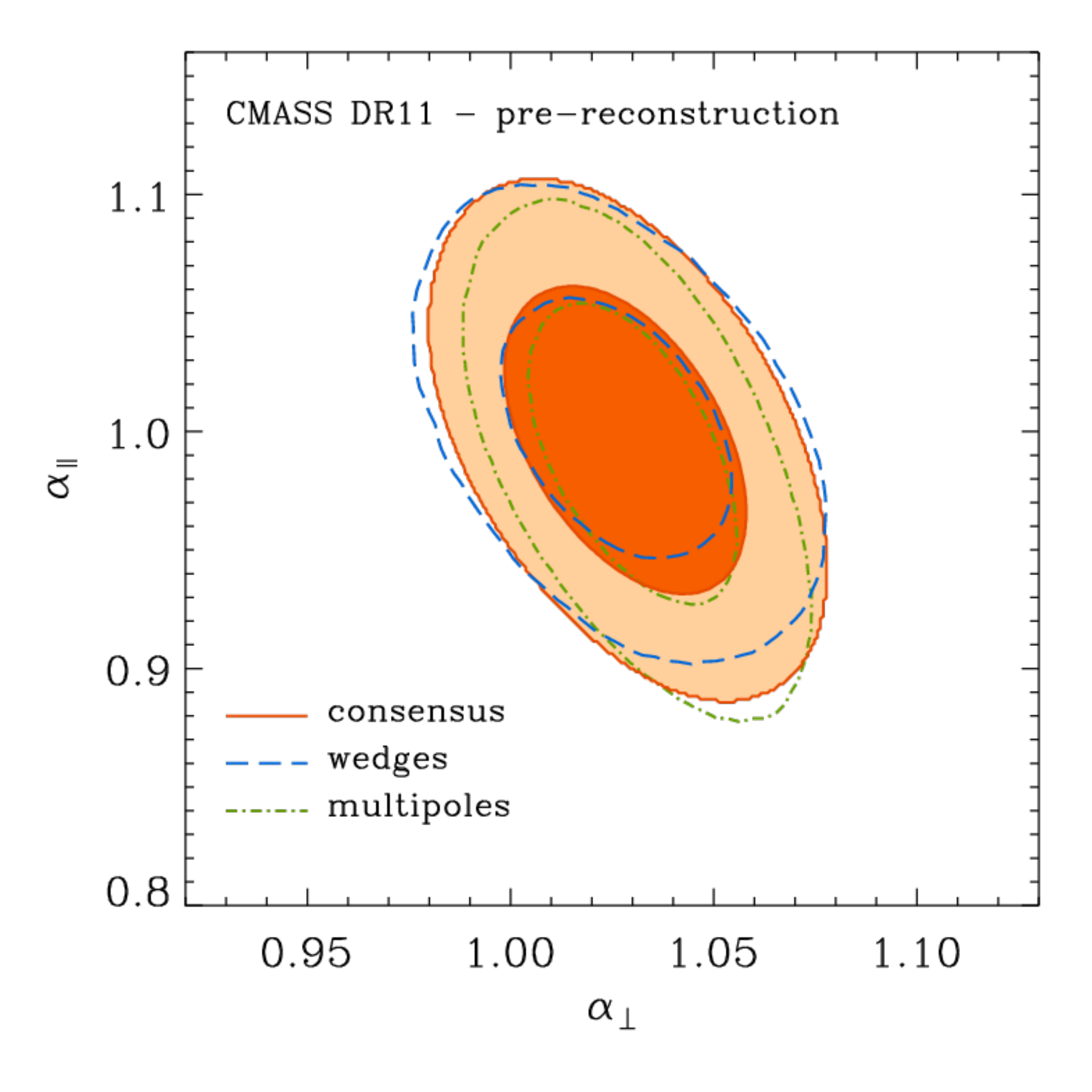}}
{\includegraphics[width=3.in]{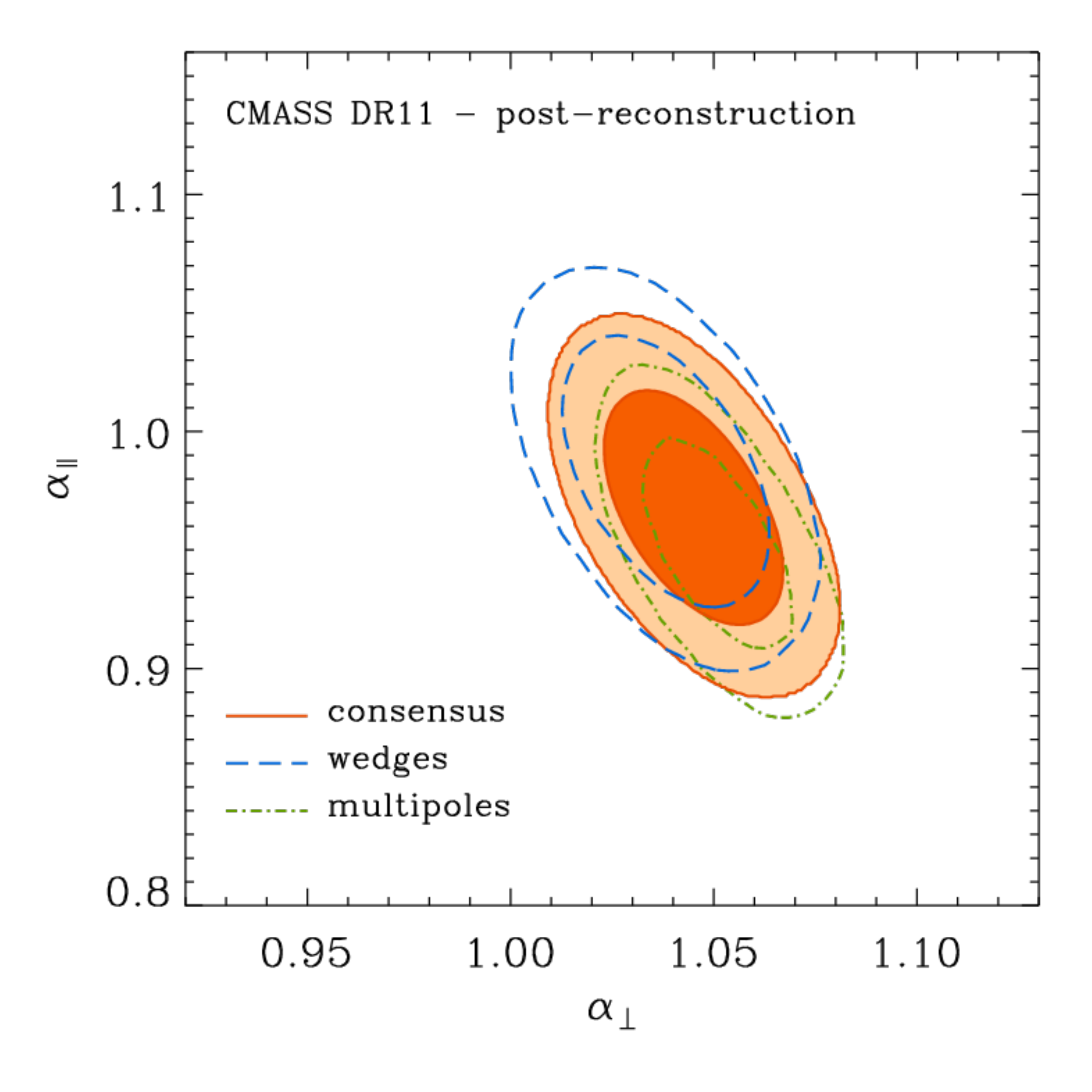}}
  \caption{
Comparison of the 68\% and 95\% constraints in the $D_A-H(z)$ plane scaled by ($r^{\rm fid}/r_d$) obtained from multipoles gridded analysis [blue short-dashed line] and wedges [green long-dashed line], for DR11 pre-reconstruction[right] and post reconstruction  [left]. The solid contours are the consensus values issues from combining the $\mathrm{log}(\chi^2)$ from both approaches. The multipoles provide slightly tighter constraints; the consensus contours follow a more elongated form  aligned with the axis of constant $\alpha$.
}
\label{fig:DaH}
\end{figure*}

Fig.~\ref{fig:DaH} displays the comparison 
of the 68\% and 95\% constraints in the $\alpha_{\perp}$ and $\alpha_{||}$ plane 
using the two methods: Multipoles Gridded (our fiducial methodology) and Wedges. The size of the contours from both methods agree well, with a more elongated contour from Multipoles.  Post-reconstruction we observe that Multipoles method favors smaller values of $\alphapar$ in both data releases while Wedges tends to higher values. 
However, the contours show a fairly good agreement between the two methodologies. Pre-reconstruction contours are well matched one to the other, but surprisingly in the DR10 pre-reconstruction the contours from Multipoles are a bit larger than wedges contours and also are favoring to even more lower values.
 

\begin{table*}
\caption{CMASS fitting results using various dataset (DR10, DR11, pre- and post-reconstruction) using different methodologies. A few notes: the isotropic fit quoted here is the single fit in correlation function with  De-Wiggled template, thus it should be most comparable to Multipoles-DeW anisotropic fit. The Multipoles-Dew anisotropic fits are most likely to be similar to the isotropic fit particularly post-reconstruction. All quoted error includes only statistical errors, but not systematic errors. Note that both Multipoles-DeW and Multipoles-RPT employs gridded approach, and thus we omit the designating "Grid" here for simplicity.}
\label{tab:datametho}

\begin{tabular}{@{}lccccccc}

Model&$\alpha$&$\epsilon$&$\rho_{\alpha, \epsilon}$&$\alpha_{||}$&$\alpha_\perp$&$ \rho_{||, \perp}$&$\chi^2/d.o.f$\\
\hline
DR11 post-reconstruction\\
{\bf Consensus}&
{\bf 1.0186}$\pm${\bf 0.0104}&
{\bf -0.0252}$\pm${\bf 0.0142}&
{\bf 0.390}&
{\bf 0.9678}$\pm${\bf 0.0329}&
{\bf 1.0449}$\pm${\bf 0.0145}&
{\bf -0.523}\\

Multipoles-DeW&
$ 1.0166\pm 0.0089$
&$-0.0324\pm  0.013$
&$  0.512$
&$ 0.9518\pm  0.031$
&$ 1.0507\pm  0.013$
&$ -0.297$
&$    21./30$\\
Multipoles-RPT&
$ 1.0168\pm 0.0088$
&$-0.0326\pm  0.013$
&$  0.505$
&$ 0.9517\pm  0.031$
&$ 1.0511\pm  0.013$
&$ -0.304$
&$    21./30$\\
Multipoles MCMC
&$1.017\pm0.010$ 
& $-0.028\pm0.012$  
&  $0.363$                   
&  $0.962\pm0.028$
& $1.047\pm0.013$      
&  $-0.439$            
& $18/30$     \\
Wedges&
$1.019\pm 0.010$&
$-0.018\pm 0.013$&
0.389&
$0.982\pm0.0312$&
$1.038\pm0.014$&
-0.501&
21./30\\
Post-Rec. Isotropic&
$1.021\pm0.009$  &
-- &
--&
--&
--&
-&
$16/17$\\
\hline
DR11 pre-reconstruction\\
Multipoles-DeW&
$ 1.0245\pm 0.0142$
&$-0.0101\pm  0.019$
&$  0.572$
&$ 1.0039\pm  0.049$
&$ 1.0350\pm  0.017$
&$ -0.144$
&$    33./30$\\
Multipoles-RPT&
$ 1.0170\pm 0.0133$
&$-0.0122\pm  0.019$
&$  0.495$
&$ 0.9923\pm  0.046$
&$ 1.0296\pm  0.018$
&$ -0.241$
&$    35./30$\\
Multipoles-MCMC
&$1.015\pm0.015$   
& $-0.016\pm0.018$  
&  $0.423$                   
 &  $0.983\pm0.044$
 & $1.033\pm0.019$     
 &  $-0.406$            
 & $31/30$     \\
 Wedges&
$1.018\pm$0.015&
$-0.008\pm$0.018&
0.236&
$1.001\pm0.043$&
$1.027\pm0.021$&
-0.453&
33/30\\
Isotropic&
$1.031\pm0.013$  &
- &
-&
-&
-&
-&
$14/17$\\
\hline
DR10 post-reconstruction\\
{\bf Consensus}&

{\bf 1.0187}$\pm${\bf 0.0151}&
{\bf -0.0123}$\pm${\bf 0.0202}&
{\bf 0.502}&
{\bf 0.9937}$\pm${\bf 0.0495}&
{\bf 1.0314}$\pm${\bf 0.0187}&
{\bf -0.501}\\
Multipoles-DeW&
$ 1.0151\pm 0.0155$
&$-0.0203\pm  0.023$
&$  0.669$
&$ 0.9744\pm  0.057$
&$ 1.0361\pm  0.019$
&$ -0.158$
&$    16./30$\\
Multipoles-RPT&
$ 1.0155\pm 0.0157$
&$-0.0203\pm  0.023$
&$  0.683$
&$ 0.9747\pm  0.058$
&$ 1.0365\pm  0.018$
&$ -0.136$
&$    16./30$\\

Multipoles MCMC
&$1.016\pm0.015$   
& $-0.019\pm0.018$  
&  $0.484$                   
&  $0.979\pm0.045$    
& $1.035\pm0.018$   
&  $-0.445$            
& $16/30$     \\
Wedges&
$1.020\pm$0.015&
$-0.006\pm$0.019&
0.513&
$1.009\pm0.049$&
$1.027\pm0.018$&
-0.474&
17/30\\
 Isotropic&
$1.022\pm 0.013$  &
-- &
--&
--&
--&
-&
$14/17$\\
\hline
DR10 pre-reconstruction\\
Multipoles-DeW&
$ 1.0123\pm 0.0177$
&$-0.0215\pm  0.026$
&$  0.555$
&$ 0.9693\pm  0.063$
&$ 1.0345\pm  0.023$
&$ -0.233$
&$    35./30$\\
Multipoles-RPT&
$ 1.0043\pm 0.0162$
&$-0.0244\pm  0.025$
&$  0.439$
&$ 0.9560\pm  0.057$
&$ 1.0294\pm  0.024$
&$ -0.344$
&$    36./30$\\

Multipoles MCMC
&$1.000\pm0.018$   
& $-0.023\pm0.022$  
&  $0.388$                      
&  $0.955\pm0.051$ 
& $1.024\pm0.024$   
&  $-0.458$            
& $32/30$     \\
Wedges&
$1.004\pm$0.018&
$-0.015\pm$0.022&
0.104&
$0.975\pm0.049$&
$1.020\pm0.028$&
-0.482&
30/30\\
Isotropic&
$1.022\pm 0.017$  &
-- &
--&
--&
--&
-&
$16/17$\\
\hline

\end{tabular}
\end{table*}

The solid contours are the consensus values from combining the $\mathrm{log}(\chi^2)$ from both approaches. The consensus anisotropic contours are larger than each of the individual methods  (Multipoles and Wedges). This is due to the fact that there is slight difference in the central values of the contours  fit by Multipoles and Wedges, thus enlarging the consensus contour. This effect can be clearly seen in Fig.~\ref{fig:DaH}.

\subsection{Isotropic vs Anisotropic results}

\begin{figure}
{\includegraphics[width=3.5in]{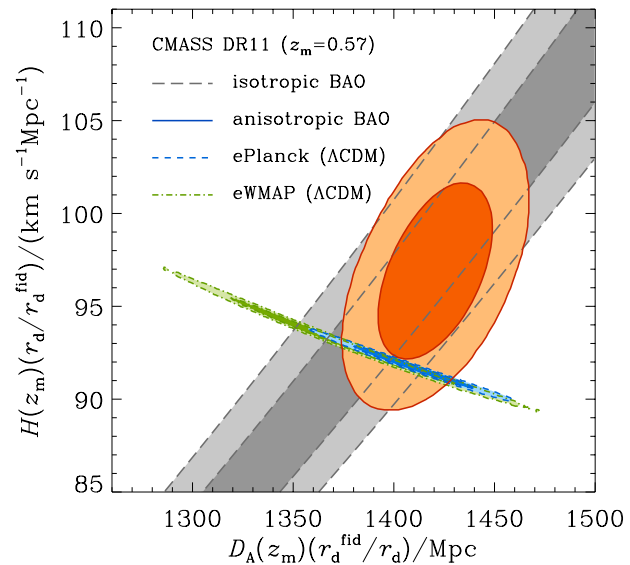}}
\caption{Comparison of the 68\% and 95\% constraints in the
$D_A(z=0.57)(r_d^{\rm fid}/r_d)-H(z=0.57)(r_d^{\rm fid}/r_d)$ plane from
CMASS consensus anisotropic and isotropic BAO constraints.
The Planck contours correspond to Planck+WMAP polarization (WP) and no lensing.
The green contours show the constraints from WMAP9.}
\label{fig:DaH_CMB}
\end{figure}

We find consistent results between isotropic and anisotropic fitting results for all datasets (DR10, DR11) pre- and post-reconstruction as described in Table.~\ref{tab:datametho}.
Post-reconstruction, the central values of $\alpha$ measured from isotropic and anisotropic clustering are consistent to well within 1-$\sigma$.  Pre-reconstruction, the fits to the isotropic correlation function are approximately 1-$\sigma$ higher than the joint fits to both the monopole and quadrupole, for both DR10 and DR11. 
Part of this difference can be explained by the different correlation function templates used for the isotropic and anisotropic analyses. 
The isotropic fitting uses a non-linear power-spectrum ``De-Wiggled template" \citep{And12,And13}, while the anisotropic fitting uses $P_{pt}(k)$ as described in equation~\ref{eqn:RPT}.

While the isotropic fits yield $\alpha$ measurements that agree with the correlation function anisotropic fits, they are in general higher than the anisotropic consensus value, and the effect of this is noticeable when the measurements are combined with the CMB data and turned into cosmological constraints (see Fig.~\ref{fig:DaH_CMB}).

Finally, we compare the anisotropic and isotropic clustering by examining the 65\% and 95\%  constraints in the $D_A(z=0.57)(r_d^{\rm fid}/r_d)-H(z=0.57)(r_d/r_d^{\rm fid})$ plane from CMASS consensus anisotropic and isotropic BAO constraints in Fig.~\ref{fig:DaH_CMB}. 
In general the isotropic and anisotropic central values align well (along the constant $\alpha$ axis), but anisotropic clustering has a smaller contour as it provides constraints in both perpendicular and parallel to line of sight direction. 
Also shown in Fig.~\ref{fig:DaH_CMB} are the flat $\Lambda$CDM, $\sum m_{\nu} = 0.06$ eV predictions from the ePlanck and eWMAP CMB data sets detailed in \cite{Aar13}.  
Our 68\% and 95\%  constraints in the $D_A(z=0.57)(r_d^{\rm
fid}/r_d)-H(z=0.57)(r_d/r_d^{\rm fid})$ plane from CMASS consensus anisotropic
measurements are highlighted in orange in Fig.~\ref{fig:DaH_CMB}.   The grey are one-dimensional one and two-sigma contours of our consensus isotropic
BAO fit (which is a combination of $P(k)$ fits and correlation function fits).  
This figure illustrates
the 0.5\% increase in the best-fit $\alpha$ from the anisotropic fits compared
with the isotropic ones.
The CMASS isotropic BAO constraints are consistent with both CMB predictions
shown here, while the anisotropic constraints show more overlap with Planck,
indicating a mild preference for higher values of $\Omega_c h^2$ preferred by
Planck.

\section{Discussion and Conclusions}\label{sec:discussion}
We test the multipoles fitting methodology with mock catalogs and data for DR10/DR11 CMASS galaxy catalogs. The methodological changes tested are enumerated in Section~\ref{sec:Fitting}.  We summarize and discuss our findings in this section. We also limit our discussion to one parameterization $\alpha-\epsilon$ and only for DR11 for ease of discussion.

\begin{itemize}

\item With a large array of robustness tests, our fiducial methodology shows a maximum  shift of 0.5\%  in best fit  $\alpha$ value and 0.3\% in best fit $\epsilon$ value pre-reconstruction. 
Post reconstruction, the maximum shifts are further reduced to $\le$ 0.1\% in $\alpha$ and  $\le 0.5\%$ in $\epsilon$.  

\item We list methodology changes that give the larger variations pre-reconstruction: (i) changing De-Wiggled Template to $P_{\rm PT}$ template (0.5\% in best fit $\alpha$) ; (ii) changing the fitting range from  $[50,200]$ $h^{-1}$Mpc to $[30,200]$ $h^{-1}$Mpc  (0.3\% in best fit $\alpha$); (iii) Changing $\Sigma_{||,\perp}$ to  $\Sigma_{||}=\Sigma_\perp=8.0$Mpc$h^{-1}$ (0.3\% in best fit $\epsilon$).

\item Post-reconstruction, the variations in $\alpha$ and $\epsilon$ are impressively small, in all cases $\Delta \alpha,\Delta \epsilon \le 0.1\%$
 except when we use polynomials of lower/higher order than the fiducial case to describe the broadband terms, we observe changes of in $\Delta \epsilon \approx 0.2-0.3\%$.

\item  A large array of robustness tests,  show only small effects in $\sigma_\alpha$ (a maximum shift of 0.1\%), while the various changes in the methodology mostly affects $\sigma_\epsilon$ (a maximum shift of 0.2\%)  pre-reconstruction, if at all. 
Post-reconstruction, the errors have similar shifts as in pre-reconstruction. 
However, there is significant scatter in the mocks when we do not include any priors on $B_0, \beta$ in both pre- and post-reconstruction, while allowing the integration interval to calculate the errors to exceed the physical range that is used actually fit the data. 

\item The effects of priors are stronger when the integration interval is not limited (which we use to calculate the errors) to a fixed $\alpha-\epsilon$ range. The range is chosen so that are not explored regions in $\alpha-\epsilon$ space that do not correspond to any physical ranges used to fitted data. 

\end{itemize}

We discuss the following changes in the fitting methodology that cause changes in the best fit values: 
\begin{itemize}

\item {\bf Templates}. The choice of templates affect mostly the best fit values and has a small effect on the fitted errors.  
In addition,  the bias related to De-Wiggled template is in the opposite direction to the biased shift when we use RPT-inspired $P_{\rm PT}(k)$  template. 
This inverted trend observed in Pre-reconstruction data generates quite different results in term of $\alpha_{||}-\alpha_{\perp}$ parametrization, as the de-wiggled result becomes heavily biased, reaching a 0.9-1.2\% shift in $\alpha_{||}$ and a 0.2\% shift in $\alpha_\perp$. In the case of RPT the shift in the parallel direction is only 0.2 \% while it is a little bit larger for the perpendicular direction,  0.2-0.3 \%. 
After reconstruction, for both DR10/DR11 consistent results are obtained using either templates, the shift on $\alpha$ reduces to $\le$1\% and  a slightly larger bias with de-wiggled template of 0.1\% compared to no bias for $P_{\rm PT}(k)$ template when we consider $\epsilon$. This result is a reason why we choose to adopt $P_{\rm PT}(k)$ template in \cite{Aar13} as the fiducial template. 
\item {\bf Priors}. The elimination of priors produces a large effect pre- and post-reconstruction when a physically meaningful integration interval is not applied when we calculate the errors for $\alpha$ and $\epsilon$. 
These changes do not actually affect the cosmological constraints provided by the anisotropic clustering measurement, as 
cosmological constraints are derived directly from using the full likelihood surface, and thus are not affected by the integration interval we adopt when we calculate the error on $\alpha$ or $\epsilon$. 
These large variation in the $\sigma$s are related to non-zero likelihood at extreme $\epsilon$. These cases correspond to situations where the  acoustic peak along the line of sight has been shifted out of the 50-200 Mpc fitting range. 
A prior on $\epsilon$ of 0.15 corresponds to a maximum dilation of $\alphapar$ of $0.85^{-3}= 1.63$, which should force the peak to be contained within in our domain. Constraining the grid to be in the $\epsilon$ range [-0.15,0.15] avoid these unphysical tails at low $\epsilon$ values.

\end{itemize}

We find the following when we compare significantly different methodologies (Multipoles-Gridded, Multipoles-MCMC, Clustering Wedges), we switch to $\alphapar -\alphaper$ for the discussion and we concentrate on DR11 for ease of discussion:

\begin{itemize}
\item There is agreement between the MCMC and the Grid version when we use Multipoles, though there is a small level of scatter.   
The median difference for pre-reconstruction for DR10/DR11 is $|\widetilde{\alpha_{||,\perp}}|\le-0.3\%$. 
The median difference  post-reconstruction reduces to $|\widetilde{\alpha_{||,\perp}}|\le-0.1\%$. 

\item The fitted errors are well correlated for the two methodologies  (Multip. MCMC and Multip. Grid) with a small  level of dispersion.
The variation post-reconstruction is $|\widetilde{\sigma_{\alphapar}} |= 0.2\% $ and $|\widetilde{\sigma_{\alphaper}} |= 0.1\%$. 
The dispersion is mostly observed in the parallel direction and decreases significantly post-reconstruction.  

\item We compare Multipoles-Gridded against Clustering Wedges.  There are small variations in the bias between methodologies but we do not find any indication that any methodology is more biased with respect to the others.  
The median differences in the best fit values are small. 
With  pre-reconstructed mocks, we find  $\Delta \alphapar = 0.7\%$,  $\Delta \alphaper = 0.4\%$; 

Post-reconstruction the discrepancies are even smaller, $\Delta \alpha_{||, \perp} = 0.3\%, 0.1\%$. 

\item When fitting using Multipoles-Gridded and Clustering Wedges, he differences in  fitted  errors are also small:  $\Delta \sigma_{\alphapar} =0.8\%$, $\Delta\sigma_{\alphaper} = 0.3\%$.
 Post-reconstruction the discrepancies are even smaller $\Delta \sigma_{\alphapar, \alphaper}\le 0.5\%$. 
The errors are well correlated between both  methodologies. 

\item The results suggests that there are  two components in the median variation and  dispersions of the fitted values and corresponding errors. 
The variation is produced bycomes from the estimator itself (wedges vs multipoles), and from the different implementations. 

\end{itemize}

To conclude, we  demonstrate the robustness of the multipoles fitting method at the 0.1-0.2\% level in $\alpha$ and $\epsilon$ .
We quote a systematic error from these fitting techniques of  0.2\% + 0.2\% in quadrature, which is 0.3\%. 
Being more conservative, since we have only shown the result of one variation direction at a time; adding several uncertainties in quadrature could accumulate slightly more error.
We highlight, however, that given the current precision on the measurements, assuming the mean error in  $\alpha$ and $\epsilon$ characterizes the full error budget, any variation observed in the $\alpha$s lies perfectly within this error. 
There are possible systematics beyond this study, e.g.,
due to possible imperfections in the mocks: light cone effects,
incorrect cosmology, incorrect dynamics in the mocks, and incorrect galaxy assignment
model.  The impacts of these issues are harder to assess, but might be at the 0.2-0.3\% level
in $\epsilon$.  These are however beyond the scope of our current paper. 
We can  conclude that for the current and near future data sets,  the fitting model is still robust against changes in the methodology.

%

\section{Acknowledgement}\label{sec:acknow}
S. H. is partially supported by the New Frontiers in Astronomy and Cosmology program at the John Templeton Foundation and was partially supported by RESCEU fellowship, and the Seaborg and Chamberlain Fellowship (via Lawrence Berkeley National Laboratory) during the preparation of this manuscript. 



Numerical computations for the PTHalos mocks were done on the Sciama
High Performance Compute (HPC) cluster which is supported by the ICG,
SEPNet and the University of Portsmouth

Funding for SDSS-III has been provided by the Alfred P. Sloan Foundation, the Participating
Institutions, the National Science Foundation, and the U.S. Department of Energy.

AGS acknowledges support from the Trans-regional Collaborative Research
Centre TR33 `The Dark Universe' of the German Research Foundation (DFG).

SDSS-III is managed by the Astrophysical Research Consortium for the
Participating Institutions of the SDSS-III Collaboration including the
University of Arizona,
the Brazilian Participation Group,
Brookhaven National Laboratory,
University of Cambridge,
Carnegie Mellon University,
University of Florida,
the French Participation Group,
the German Participation Group,
Harvard University,
the Instituto de Astrofisica de Canarias,
the Michigan State/Notre Dame/JINA Participation Group,
Johns Hopkins University,
Lawrence Berkeley National Laboratory,
Max Planck Institute for Astrophysics,
Max Planck Institute for Extraterrestrial Physics,
New Mexico State University,
New York University,
Ohio State University,
Pennsylvania State University,
University of Portsmouth,
Princeton University,
the Spanish Participation Group,
University of Tokyo,
University of Utah,
Vanderbilt University,
University of Virginia,
University of Washington,
and Yale University.

\appendix{DR11 versus DR10 mocks}
\section{DR11 versus DR10 mocks}

\subsection{DR11 versus DR10 mocks}

The results for DR10 are shown in Tables.~\ref{tab:dr10norec}, and~\ref{tab:dr10rec} for the best fitting values, pre and post-reconstruction and in~\ref{tab:sigmadr10norec}, and~\ref{tab:sigmadr10rec} for the uncertainties pre and post reconstruction.
For the best fitting values we observe the same trends in both data releases, however we can notice two differences, the median $\alpha$'s in DR11 are for the most part of them lower by 0.001-0.002 in $\alpha_\parallel$ and $\alpha_\perp$ (with 4 cases showing a difference 0.003). 
This differences in the bias could be associated to the correlation between mocks generated by the overlapping regions. These correlation is larger in DR10 is r=0.49 while in DR10, r=0.33. 
The second difference is the dispersion quantified by the percentiles of the distributions of the best fit values and their uncertainties for all cases are lower in DR11, the percentiles are also more symmetric for $\alpha_\parallel$ and $\alpha_\perp$.  The dispersion reduction is  related to the signal to noise increase with the larger volume covered  by DR11 
Concerning the best fitting values, pre-reconstruction, we observe two cases that appears only in one data release: 1)A $\Delta \alpha =0.1\%$ is observed when using second order polynomials  in DR11 that is not observed in DR10; 2)a $\Delta \epsilon=0.1\%$ observed using smaller bins in DR10 is not observed in DR11. Post-reconstruction two cases shows a $\Delta \alpha =0.1\%$ in DR10 not reproduced in DR11: using 4-mPc bins and higher order polynomials. One case in DR11 shows $\Delta \epsilon =0.1\%$ non observed in DR10, the larger fitting range.

For the uncertainties we expect that the mean error decreases for DR11compared with DR10 results because the increasing volume surveyed. The DR11 fiducial case shows median uncertainty of $\widetilde{\sigma_{||}}=0.044$ compared with DR10 $\widetilde{\sigma_{||}}=0.057$. For the perpendicular direction $\widetilde{\sigma_{\perp}}= 0.021$  compared with $\widetilde{\sigma_{\perp}}= 0.027$. Post-reconstruction, $\widetilde{\sigma_{||}}=0.036 \rightarrow 0.028$ and $\widetilde{\sigma_{\perp}}= 0.019\rightarrow 0.015$.

In the case of the errors there are also some differences between data releases. Pre-reconstruction, two cases shows a variation $\Delta \sigma_\alpha =0.001$, 4 Mpc bins and 2order polynomials in DR11, that is not observed in DR10. Also a  $\Delta \sigma_\alpha =0.001$ is observed in DR10 when using RPT templates with fixed $\beta$ that is not reproduced in DR11. In the $\sigma_\epsilon$, there three cases showing a 0.001 difference in DR10 not observed in DR11: large fitting range, 2-order polynomials and RPT templates with fixed $\beta$. Post-reconstruction, we observe a $\Delta \sigma_\alpha, \Delta \sigma_\epsilon=0.001$ when using PRT template with floating $\beta$ in DR11 non observed in DR10 and vice-versa, we observe a $\Delta \sigma_\epsilon=0.001$  in DR11 when using 2-order polynomials non observed in DR11.

\begin{table*}
\caption{Fitting results from mocks from numerous variations of our fiducial fitting methodology for DR10 mock galaxy catalogs pre-reconstruction without covariance corrections for the overlapping mock regions. First line shows the median and 16th and 84th percentiles for the $\alpha, \epsilon, \alpha_{||, \perp}$. The rest of the lines shows the median bias and median  variations $\Delta v=v_{i}-v_{f}$ with their corresponding 16th and 84th percentiles. The median bias $\widetilde{b}$, median variations $\widetilde{\Delta v}$, and percentiles are multiplied by 100. }
\label{tab:dr10norec}

\begin{tabular}{@{}lrrrrrrrr}

\hline
Model&
$\widetilde{\alpha}$&
-&
$\widetilde{\epsilon}$&
-&
$\widetilde{\alpha_{\parallel}}$&
-&
$\widetilde{\alpha_{\perp}}$&
-\\
\hline
\multicolumn{9}{c}{DR10 Pre-Reconstruction}\\
\hline
\\[-1.5ex]

Fiducial&
$1.0033^{+0.0186}_{-0.0177}$&-&
$0.0029^{+0.0201}_{-0.0239}$&-&
$1.0092^{+0.0469}_{-0.0512}$&-&
$1.0022^{+0.026}_{-0.026}$&-\\
\hline
Model&
$\widetilde{b_\alpha}$&
$\widetilde{\Delta\alpha}$&
$\widetilde{b_\epsilon}$&
$\widetilde{\Delta\epsilon}$&
$\widetilde{b_{\parallel}}$&
$\widetilde{\Delta\alpha_{\parallel}}$&
$\widetilde{b_{\perp}}$&
$\widetilde{\Delta\alpha_{\perp}}$\\
\hline
\\
$30<r<200\mathrm{Mpc}/h$&
$-0.00^{+1.96}_{-1.73}$&
$-0.29^{+0.25}_{-0.30}$&
$-0.10^{+2.01}_{-2.52}$&
$-0.20^{+0.20}_{-0.25}$&
$0.16^{+4.63}_{-4.89}$&
$-0.70^{+0.57}_{-0.73}$&
$0.16^{+2.63}_{-2.62}$&
$-0.06^{+0.20}_{-0.22}$\\
\\[-1.5ex]

$\mathrm{2-term} \; A_\ell(r)$&
$0.26^{+1.88}_{-1.68}$&
$-0.01^{+0.16}_{-0.18}$&
$-0.28^{+1.95}_{-2.27}$&
$0.01^{+0.31}_{-0.28}$&
$0.79^{+4.66}_{-4.60}$&
$0.00^{+0.77}_{-0.71}$&
$0.23^{+2.41}_{-2.53}$&
$-0.02^{+0.14}_{-0.17}$\\
\\[-1.5ex]

$\mathrm{4-term} \; A_ell(r)$&
$0.32^{+1.83}_{-1.89}$&
$-0.04^{+0.18}_{-0.19}$&
$-0.30^{+1.97}_{-2.44}$&
$-0.03^{+0.23}_{-0.26}$&
$0.87^{+4.70}_{-5.23}$&
$-0.09^{+0.54}_{-0.59}$&
$0.21^{+2.61}_{-2.55}$&
$-0.03^{+0.26}_{-0.19}$\\
\\[-1.5ex]

$\mathrm{Fixed \;} \beta=0.4$&
$0.32^{+1.88}_{-1.76}$&
$-0.00^{+0.01}_{-0.03}$&
$-0.29^{+1.98}_{-2.50}$&
$0.00^{+0.07}_{-0.07}$&
$0.90^{+4.70}_{-5.08}$&
$0.00^{+0.11}_{-0.14}$&
$0.22^{+2.62}_{-2.60}$&
$-0.00^{+0.09}_{-0.09}$\\
\\[-1.5ex]

$\spar=\sperp=8\mathrm{Mpc}/h$&
$0.32^{+1.88}_{-1.65}$&
$0.00^{+0.26}_{-0.28}$&
$-0.03^{+2.03}_{-2.56}$&
$-0.30^{+0.35}_{-0.35}$&
$0.35^{+5.00}_{-5.30}$&
$-0.57^{+0.86}_{-0.82}$&
$0.57^{+2.67}_{-2.61}$&
$0.32^{+0.34}_{-0.37}$\\
\\[-1.5ex]

$\Sigma_s \rightarrow 3.0$&
$0.38^{+1.87}_{-1.75}$&
$0.04^{+0.05}_{-0.05}$&
$-0.37^{+2.03}_{-2.45}$&
$0.06^{+0.07}_{-0.08}$&
$1.04^{+4.86}_{-5.08}$&
$0.17^{+0.18}_{-0.20}$&
$0.23^{+2.59}_{-2.56}$&
$-0.02^{+0.04}_{-0.04}$\\
\\[-1.5ex]

$\mathrm{FoG \; model} \rightarrow exp$&
$0.33^{+1.87}_{-1.77}$&
$-0.01^{+0.01}_{-0.01}$&
$-0.28^{+2.02}_{-2.40}$&
$-0.01^{+0.01}_{-0.01}$&
$0.91^{+4.68}_{-5.09}$&
$-0.02^{+0.03}_{-0.03}$&
$0.22^{+2.61}_{-2.56}$&
$0.00^{+0.01}_{-0.01}$\\
\\[-1.5ex]

$\mathrm{FoG \; model} \rightarrow gauss$&
$0.33^{+1.86}_{-1.77}$&
$0.00^{+0.00}_{-0.00}$&
$-0.29^{+2.01}_{-2.39}$&
$0.00^{+0.00}_{-0.00}$&
$0.92^{+4.69}_{-5.12}$&
$0.00^{+0.00}_{-0.00}$&
$0.22^{+2.62}_{-2.57}$&
$-0.00^{+0.00}_{-0.00}$\\
\\[-1.5ex]

$\mathrm{No \; Priors}$&
$0.43^{+1.96}_{-1.87}$&
$0.04^{+0.19}_{-0.11}$&
$-0.29^{+2.04}_{-2.40}$&
$-0.01^{+0.35}_{-0.33}$&
$0.97^{+4.95}_{-5.12}$&
$-0.02^{+0.73}_{-0.52}$&
$0.21^{+2.79}_{-2.51}$&
$0.02^{+0.47}_{-0.33}$\\
\\[-1.5ex]

$\mathrm{Only} \; \beta \; \mathrm{prior}$&
$0.36^{+1.93}_{-1.80}$&
$0.00^{+0.13}_{-0.07}$&
$-0.29^{+2.00}_{-2.37}$&
$-0.00^{+0.07}_{-0.06}$&
$0.96^{+4.75}_{-5.01}$&
$0.01^{+0.20}_{-0.15}$&
$0.27^{+2.63}_{-2.52}$&
$0.01^{+0.12}_{-0.06}$\\
\\[-1.5ex]

$\mathrm{Only} \; \log (B_0^2) \; \mathrm{prior}$&
$0.36^{+1.98}_{-1.73}$&
$0.01^{+0.16}_{-0.05}$&
$-0.27^{+2.04}_{-2.34}$&
$-0.00^{+0.32}_{-0.30}$&
$0.87^{+4.96}_{-5.14}$&
$-0.01^{+0.59}_{-0.50}$&
$0.21^{+2.61}_{-2.52}$&
$0.00^{+0.43}_{-0.34}$\\
\\[-1.5ex]

$\mathrm{P_{pt}(k) \; floating \;} \beta$&
$-0.24^{+1.90}_{-1.71}$&
$-0.58^{+0.21}_{-0.18}$&
$-0.26^{+2.05}_{-2.41}$&
$-0.04^{+0.25}_{-0.21}$&
$0.25^{+4.80}_{-4.96}$&
$-0.70^{+0.58}_{-0.43}$&
$-0.31^{+2.68}_{-2.51}$&
$-0.54^{+0.32}_{-0.26}$\\
\\[-1.5ex]

$\mathrm{P_{pt}(k) \;} \beta=0.0$&
$-0.25^{+1.90}_{-1.71}$&
$-0.58^{+0.22}_{-0.18}$&
$-0.25^{+2.06}_{-2.52}$&
$-0.07^{+0.25}_{-0.25}$&
$0.17^{+4.85}_{-5.11}$&
$-0.74^{+0.59}_{-0.48}$&
$-0.27^{+2.75}_{-2.58}$&
$-0.51^{+0.36}_{-0.30}$\\
\\[-1.5ex]

\hline

\end{tabular}
\end{table*}

\begin{table*}
\caption{Fitting results from mocks from numerous variations of our fiducial fitting methodology for DR10 mock galaxy catalogs post-reconstruction without covariance corrections for the overlapping mock regions.  First line shows the median and 16th and 84th percentiles for the $\alpha, \epsilon, \alpha_{||, \perp}$. The rest of the lines shows the median bias and median  variations $\Delta v=v_{i}-v_{f}$ with their corresponding 16th and 84th percentiles. The median bias $\widetilde{b}$, median variations $\widetilde{\Delta v}$, and percentiles are multiplied by 100.}
\label{tab:dr10rec}

\begin{tabular}{@{}lrrrrrrrr}

\hline
Model&
$\widetilde{\alpha}$&
-&
$\widetilde{\epsilon}$&
-&
$\widetilde{\alpha_{\parallel}}$&
-&
$\widetilde{\alpha_{\perp}}$&
-\\


\hline
\multicolumn{9}{c}{DR10 Post-Reconstruction}\\
\hline
\\[-1.5ex]

Fiducial&
$1.0012^{+0.0123}_{-0.0121}$&-&
$0.0007^{+0.0154}_{-0.0138}$&-&
$1.0021^{+0.0367}_{-0.0283}$&-&
$1.0000^{+0.020}_{-0.019}$&-\\
\hline
Model&
$\widetilde{b_\alpha}$&
$\widetilde{\Delta\alpha}$&
$\widetilde{b_\epsilon}$&
$\widetilde{\Delta\epsilon}$&
$\widetilde{b_{\parallel}}$&
$\widetilde{\Delta\alpha_{\parallel}}$&
$\widetilde{b_{\perp}}$&
$\widetilde{\Delta\alpha_{\perp}}$\\
\hline
$30<r<200\mathrm{Mpc}/h$&
$0.28^{+1.25}_{-1.19}$&
$0.15^{+0.17}_{-0.14}$&
$-0.12^{+1.48}_{-1.39}$&
$0.02^{+0.15}_{-0.10}$&
$0.49^{+3.40}_{-2.65}$&
$0.17^{+0.40}_{-0.26}$&
$0.13^{+1.97}_{-1.92}$&
$0.10^{+0.16}_{-0.14}$\\
\\[-1.5ex]

$\mathrm{2-term} \; A_\ell(r)$&
$0.06^{+1.22}_{-1.15}$&
$-0.05^{+0.07}_{-0.10}$&
$-0.33^{+1.49}_{-1.24}$&
$0.27^{+0.19}_{-0.17}$&
$0.73^{+3.34}_{-2.72}$&
$0.47^{+0.35}_{-0.28}$&
$-0.33^{+1.85}_{-1.79}$&
$-0.32^{+0.19}_{-0.29}$\\
\\[-1.5ex]

$\mathrm{4-term} \; A_\ell(r)$&
$-0.04^{+1.23}_{-1.18}$&
$-0.14^{+0.14}_{-0.15}$&
$0.05^{+1.62}_{-1.39}$&
$-0.12^{+0.14}_{-0.15}$&
$-0.09^{+3.43}_{-3.02}$&
$-0.39^{+0.36}_{-0.35}$&
$0.02^{+1.97}_{-1.89}$&
$-0.01^{+0.15}_{-0.17}$\\
\\[-1.5ex]

$\spar=4\; \& \; \sperp=2(\mathrm{Mpc}/h)$&
$0.12^{+1.24}_{-1.24}$&
$-0.01^{+0.04}_{-0.04}$&
$-0.13^{+1.53}_{-1.40}$&
$0.05^{+0.06}_{-0.06}$&
$0.30^{+3.56}_{-2.82}$&
$0.09^{+0.11}_{-0.12}$&
$-0.06^{+2.07}_{-1.89}$&
$-0.06^{+0.07}_{-0.07}$\\
\\[-1.5ex]

$\Sigma_s \rightarrow 3.0$&
$0.15^{+1.20}_{-1.15}$&
$0.04^{+0.06}_{-0.05}$&
$-0.15^{+1.57}_{-1.41}$&
$0.07^{+0.09}_{-0.08}$&
$0.33^{+3.72}_{-2.81}$&
$0.18^{+0.21}_{-0.22}$&
$-0.02^{+2.02}_{-1.90}$&
$-0.03^{+0.05}_{-0.05}$\\
\\[-1.5ex]

$\mathrm{FoG \; model} \rightarrow exp$&
$0.11^{+1.23}_{-1.21}$&
$-0.01^{+0.01}_{-0.01}$&
$-0.06^{+1.55}_{-1.38}$&
$-0.01^{+0.01}_{-0.01}$&
$0.20^{+3.68}_{-2.88}$&
$-0.02^{+0.04}_{-0.04}$&
$0.00^{+2.03}_{-1.87}$&
$0.00^{+0.01}_{-0.01}$\\
\\[-1.5ex]

$\mathrm{FoG \; model} \rightarrow gauss$&
$0.12^{+1.23}_{-1.21}$&
$0.00^{+0.00}_{-0.00}$&
$-0.07^{+1.54}_{-1.38}$&
$0.00^{+0.00}_{-0.00}$&
$0.22^{+3.67}_{-2.83}$&
$0.00^{+0.00}_{-0.00}$&
$0.00^{+2.02}_{-1.88}$&
$-0.00^{+0.00}_{-0.00}$\\
\\[-1.5ex]

$\mathrm{No} \; \mathrm{priors}$&
$0.15^{+1.24}_{-1.22}$&
$0.01^{+0.07}_{-0.05}$&
$-0.06^{+1.60}_{-1.43}$&
$-0.02^{+0.25}_{-0.19}$&
$0.26^{+3.82}_{-3.00}$&
$-0.05^{+0.53}_{-0.37}$&
$0.02^{+2.12}_{-1.93}$&
$0.00^{+0.22}_{-0.20}$\\
\\[-1.5ex]

$\mathrm{Only} \; \beta \; \mathrm{prior}$&
$0.12^{+1.22}_{-1.22}$&
$-0.02^{+0.04}_{-0.03}$&
$-0.07^{+1.58}_{-1.41}$&
$0.00^{+0.03}_{-0.04}$&
$0.21^{+3.75}_{-2.93}$&
$-0.00^{+0.07}_{-0.09}$&
$-0.00^{+2.04}_{-1.88}$&
$-0.00^{+0.04}_{-0.05}$\\
\\[-1.5ex]

$\mathrm{Only} \; B_0 \mathrm{prior}$&
$0.15^{+1.25}_{-1.20}$&
$0.01^{+0.06}_{-0.02}$&
$-0.07^{+1.57}_{-1.43}$&
$-0.02^{+0.24}_{-0.17}$&
$0.25^{+3.76}_{-2.89}$&
$-0.04^{+0.49}_{-0.32}$&
$0.01^{+2.11}_{-1.91}$&
$0.01^{+0.21}_{-0.20}$\\
\\[-1.5ex]

$P_{pt}(k) \mathrm{floating \;} \beta$&
$0.14^{+1.23}_{-1.23}$&
$0.01^{+0.04}_{-0.04}$&
$-0.05^{+1.54}_{-1.37}$&
$-0.02^{+0.05}_{-0.06}$&
$0.26^{+3.63}_{-3.00}$&
$-0.02^{+0.13}_{-0.14}$&
$0.03^{+2.02}_{-1.87}$&
$0.03^{+0.05}_{-0.05}$\\
\\[-1.5ex]

$P_{pt}(k)\; \beta=0.0$&
$0.12^{+1.25}_{-1.25}$&
$-0.00^{+0.05}_{-0.05}$&
$-0.05^{+1.54}_{-1.36}$&
$-0.01^{+0.09}_{-0.11}$&
$0.28^{+3.55}_{-2.98}$&
$-0.02^{+0.19}_{-0.25}$&
$0.01^{+2.04}_{-1.89}$&
$0.02^{+0.11}_{-0.10}$\\
\\[-1.5ex]

\hline

\end{tabular}

\end{table*}

\begin{table*}
\caption{Fitted errors from mock from numerous variations of our fiducial fitting methodology for DR10 mock galaxy catalogs pre-reconstruction without covariance corrections for the overlapping mock regions. The columns shows the median and 16th and 84th percentiles of the $\sigma_{\alpha,\epsilon, ||, \perp}$ and variations $\Delta v=v_{i}-v_{f}$. Except for the fiducial case all quantities are multiplied by 100.}
\label{tab:sigmadr10norec}

\begin{tabular}{@{}lrrrrrrrr}

\hline
Model&
$\widetilde{\sigma_{\alpha}}$&
$\widetilde{\Delta \sigma_{\alpha}}$&
$\widetilde{\sigma_{\epsilon}}$&
$\widetilde{\Delta \sigma_{\epsilon}}$&
$\widetilde{\sigma_{\alpha_{\parallel}}}$&
$\widetilde{\Delta \sigma_{\alpha_{\parallel}}}$&
$\widetilde{\sigma_{\alpha_{\perp}}}$&
$\widetilde{\Delta \sigma_{\alpha_{\perp}}}$\\

\hline
\multicolumn{9}{c}{DR10 Pre-Reconstruction}\\
\hline
\\[-1.5ex]

Fiducial&
$0.0196^{+0.0066}_{-0.0037}$&-&
$0.0249^{+0.0100}_{-0.0048}$&-&
$0.0574^{+0.0271}_{-0.0127}$&-&
$0.0269^{+0.0059}_{-0.0034}$&-\\
\\[-1.5ex]

$30<r<200\mathrm{Mpc}/h$&
$2.00^{+0.82}_{-0.40}$&
$0.02^{+0.15}_{-0.08}$&
$2.59^{+1.22}_{-0.52}$&
$0.06^{+0.29}_{-0.13}$&
$6.08^{+3.00}_{-1.45}$&
$0.15^{+0.77}_{-0.37}$&
$2.72^{+0.58}_{-0.39}$&
$0.00^{+0.08}_{-0.05}$\\
\\[-1.5ex]

$\mathrm{2-term} \; A_\ell(r)$&
$1.95^{+0.63}_{-0.33}$&
$0.00^{+0.08}_{-0.09}$&
$2.39^{+0.94}_{-0.43}$&
$-0.07^{+0.08}_{-0.16}$&
$5.47^{+2.63}_{-1.10}$&
$-0.16^{+0.18}_{-0.39}$&
$2.66^{+0.55}_{-0.35}$&
$-0.03^{+0.07}_{-0.10}$\\
\\[-1.5ex]

$\mathrm{4-term} \; A_\ell(r)$&
$1.95^{+0.69}_{-0.36}$&
$0.01^{+0.05}_{-0.06}$&
$2.54^{+1.03}_{-0.49}$&
$0.05^{+0.12}_{-0.10}$&
$5.85^{+2.73}_{-1.31}$&
$0.08^{+0.28}_{-0.19}$&
$2.72^{+0.60}_{-0.35}$&
$0.03^{+0.06}_{-0.05}$\\
\\[-1.5ex]

$\mathrm{Fixed \;} \beta=0.4$&
$1.93^{+0.64}_{-0.36}$&
$-0.01^{+0.01}_{-0.03}$&
$2.45^{+0.96}_{-0.45}$&
$-0.03^{+0.02}_{-0.03}$&
$5.62^{+2.58}_{-1.16}$&
$-0.06^{+0.11}_{-0.14}$&
$2.66^{+0.57}_{-0.32}$&
$-0.03^{+0.04}_{-0.04}$\\
\\[-1.5ex]

$\spar=\sperp=8\mathrm{Mpc}/h$&
$1.89^{+0.64}_{-0.33}$&
$-0.04^{+0.07}_{-0.09}$&
$2.38^{+1.10}_{-0.48}$&
$-0.09^{+0.17}_{-0.14}$&
$5.19^{+2.93}_{-1.25}$&
$-0.52^{+0.38}_{-0.38}$&
$2.93^{+0.65}_{-0.36}$&
$0.23^{+0.14}_{-0.09}$\\
\\[-1.5ex]

$\Sigma_s \rightarrow 3.0$&
$1.99^{+0.68}_{-0.37}$&
$0.03^{+0.02}_{-0.01}$&
$2.57^{+1.00}_{-0.49}$&
$0.07^{+0.04}_{-0.03}$&
$5.94^{+2.69}_{-1.26}$&
$0.20^{+0.11}_{-0.07}$&
$2.70^{+0.59}_{-0.35}$&
$0.02^{+0.02}_{-0.02}$\\
\\[-1.5ex]

$\mathrm{FoG \; model} \rightarrow exp$&
$1.96^{+0.65}_{-0.37}$&
$-0.00^{+0.00}_{-0.00}$&
$2.47^{+1.01}_{-0.47}$&
$-0.01^{+0.00}_{-0.01}$&
$5.71^{+2.71}_{-1.27}$&
$-0.03^{+0.01}_{-0.01}$&
$2.69^{+0.59}_{-0.35}$&
$-0.00^{+0.00}_{-0.00}$\\
\\[-1.5ex]

$\mathrm{FoG \; model} \rightarrow gauss$&
$1.96^{+0.66}_{-0.37}$&
$0.00^{+0.00}_{-0.00}$&
$2.49^{+1.00}_{-0.48}$&
$0.00^{+0.00}_{-0.00}$&
$5.74^{+2.71}_{-1.27}$&
$0.00^{+0.00}_{-0.00}$&
$2.69^{+0.59}_{-0.34}$&
$0.00^{+0.00}_{-0.00}$\\
\\[-1.5ex]

$\mathrm{No \; Priors}$&
$2.68^{+1.69}_{-0.81}$&
$0.60^{+1.19}_{-0.44}$&
$3.41^{+2.19}_{-1.12}$&
$0.74^{+1.21}_{-0.53}$&
$7.54^{+6.87}_{-2.80}$&
$1.40^{+3.97}_{-1.25}$&
$3.19^{+1.93}_{-0.67}$&
$0.42^{+1.28}_{-0.31}$\\
\\[-1.5ex]

$\mathrm{Only} \; \beta \; \mathrm{prior}$&
$2.19^{+1.41}_{-0.57}$&
$0.17^{+0.67}_{-0.16}$&
$2.78^{+1.82}_{-0.70}$&
$0.18^{+0.87}_{-0.15}$&
$6.61^{+5.14}_{-1.99}$&
$0.53^{+2.63}_{-0.44}$&
$2.77^{+0.79}_{-0.39}$&
$0.08^{+0.20}_{-0.05}$\\
\\[-1.5ex]

$\mathrm{Only} \; \log (B_0^2) \; \mathrm{prior}$&
$2.38^{+1.30}_{-0.60}$&
$0.31^{+0.77}_{-0.22}$&
$3.07^{+1.80}_{-0.87}$&
$0.48^{+0.87}_{-0.33}$&
$6.60^{+5.67}_{-1.99}$&
$0.78^{+2.52}_{-0.78}$&
$3.01^{+1.36}_{-0.58}$&
$0.27^{+0.92}_{-0.22}$\\
\\[-1.5ex]

$\mathrm{P_{pt}(k) \; floating \;} \beta$&
$1.90^{+0.59}_{-0.34}$&
$-0.05^{+0.07}_{-0.09}$&
$2.46^{+1.01}_{-0.46}$&
$-0.02^{+0.11}_{-0.12}$&
$5.47^{+2.72}_{-1.21}$&
$-0.24^{+0.28}_{-0.34}$&
$2.83^{+0.60}_{-0.36}$&
$0.12^{+0.10}_{-0.10}$\\
\\[-1.5ex]

$\mathrm{P_{pt}(k) \;} \beta=0.0$&
$1.88^{+0.57}_{-0.33}$&
$-0.06^{+0.07}_{-0.10}$&
$2.42^{+0.98}_{-0.45}$&
$-0.05^{+0.10}_{-0.13}$&
$5.38^{+2.57}_{-1.16}$&
$-0.33^{+0.24}_{-0.41}$&
$2.82^{+0.59}_{-0.36}$&
$0.11^{+0.10}_{-0.11}$\\
\\[-1.5ex]

\hline

\end{tabular}
\end{table*}

\begin{table*}
\caption{We show the fitting errors from DR10 mock galaxy catalogs post-reconstruction for variations of the fiducial models without covariance corrections for the overlapping mock regions. The columns shows the median and 16th and 84th percentiles of the $\sigma_{\alpha,\epsilon, ||, \perp}$ and variations $\Delta v=v_{i}-v_{f}$. Except for the fiducial case all quantities are multiplied by 100.}
\label{tab:sigmadr10rec}
\begin{tabular}{@{}lrrrrrrrr}

\hline
Model&
$\widetilde{\sigma_{\alpha}}$&
$\widetilde{\Delta \sigma_{\alpha}}$&
$\widetilde{\sigma_{\epsilon}}$&
$\widetilde{\Delta \sigma_{\epsilon}}$&
$\widetilde{\sigma_{\alpha_{\parallel}}}$&
$\widetilde{\Delta \sigma_{\alpha_{\parallel}}}$&
$\widetilde{\sigma_{\alpha_{\perp}}}$&
$\widetilde{\Delta \sigma_{\alpha_{\perp}}}$\\

\hline
\multicolumn{9}{c}{DR10 Post-Reconstruction}\\
\hline
\\[-1.5ex]

Fiducial&
$0.0126^{+0.0028}_{-0.0019}$&-&
$0.0162^{+0.0045}_{-0.0024}$&-&
$0.0361^{+0.0127}_{-0.0068}$&-&
$0.0191^{+0.0035}_{-0.0019}$&-\\
\\[-1.5ex]

$30<r<200\mathrm{Mpc}/h$&
$1.26^{+0.24}_{-0.19}$&
$-0.01^{+0.03}_{-0.04}$&
$1.61^{+0.40}_{-0.22}$&
$-0.01^{+0.05}_{-0.08}$&
$3.60^{+1.08}_{-0.63}$&
$-0.01^{+0.16}_{-0.24}$&
$1.89^{+0.36}_{-0.20}$&
$-0.01^{+0.03}_{-0.03}$\\
\\[-1.5ex]

$\mathrm{2-term} \; A_\ell(r)$&
$1.27^{+0.25}_{-0.19}$&
$0.02^{+0.02}_{-0.04}$&
$1.54^{+0.33}_{-0.21}$&
$-0.07^{+0.06}_{-0.13}$&
$3.51^{+0.97}_{-0.62}$&
$-0.09^{+0.16}_{-0.35}$&
$1.82^{+0.29}_{-0.20}$&
$-0.07^{+0.04}_{-0.07}$\\
\\[-1.5ex]

$\mathrm{4-term} \; A_\ell(r)$&
$1.27^{+0.33}_{-0.21}$&
$-0.00^{+0.05}_{-0.02}$&
$1.67^{+0.52}_{-0.27}$&
$0.03^{+0.13}_{-0.05}$&
$3.67^{+1.52}_{-0.73}$&
$0.03^{+0.33}_{-0.13}$&
$1.94^{+0.36}_{-0.20}$&
$0.03^{+0.04}_{-0.03}$\\
\\[-1.5ex]

$\spar=4\; \& \; \sperp=2(\mathrm{Mpc}/h)$&
$1.27^{+0.28}_{-0.20}$&
$0.00^{+0.02}_{-0.01}$&
$1.63^{+0.45}_{-0.24}$&
$0.01^{+0.02}_{-0.02}$&
$3.67^{+1.30}_{-0.68}$&
$0.07^{+0.06}_{-0.05}$&
$1.87^{+0.35}_{-0.20}$&
$-0.03^{+0.02}_{-0.02}$\\
\\[-1.5ex]

$\Sigma_s \rightarrow 3.0$&
$1.29^{+0.27}_{-0.20}$&
$0.02^{+0.02}_{-0.01}$&
$1.69^{+0.46}_{-0.24}$&
$0.07^{+0.04}_{-0.03}$&
$3.83^{+1.28}_{-0.70}$&
$0.21^{+0.10}_{-0.08}$&
$1.92^{+0.34}_{-0.20}$&
$0.02^{+0.02}_{-0.02}$\\
\\[-1.5ex]

$\mathrm{FoG \; model} \rightarrow exp$&
$1.26^{+0.27}_{-0.19}$&
$-0.00^{+0.00}_{-0.00}$&
$1.61^{+0.45}_{-0.24}$&
$-0.01^{+0.01}_{-0.01}$&
$3.58^{+1.29}_{-0.67}$&
$-0.03^{+0.02}_{-0.01}$&
$1.90^{+0.35}_{-0.19}$&
$-0.00^{+0.00}_{-0.00}$\\
\\[-1.5ex]

$\mathrm{FoG \; model} \rightarrow gauss$&
$1.26^{+0.28}_{-0.19}$&
$0.00^{+0.00}_{-0.00}$&
$1.62^{+0.45}_{-0.24}$&
$0.00^{+0.00}_{-0.00}$&
$3.61^{+1.26}_{-0.68}$&
$0.00^{+0.00}_{-0.00}$&
$1.91^{+0.35}_{-0.19}$&
$0.00^{+0.00}_{-0.00}$\\
\\[-1.5ex]

$\mathrm{No} \; \mathrm{priors}$&
$3.02^{+4.33}_{-1.58}$&
$1.69^{+4.25}_{-1.40}$&
$3.63^{+5.69}_{-1.86}$&
$2.03^{+5.15}_{-1.70}$&
$9.68^{+15.99}_{-5.75}$&
$6.06^{+14.76}_{-5.21}$&
$2.24^{+0.94}_{-0.37}$&
$0.30^{+0.71}_{-0.17}$\\
\\[-1.5ex]

$\mathrm{Only} \; \beta \; \mathrm{prior}$&
$2.53^{+3.87}_{-1.24}$&
$1.19^{+3.79}_{-1.05}$&
$3.18^{+4.98}_{-1.52}$&
$1.48^{+4.54}_{-1.29}$&
$8.33^{+14.16}_{-4.47}$&
$4.60^{+12.88}_{-3.95}$&
$2.05^{+0.78}_{-0.29}$&
$0.07^{+0.52}_{-0.06}$\\
\\[-1.5ex]

$\mathrm{Only} \; B_0 \mathrm{prior}$&
$1.71^{+2.30}_{-0.51}$&
$0.40^{+2.05}_{-0.33}$&
$2.22^{+3.08}_{-0.72}$&
$0.55^{+2.60}_{-0.48}$&
$5.25^{+9.29}_{-2.19}$&
$1.57^{+7.67}_{-1.54}$&
$2.09^{+0.54}_{-0.28}$&
$0.16^{+0.27}_{-0.09}$\\
\\[-1.5ex]

$P_{pt}(k) \mathrm{\;floating \;} \beta$&
$1.26^{+0.27}_{-0.20}$&
$-0.01^{+0.01}_{-0.01}$&
$1.57^{+0.46}_{-0.24}$&
$-0.05^{+0.02}_{-0.02}$&
$3.49^{+1.33}_{-0.68}$&
$-0.11^{+0.06}_{-0.05}$&
$1.88^{+0.35}_{-0.19}$&
$-0.03^{+0.02}_{-0.01}$\\
\\[-1.5ex]

$P_{pt}(k)\; \beta=0.0$&
$1.22^{+0.24}_{-0.19}$&
$-0.03^{+0.01}_{-0.04}$&
$1.52^{+0.38}_{-0.21}$&
$-0.09^{+0.04}_{-0.10}$&
$3.38^{+0.94}_{-0.56}$&
$-0.21^{+0.14}_{-0.34}$&
$1.86^{+0.33}_{-0.19}$&
$-0.05^{+0.04}_{-0.04}$\\
\\[-1.5ex]

\hline

\end{tabular}

\end{table*}
 
\end{document}